\documentclass[12pt,twoside]{report}

\newcommand{\reporttitle}{Dynamic Graph Communication for Decentralised Multi-Agent Reinforcement Learning \\[3pt]}

\newcommand{\reportauthor}{Ben McClusky}
\newcommand{\supervisor}{Professor Kin Leung}
\newcommand{\degreetype}{MSc Computing (Artificial Intelligence and Machine Learning)}

\usepackage[a4paper,hmargin=2.8cm,vmargin=2.0cm,includeheadfoot]{geometry}
\usepackage{textpos}
\usepackage[numbers]{natbib}
\usepackage{tabularx,longtable,multirow,caption}
\usepackage{fancyhdr} 
\usepackage{url} 
\usepackage[english]{babel}
\usepackage{amsmath}
\usepackage{amssymb}
\usepackage{graphicx}
\usepackage{dsfont}
\usepackage{epstopdf} 
\usepackage{backref} 
\usepackage{array}
\usepackage{latexsym}
\usepackage[ruled,vlined]{algorithm2e}
\usepackage{amsfonts}
\usepackage[pagebackref,hypertexnames=false,colorlinks]{hyperref} 

\usepackage{setspace}

\hypersetup{pdftitle={},
  pdfsubject={}, 
  pdfauthor={},
  pdfkeywords={}, 
  pdfstartview=FitH,
  pdfpagemode={UseOutlines},
  bookmarksnumbered=true, bookmarksopen=true, colorlinks,
    citecolor=black,%
    filecolor=black,%
    linkcolor=black,%
    urlcolor=black}

\usepackage[all]{hypcap}

\usepackage{float}
\newtheorem{definition}{Definition} 
\usepackage{booktabs} 
\usepackage{makecell, multirow} 
\usepackage{subfig}
\usepackage{longtable} 
\usepackage{xcolor, colortbl}

\newcolumntype{C}[1]{>{\centering\arraybackslash}p{#1}}
\newcolumntype{M}[1]{>{\centering\arraybackslash}m{#1}}

\def\@makechapterhead#1{%
  \vspace*{10\p@}%
  {\parindent \z@ \raggedright \sffamily
    \interlinepenalty\@M
    \Huge\bfseries \thechapter \space\space #1\par\nobreak
    \vskip 30\p@
  }}

\def\@makeschapterhead#1{%
  \vspace*{10\p@}%
  {\parindent \z@ \raggedright
    \sffamily
    \interlinepenalty\@M
    \Huge \bfseries  #1\par\nobreak
    \vskip 30\p@
  }}

\allowdisplaybreaks

\renewcommand{\vec}[1]{{\boldsymbol{{#1}}}} 

\date{September 2024}

\begin{document}

\begin{titlepage}

\newcommand{\HRule}{\rule{\linewidth}{0.5mm}} 


\includegraphics[width = 4cm]{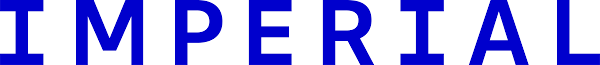}\\[0.5cm] 

\center 


\textsc{\Large Imperial College London}\\[0.5cm] 
\textsc{\large Department of Computing}\\[0.5cm] 


\HRule \\[0.0cm]
{ \huge \bfseries \reporttitle} 
\HRule \\[1.5cm]
 

\begin{minipage}[t]{0.4\textwidth}
\begin{flushleft} \large
\emph{Author:}\\
\reportauthor 
\end{flushleft}
\end{minipage}
~
\begin{minipage}[t]{0.4\textwidth}
\begin{flushright} \large
\emph{Supervisor:} \\
\supervisor  \\ 
\vspace{0.5cm} 
\emph{Second Marker:} \\
Professor Andrew Davison 
\end{flushright}
\end{minipage}\\[4cm]

\vfill 
Submitted in partial fulfillment of the requirements for the MSc degree in
\degreetype~of Imperial College London\\[0.5cm]

\makeatletter
\@date 
\makeatother

\end{titlepage}

\pagenumbering{roman}
\clearpage{\pagestyle{empty}\cleardoublepage}
\setcounter{page}{1}
\pagestyle{fancy}

\section*{\centering Abstract}
The deployment of decentralised multi-agent systems in dynamic networks introduces substantial challenges in communication and coordination. Agents must continuously adapt their communication strategies as the network topology evolves, ensuring that only the most relevant information is shared to strengthen the network's collective knowledge and to support efficient decision-making.

\vspace{12pt}

This thesis presents a novel communication framework for decentralised multi-agent systems in dynamic networks. Building on the work of Weil et al. \cite{weil2024towards} on decentralised multi-agent reinforcement learning (MARL) in static networks, this research extends their recurrent message-passing model to dynamic environments, optimising communication efficiency to improve decision-making while reducing communication overhead.

\vspace{12pt}

Key contributions of this work include transforming a static network packet routing environment into a dynamic network by introducing node failures, integrating a Graph Attention Network (GAT) layer into the recurrent message-passing framework, and developing a novel multi-round communication targeting mechanism. To the best of our knowledge, this is the first successful application of an attention-based aggregation mechanism in a sparse-reward, dynamic network packet routing environment using only reinforcement learning. The proposed communication system achieved substantial improvements in routing performance, with 9.5\% higher rewards and 6.4\% lower communication overhead compared to the original recurrent message-passing system. These results demonstrate the potential of the system for more efficient and scalable routing in dynamic, real-world networks.

\vspace{12pt}

This study provides a thorough background on reinforcement learning and multi-agent systems, along with a comprehensive literature review of communication-based MARL systems. It outlines the design process in detail, including descriptions of the testing environments and an evaluation of the system’s performance compared to leading MARL approaches. Ablation studies were performed to isolate the specific contributions of the GAT layer and the multi-round targeting mechanism. Additionally, the research addresses the ethical and legal considerations surrounding the deployment of such systems in critical infrastructure and military applications, as well as the limitations of the current work and potential directions for future research.

\vspace{90pt}

\textbf{Keywords:} Multi-Agent Reinforcement Learning, Decentralised Systems, Dynamic Networks, Graph Neural Networks, Graph Attention Networks, Multi-Round Communication, Network Packet Routing

\section*{\centering Acknowledgements}
I would like to sincerely thank Professor Kin Leung for introducing me to the exciting world of multi-agent learning and for his support, guidance, and valuable feedback throughout this project.

\vspace{12pt}

I would also like to extend my heartfelt thanks to my friends and family for their constant support and encouragement during this process.

\clearpage{\pagestyle{empty}\cleardoublepage}

\fancyhead[RE,LO]{\sffamily {Table of Contents}}
\tableofcontents

\clearpage{\pagestyle{empty}\cleardoublepage}
\pagenumbering{arabic}
\setcounter{page}{1}
\fancyhead[LE,RO]{\slshape \rightmark}
\fancyhead[LO,RE]{\slshape \leftmark}

\renewcommand{\arraystretch}{1.5} 

\chapter{Introduction}
\label{cha:introduction}

\section{Motivation Of The Thesis}

The accelerating deployment of intelligent and autonomous systems reveals the critical need to manage interactions between multiple autonomous agents. Multi-agent systems are expected to operate at nearly every level of society, with many systems requiring agents to cooperate, compete, and coordinate their actions to achieve complex goals that would be unattainable by a single agent.

\vspace{12pt}

For these systems to function effectively on a large scale, agents must be capable of making independent decisions, with communication typically constrained to interactions with nearby agents. The challenge, therefore, lies in how to efficiently propagate information among agents, ensuring that each agent can make informed decisions based on the collective knowledge of the network. 

\vspace{12pt}

Networks in real-world applications, such as network packet routing, are rarely static, with nodes often failing or moving positions. This adds another layer of complexity, requiring agents to not only determine what information to share but also to dynamically identify the most appropriate agents with whom to share it.

\vspace{12pt}

In these large, dynamic networks, efficient communication is crucial. Agents must prioritise sharing only relevant information to avoid overwhelming the receiving agents and to conserve communication resources. This thesis introduces a novel communication system that adapts to changing topologies and communication needs, facilitating more effective collaboration among agents and enabling them to achieve complex objectives more efficiently.

\clearpage 

\section{Research Aim and Objectives}

This project aims to extend the leading decentralised, networked multi-agent reinforcement learning approach for packet routing in static networks to dynamic networks, thereby better reflecting the challenges encountered in real-world applications. Specifically, this research seeks to address the following questions:

\begin{itemize}
    \item \textbf{Q1:} What is the most effective method for identifying critical information from neighbouring agents in a large-scale, dynamic system?
    \item \textbf{Q2:} How can communication be dynamically controlled to reduce overhead within a multi-round communication system?
    \item \textbf{Q4:} How can a dynamic communication control system be trained end-to-end using only reinforcement learning?
    \item \textbf{Q3:} Does more effective and efficient communication lead to improved packet routing decisions?
\end{itemize}

\section{Contributions}

This thesis presents the following key contributions:

\begin{enumerate}
    \item Creation of a dynamic network packet routing environment, building upon the static routing environment developed by Weil et al. \cite{weil2024towards}. This environment features randomised node failures with varying probabilities and durations, increasing its resemblance to real-world network scenarios.

    \item Demonstrated that incorporating a single Graph Attention Network (GAT) layer \cite{velivckovic2017graph} as an aggregation mechanism within a recurrent message-passing system improves graph representations and enhances routing performance by 4.8\% in dynamic environments.

    \item To the best of our knowledge, this is the first work to successfully train an attention-based aggregation mechanism in a sparse-reward, dynamic network packet routing environment using only reinforcement learning.

   \item Designed a novel targeting mechanism that leverages multi-round communication, using previously received states from neighbouring agents to identify the most relevant agents for communication in subsequent rounds.

    \item Demonstrated that when isolated the proposed targeting mechanism reduced communication overhead by 5.4\% and improved routing performance by 9.1\% when trained online in dynamic networks.

    \item Compared the proposed communication system with leading communication-based MARL approaches in a dynamic routing environment, achieving 9.5\% better performance while using 6.4\% less communication than the base recurrent message-passing system.

\end{enumerate}

\clearpage 

\section{Research Outline}

\hyperref[cha:introduction]{Chapter \ref{cha:introduction}} discusses the importance of efficient communication in decentralised multi-agent systems operating in dynamic environments. It outlines the research aims, objectives, key contributions, and ethical and legal considerations, and concludes by defining essential terminology used throughout the thesis.

\vspace{12pt}

\hyperref[cha:background]{Chapter \ref{cha:background}} covers the core concepts of reinforcement learning, multi-agent reinforcement learning, and network packet routing. It includes a comprehensive literature review of MARL approaches, with a focus on communication protocols and their applications in network packet routing.

\vspace{12pt}

\hyperref[cha:system-design]{Chapter \ref{cha:system-design}} details the design and development of the proposed communication system, covering the objectives, assumptions, constraints, key parameters, and the training process crucial to the system's effectiveness.

\vspace{12pt}

\hyperref[cha:environment-implementation]{Chapter \ref{cha:environment-implementation}} describes the implementation of testing environments used to evaluate the proposed communication system. It includes the graph generation process, observation spaces for both node and agent models, as well as the experimental configurations and evaluation metrics.

\vspace{12pt}

\hyperref[cha:aggregation-mechanism]{Chapter \ref{cha:aggregation-mechanism}} assesses the isolated impact of GAT as an aggregation mechanism within the distributed message-passing system. It discusses the motivations, related work, methodology, challenges encountered, and results across the two environments.

\vspace{12pt}

\hyperref[cha:iteration-controller]{Chapter \ref{cha:iteration-controller}} examines the isolated impact of the Iteration Controller within the distributed message-passing system. It covers the motivations, related work, methodology, challenges encountered, and results in both environments.

\vspace{12pt}

\hyperref[cha:integrated-dynamic-communication]{Chapter \ref{cha:integrated-dynamic-communication}} evaluates the integrated communication system, combining the GAT and Iteration Controller, against leading MARL communication approaches within the dynamic routing environment.

\vspace{12pt}

Chapter \ref{cha:conclusion} concludes the thesis by summarising the key contributions, discussing the study's limitations, and suggesting potential directions for future work.

\section{Ethical and Legal Considerations}

The development of dynamic decentralised multi-agent systems, as explored in this research, raises significant ethical and legal challenges. These systems have potential applications in critical infrastructure, such as communication networks and energy grids, as well as in military contexts, including autonomous drones and missile systems. While the ability of intelligent autonomous systems to adapt to changing conditions offers substantial benefits, it is crucial to incorporate manual safeguards such as human oversight or stop conditions to prevent deviations that could lead to catastrophic outcomes.

\vspace{12pt}

Additionally, the decentralised nature of these systems raises concerns about accountability, as it may become difficult to pinpoint responsibility for system failures or harmful actions. Furthermore, privacy risks are inherent in the data-sharing mechanisms of multi-agent systems, where sensitive information could be intercepted or misused. Security vulnerabilities must also be addressed, as malicious actors may exploit weaknesses in the system’s communication protocols to compromise critical infrastructure.

\vspace{12pt}

Ensuring transparency and accountability in the deployment of these systems is essential to prioritise human safety, maintain public trust, and comply with legal and ethical standards across all sectors.

\section{Terminology Used Throughout The Report}

\begin{itemize}

    \item \textbf{Markov Decision Process (MDP):} A mathematical framework used to model decision-making in environments with stochastic outcomes, defined by a set of states, actions, transition probabilities, rewards, and a discount factor.
    
    \item \textbf{Partially Observable Markov Decision Process (POMDP):} A generalisation of MDPs where agents have limited visibility of the environment’s state, making decisions based on observations rather than full knowledge of the state.
    
    \item \textbf{Decentralised Partially Observable Markov Decision Process (Dec-POMDP):} An extension of POMDPs where multiple agents operate based on local observations without centralised control, typically in distributed environments where full observability is not possible.

    \item \textbf{Centralised Training with Decentralised Execution (CTDE):} A training paradigm where agents are trained with centralised coordination and global knowledge but execute policies independently during deployment, often in environments with dynamic network topologies.
    
    \item \textbf{Distributed:} A system where tasks and computations are performed by multiple agents or nodes, each operating independently while collectively contributing to a shared goal. "Distributed" is often used interchangeably with "decentralised.

    \item \textbf{Communication Overhead:} The additional cost in terms of time, bandwidth, or computational resources associated with exchanging information between agents. 

    \item \textbf{Experience Replay:} A technique used in RL where past experiences are stored in a memory buffer and sampled randomly for training, helping to break correlations between consecutive experiences.
    
    \item \textbf{Epsilon-Greedy Strategy:} A method used in RL to balance exploration and exploitation, where the agent selects random actions with probability \(\epsilon\) and the best-known action otherwise.

    \item \textbf{Graph Neural Network (GNN):} A class of neural networks designed to operate on graph-structured data, capable of capturing dependencies between nodes in graphs through message passing and aggregation mechanisms, often useful in analysing network topologies.

    \item \textbf{Recurrent Neural Network (RNN):} A type of neural network where connections between nodes form a directed graph along a temporal sequence, enabling the network to capture dynamic temporal behaviour.

    \item \textbf{Gated Recurrent Unit (GRU):} A type of recurrent neural network that is simpler than an LSTM, often used in sequence processing tasks due to its computational efficiency.
    
    \item \textbf{Multi-Layer Perceptron (MLP):} A class of feedforward artificial neural networks consisting of multiple layers of nodes. Each node in a layer is fully connected to the nodes in the next layer.

    \item \textbf{Unroll Depth:} The number of consecutive time steps that a recurrent neural network processes in sequence before making a prediction or taking an action.

    \item \textbf{Multi-Head Attention (MHA):} A mechanism in neural networks that applies attention functions multiple times in parallel, each with different learned parameters, to capture different aspects of the input information.

    \item \textbf{Soft Attention:} A type of attention mechanism that assigns a probability distribution over all possible input elements, allowing the model to attend to all elements but with varying degrees of focus.

    \item \textbf{Hard Attention:} A type of attention mechanism that selects a subset of input elements to focus on, rather than distributing attention over all elements. It is more computationally efficient but harder to train.

    \item \textbf{Software-Defined Networking (SDN):} A networking architecture that decouples the control plane from the data plane, allowing for more flexible and dynamic management of network traffic and routing decisions.

\end{itemize}

\chapter{Background \& Literature Review}
\label{cha:background}
\section{Reinforcement Learning}
\label{sec:reinforcement-learning}

The origins of reinforcement learning (RL) trace back to B.F. Skinner's behaviourist theory of learning \cite{skinner1938behavior}. This theory suggests that learning is a process of conditioning an animal's behaviour within an environment of stimuli, rewards, and punishments. RL extends this concept to computational agents, allowing computers to emulate animal learning processes. This powerful mechanism has proven exceptionally effective in complex sequential decision-making tasks, achieving superhuman performance in both board games \cite{silver2016mastering} and computer games \cite{mnih2013playing}, whilst also making significant strides in real-world applications, particularly in fields such as robotics \cite{kober2013reinforcement}, autonomous driving \cite{zhang2023high}, and natural language processing \cite{christiano2017deep}.

\begin{figure}[H]
    \centering
    \includegraphics[width=0.75\linewidth]{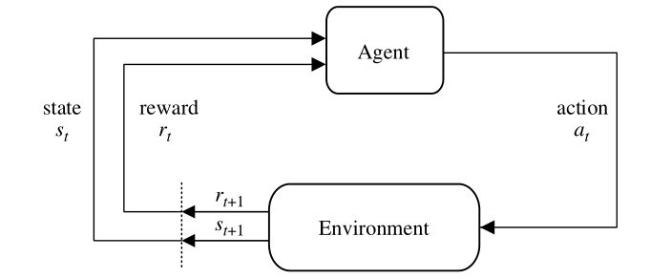}
    \caption{Agent-Environment interaction within a Markov Decision Process \cite{sutton2018reinforcement}: The agent receives the current state $s_t$ from the environment and takes an action $a_t$. The environment then provides a reward $r_t$ and the next state $s_{t+1}$.}
    \label{fig:MDP-Agent-Enviroment}
\end{figure}

At its core, RL is the interaction between an agent and its environment. The agent observes the current state of the environment and takes an action according to its learned policy (mapping of state to action). The environment then transitions to a new state and emits a scalar reward signal which reflects the consequences of the action taken. This is an iterative process, with the agent continuously learning how to improve its policy to maximise the cumulative reward over time.

\subsubsection{Markov Decision Process}

The environment is typically modelled as an infinite-horizon discounted Markov Decision Process (MDP) \cite{bellman1957markovian}. An MDP provides a mathematical framework for stochastic sequential decision-making, where the Markov property ensures that the future state depends only on the current state, not on the sequence of previous events. As noted by Puterman \cite{puterman2014markov}, an optimal policy in any MDP is deterministic and Markovian. An MDP is formally defined by the quintuple \( (S, A, R, P, \gamma) \).

\begin{itemize}
    \item \textbf{State Space \((S)\)}: A finite set of possible states.
    \item \textbf{Action Space \((A)\)}: A finite set of actions available in each state \(s\).
    \item \textbf{Transition Probability \((P)\)}: The probability \(P(s_{t+1}|s_t,a_t)\) that action \(a_t\) in state \(s_t\) will lead to state \(s_{t+1}\).
    \item \textbf{Reward Function \((R)\)}: The immediate reward \(R(s_t, a_t, s_{t+1})\) for transitioning from \(s_t\) to \(s_{t+1}\) via \(a_t\).
    \item \textbf{Discount Factor \((\gamma)\)}: A factor \(\gamma \in [0, 1]\) for weighting future rewards.
\end{itemize}

\subsubsection{Value Based Reinforcement Learning}

The goal of a RL agent is to find the optimal policy \( \pi^* \) that maximises the expected cumulative reward. Typically, this involves computing the optimal value function \( V^*(s) \) for each state \( s \), which guides a greedy policy to select actions that maximise the expected value of the next state. Using the Markovian assumption, the value \( V^\pi(s) \) can be decomposed into the expected immediate reward \( R(s,a) \) and the expected discounted value of successor states \( \gamma V^\pi(s') \), where \( s' \) is the next state. In practice, it is often more effective to calculate action values \( Q^*(s, a) \), representing the value of taking a specific action \( a \) in state \( s \).

\begin{equation}
V^\pi(s) = \mathbb{E}_\pi[R_t \mid S_t = s] = \sum_{a} \pi(s, a) \sum_{s'} P^a_{ss'} \left[ R^a_{ss'} + \gamma V^\pi(s') \right]
\end{equation}
\vspace{-6pt}
\begin{equation}
Q^\pi(s, a) = \mathbb{E}_\pi[R_t \mid S_t = s, A_t = a] = \sum_{s'} P^a_{ss'} \left[ R^a_{ss'} + \gamma \sum_{a'} \pi(s', a') Q^\pi(s', a') \right]
\end{equation}

Decomposing the value functions formulates the problem as overlapping sub-problems. Bellman proposed solving these recursively via \textit{Dynamic Programming (DP)} to find the optimal state value function \( V^*(s) \) and action value function \( Q^*(s, a) \), resulting in the \textit{Bellman Optimality Equations (BOE)} \cite{bellman1957markovian}.

\begin{equation}
\label{eq: bellman}
V^*(s) = \max_a \sum_{s'} P^a_{ss'} \left[ R^a_{ss'} + \gamma V^*(s') \right] 
\end{equation}

\begin{equation}
Q^*(s, a) = \sum_{s'} P^a_{ss'} \left[ R^a_{ss'} + \gamma \max_{a'} Q^*(s', a') \right]
\end{equation}

DP techniques require complete knowledge of the environment's dynamics, including transition probabilities and reward functions, making them \textit{model-based} methods. However, in many cases, the dynamics are either inaccessible or too complex to model. An alternative is \textit{model-free} methods, which estimate values based on accumulated experience. \textit{Monte Carlo (MC)} methods, sample trajectories through the environment, aggregate returns for each state, and calculate the average reward at the end of an episode. The MC value estimation is given by \( V(s) = \frac{1}{N} \sum_{i=1}^{N} G_i(s) \), where \( N \) is the number of episodes in which state \( s \) is visited, and \( G_i(s) \) is the return (cumulative reward) following the \( i \)-th visit to state \( s \).

\vspace{12pt}

MC methods provide unbiased value estimates but require complete episodes, limiting their use in non-episodic environments. \textit{Temporal Difference (TD)} learning \cite{sutton1988learning} overcomes this by updating values from incomplete episodes through \textit{bootstrapping} value estimates of subsequent states. Through updating from incomplete experience, TD generally results in faster, more stable convergence. However, using bootstrapped estimates of subsequent states introduces a bias. The basic TD error for TD(0) is given by \( \delta_t = R_{t+1} + \gamma V(s_{t+1}) - V(s_t) \), where \( \delta_t \) is the TD error.

\vspace{12pt}

Methods like TD learning that use intermediate value functions for policy optimisation are called \textit{value-based methods}. As an \textit{on-policy} algorithm, TD learning updates the value of the policy the agent follows during exploration, allowing policy improvement based on the states it actually visits. In contrast, \textit{Q-learning} \cite{watkins1989learning}, introduced by Watkins, extends TD learning as an \textit{off-policy} algorithm. Q-learning can learn the value of the optimal policy regardless of the agent's actions by evaluating the best possible action at each state, rather than only considering the actions taken by the agent, offering greater flexibility and robustness in learning the optimal policy. The Q-learning update rule is given by:

\begin{equation}
Q(s, a) \leftarrow Q(s, a) + \alpha \left[ r + \gamma \max_{a'} Q(s', a') - Q(s, a) \right]
\end{equation}

\begin{figure}[H]
    \centering
    \hspace*{-0.2\linewidth} 
    \includegraphics[width=0.7\linewidth]{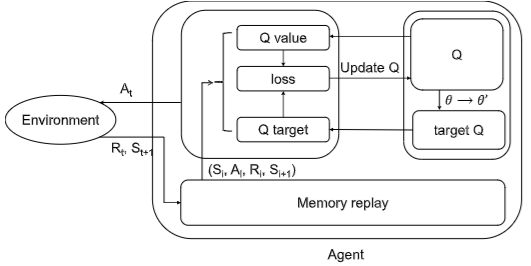}
    \caption{Deep Q-Network (DQN) Architecture \cite{cong2021deep}: The agent utilises experience replay to store and sample experiences for training. The Q-network is trained by minimising the loss between the predicted Q-value and the target Q-value, with periodic updates to the target network's weights ($\theta \rightarrow \theta'$).}
    \label{fig:DQN}
\end{figure}

\addvspace{12pt}

Traditional tabular methods like Q-learning struggle with large state and action spaces due to the substantial memory and computational resources needed, leading to inefficiency and poor generalisation to unseen states. \textit{Deep Q-Networks} (DQNs) \cite{mnih2013playing} overcome these limitations by using neural networks to approximate the Q-value function, eliminating the need for a lookup table and effectively handling large, continuous state spaces, making them suitable for complex, high-dimensional problems.

\addvspace{12pt}

DQNs achieve off-policy learning using two networks: the \textit{behaviour network}, which selects actions, and the \textit{target network}, a periodically updated copy that provides stable Q-value targets, reducing the risk of destabilising feedback loops. Additionally, DQNs use an \textit{experience replay buffer} to store and randomly sample past experiences during training, improving sampling efficiency, reducing correlation between experiences, and further stabilising learning.

\subsubsection{Policy Based Reinforcement Learning}

Value-based methods determine the optimal policy using the \(\text{argmax}\) over state values, limiting them to discrete action spaces. \textit{Policy-based methods} overcome this by directly searching the policy space, enabling use in continuous action spaces. Based on Sutton's \textit{Policy Gradient (PG) Theorem} \cite{sutton1999policy}, these methods compute the gradient of the expected return with respect to policy parameters \(\theta\), where \( \tau = (s_0, a_0, s_1, a_1, \ldots, s_T, a_T) \) represents a trajectory following policy \( \pi_{\theta} \). Here, \( T \) denotes the time horizon and \( \alpha \) is the learning rate.

\vspace{-12pt}

\begin{gather}
\label{eq: PG}
\nabla_{\theta} J(\pi_{\theta}) = \mathbb{E}_{\tau \sim \pi_{\theta}} \left[ \sum_{t=0}^{T} \nabla_{\theta} \log \pi_{\theta}(a_t \mid s_t) \cdot Q^{\pi_{\theta}}(s_t, a_t) \right] \\
\theta_{new} = \theta_{old} + \alpha \nabla_{\theta} J(\pi_{\theta})
\end{gather}

Policy-based methods offer several advantages over value-based approaches. By directly parameterising the policy, they handle high-dimensional action spaces more effectively, while typically offering more stable convergence properties. Policy-based methods can also learn stochastic policies, which are especially useful in uncertain environments or where a diverse variety of actions is beneficial. The foundational policy-based algorithm, REINFORCE \cite{williams1992simple}, is an MC policy gradient method that updates policy parameters using episodic returns. The update rule relies on \( G_t \), the return from time step \( t \) onward, and the gradient \( \nabla_{\theta} \log \pi_{\theta}(a_t \mid s_t) \), which adjusts the policy to favor actions leading to higher returns.

\begin{equation}
    \theta_{new} = \theta_{old} + \alpha \sum_{t=0}^{T} \nabla_{\theta} \log \pi_{\theta}(a_t \mid s_t) \cdot G_t
\end{equation}

\subsubsection{Actor-Critic Reinforcement Learning}

Policy-based methods have advantages but often suffer from high variance in gradient estimates, especially in sparse reward environments. Actor-Critic methods \cite{konda1999actor} address this by decoupling policy and value estimation: the \textit{critic} network estimates the value function, reducing variance in the \textit{actor} network’s policy updates. This dual approach, using separate neural networks for actor and critic, allows for faster convergence, different learning rates, and combines the strengths of both policy-based and value-based methods.

\begin{figure}[H]
    \centering
    \includegraphics[width=0.45\linewidth]{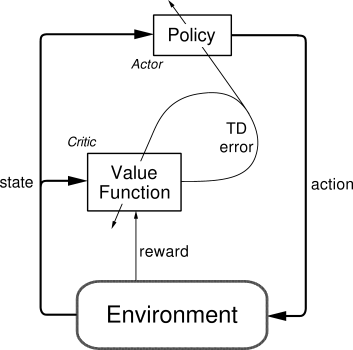}
    \caption{Actor-Critic Framework: The Actor (Policy) selects actions based on the state. The Critic (Value Function) evaluates these actions using the TD error to update both the Policy and Value Function, iteratively improving performance. \cite{sutton2018reinforcement}}
    \label{fig:actor-critic-framework}
\end{figure}

\section{Multi-Agent Systems}

As intelligent and autonomous systems become more prevalent, effectively managing interactions between multiple autonomous agents has become increasingly important. Multi-agent systems, where multiple agents operate in a shared environment, are crucial in domains like robotics \cite{yu2023asynchronous} and autonomous vehicles \cite{kiran2021deep}. These systems enable agents to cooperate, compete, and coordinate to achieve complex goals beyond the capability of a single agent.

\subsubsection{Challenges in a Multi-Agent System}

The dynamic and interactive nature of multi-agent systems introduces unique challenges not encountered in single-agent systems. These challenges stem from the complexities of coordinating and learning in an environment influenced by multiple agents, each with their own objectives and behaviours. Key challenges include partial observability, non-stationarity, credit assignment and scalability.

\vspace{12pt}
\textit{Partial Observability}
\vspace{12pt}

In a MDP, agents are assumed to have full observability of the global state. However, in many real-world situations, decisions must be made with incomplete information due to factors such as scale, privacy, or communication constraints. This gives rise to a \textit{Partially Observable MDP (POMDP)}. For a POMDP framed within a Bayesian framework, the environment exists in a true state \( s_t \), but the agent only receives an observation \( o_t \), drawn from the set of possible agent specific observations \( O \). The agent then updates its belief state \( b_t \), a probability distribution over possible states, using Bayes' rule. Based on this updated belief \( b_{t+1} \), the agent selects an action \( a_{t+1} \) according to the policy \( \pi(b_{t+1}) \). Partial observability complicates decision-making, reducing sample efficiency and increasing the computational complexity of solving POMDPs from P-Class (MDP) to PSPACE-complete \cite{mundhenk2000complexity}.

\begin{figure}[H]
    \centering
    \includegraphics[width=0.5\linewidth]{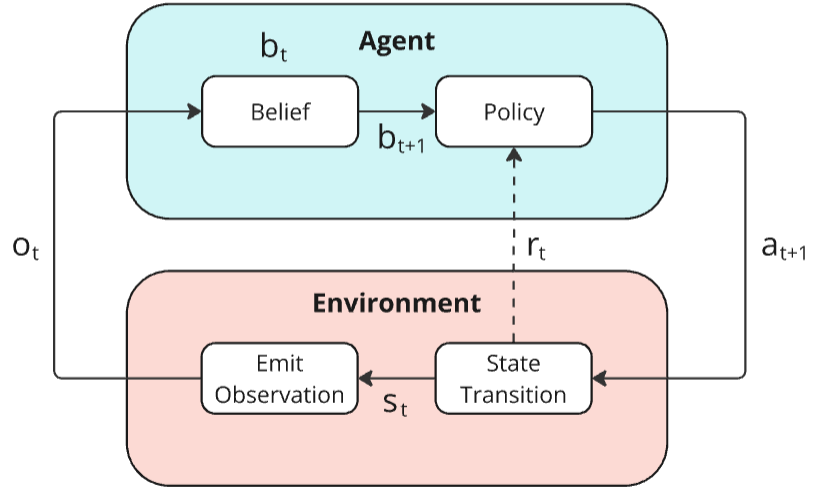}
    \caption{POMDP framework: The agent updates its belief state \( b_t \) based on observation \( o_t \) and selects action \( a_{t+1} \) according to its policy. The environment transitions to a new state \( s_{t+1} \), emits an observation, and provides a reward \( r_t \).}
    \label{fig:pompd}
\end{figure}

\vspace{12pt}
\textit{Non-Stationarity}
\vspace{12pt}

\begin{figure}[H]
\centering
\includegraphics[width=0.53\linewidth]{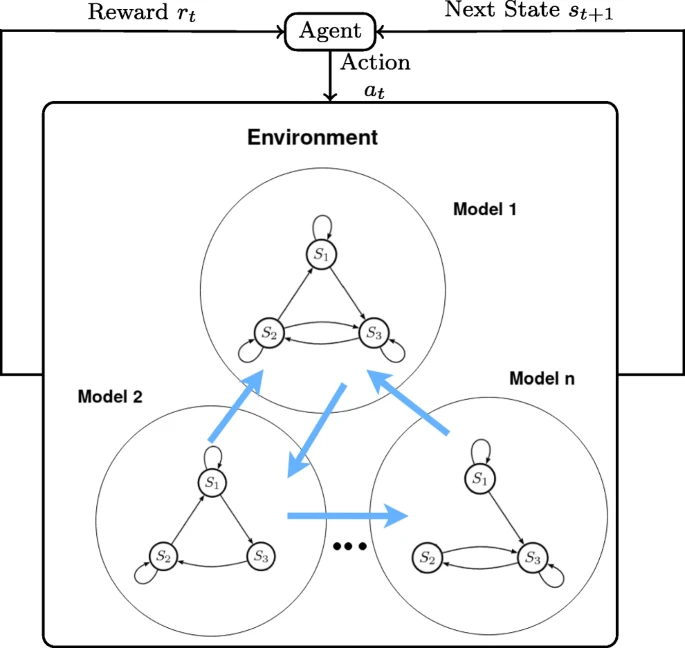}
\caption{A non-stationary environment in a multi-agent system, where each "model" represents an agent. Agents' actions \(a_t\) cause transitions between states \(S_1, S_2, S_3\), with evolving system dynamics as agents update their policies. \cite{padakandla2020reinforcement}}
\label{fig:non-stationary}
\end{figure}

In single-agent environments, dynamics are stationary with consistent transition probabilities and rewards. In contrast, multi-agent systems create a non-stationary environment for each agent, as outcomes depend on the joint actions of all agents. This interdependence breaks the Markovian assumption, invalidates Q-learning convergence guarantees \cite{hernandez2017survey}, and increases variance in the learning process. 

\vspace{12pt}
\textit{Credit Assignment}
\vspace{12pt}

In single-agent RL, distributing rewards across past actions is known as \textit{temporal credit assignment} \cite{sutton2018reinforcement}. This becomes more complex in multi-agent systems, where identifying each agent's contribution to a shared outcome is challenging, especially in cooperative tasks with joint rewards. Ineffective reward distribution, even with centralised control, can hinder cooperation and lead to the \textit{lazy agent} phenomenon \cite{sunehag2017value}, where agents avoid exploring to prevent disrupting others' effective policies.

\vspace{12pt}
\textit{Scalability}
\vspace{12pt}

As the number of agents increases, the information to be shared and processed grows exponentially, leading to the \textit{combinatorial nature} of multi-agent reinforcement learning \cite{hernandez2019survey}. This makes centralised control computationally infeasible. Decentralised and hierarchical control structures address this by distributing the computational load. Decentralised approaches allow agents to make decisions based on local information with limited communication, while hierarchical structures break the problem into smaller sub-problems, organising agents into layers with distinct decision-making roles.

\subsubsection{Interaction Types in a Multi-Agent System}

Multi-agent systems are primarily categorised by the type of interactions among agents. Understanding how individual agent objectives align or diverge from collective goals is essential for effectively shaping and distributing rewards. In the Markov games framework, the optimal solution for a multi-agent system is often the \textit{Nash Equilibrium (NE)} \cite{nash1950non}, where no agent benefits from deviating from its current policy. This concept applies to both cooperative and competitive environments, and many MARL algorithms aim to converge to this point.

\begin{definition}
\textbf{(Nash equilibrium of the MG)} \\
$\left( \mathcal{N}, \mathcal{S}, \{A^{i}\}_{i \in \mathcal{N}}, \mathcal{P}, \{R^{i}\}_{i \in \mathcal{N}}, \gamma \right)$ is a joint policy $\pi^* = 
\left(\pi^{1,*}, \pi^{2,*}, \ldots, \pi^{N,*}\right)$, such that for any $s \in \mathcal{S}$ and $i \in \mathcal{N}$, there is 
$V^{i}_{\pi^{i,*},\pi^{-i,*}}(s) \geq V^{i}_{\pi^{i},\pi^{-i,*}}(s)$, for any $\pi^{i}$.
\end{definition}

In a cooperative setting, all agents work together towards a common objective. Homogeneous agents typically perform more effectively under a shared reward function, while heterogeneous agents benefit from a reward sum system that accounts for their diverse capabilities and roles.

\begin{itemize}
    \item \textit{Reward Sum:}  \(\bar{R}(s,a,s') = N^{-1} \cdot \sum_{i \in N} R^{i}(s,a,s')\)\\ 
    Calculates the average reward, encouraging diverse agents to contribute using their unique strengths whilst ensuring the overall performance is prioritised.

    \item \textit{Shared Reward:} \( R_1 = R_2 = \dots = R_N = R \)\\ 
    A special case of the reward sum, where all agents receive the same reward, ensuring uniform motivation to achieve the group's objectives, naturally motivating agents to work together and support each other's actions.
\end{itemize}

In a competitive setting, agents compete against each other, often within a zero-sum framework where one agent's gain is equivalent to another's loss. This scenario is typical in duel-agent setups or games where only one winner can emerge.

\begin{itemize}
    \item \textit{Zero-Sum Game:} For two agents, \( R_1 + R_2 = 0 \). \\ The reward for one agent is the exact negative of the other’s reward \cite{littman1994markov}, reinforcing direct competition.
\end{itemize}

\section[Multi-Agent RL]{Multi-Agent Reinforcement Learning}

The complexities of multi-agent systems necessitate advanced learning algorithms to manage interactions and coordination among agents. Multi-agent reinforcement learning extends traditional RL to these environments, tackling the unique challenges and opportunities that arise to ensure effective learning and robust performance across a diverse range of multi-agent applications.

\begin{figure}[H]
    \centering
    \includegraphics[width=1\linewidth]{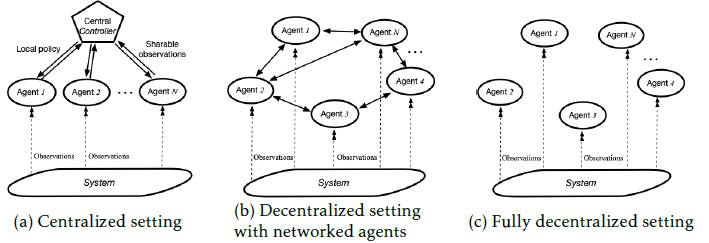}
    \caption{Categories of communication structures in Multi-Agent Systems \cite{zhang2021multi}.}
    \label{fig:marl-information-structures}
\end{figure}

Effective communication in a multi-agent system involves deciding what information to share, with whom, and when. There are three types of communication structures: centralised, where agents share local observations with a global controller that directs their actions; fully decentralised, where agents make policy decisions entirely on local information and decentralised with networked agents, where agents share information directly with each other.

\subsection{Centralised Learning}
\label{sec:centralised-learning}

A common but naive approach to multi-agent problems is using a centralised controller to determine the joint action \( a = (a_1, a_2, \ldots, a_N) \), known as \textit{Centralised Training with Centralised Execution (CTCE)}. While effective for small systems, this method becomes impractical for large-scale problems due to the exponentially growing action space, often making execution intractable. To leverage the substantial computational resources often available during offline training but not during execution, the \textit{Centralised Training with Decentralised Execution (CTDE)} framework was introduced. A central controller gathers observations from multiple agents, utilising its comprehensive view of the entire system to shape each agent's policy. After training, each agent operates independently, executing the learned policies without centralised coordination. 

\begin{table}[H]
\centering
\caption{Comparison of the Advantages and Disadvantages of Centralised Training versus Decentralised Training in Multi-Agent Systems}
\label{table:ctde_pros_cons}
\begin{tabular}{ p{0.47\textwidth}|p{0.47\textwidth} }
\hline
\textbf{Advantages} & \textbf{Disadvantages} \\
\hline
\textbf{Observability:} A central controller uses aggregated observations to better estimate the environment's true state & \textbf{Scalability:} Combinatorial nature of MARL makes centralised training expensive/impractical for large systems. \\
\hline
\textbf{Effective Credit Assignment:} A global view allows the central controller to better assign rewards based on the collective performance of agents. & \textbf{Communication Overhead:} Fast and effective communication channels are required during training, which can be difficult to maintain over large scales. \\
\hline
\textbf{Stationarity:} Coordinated policy updates minimise agents' need to adapt to independently evolving policies & \textbf{Privacy Concerns:} Aggregating information from all agents can lead to privacy issues in sensitive environments. \\
\hline
& \textbf{Robustness:} A central controller can be a bottleneck / single failure point. \\
\hline
& \textbf{Availability:} A central controller may not be feasible or available. \\
\hline
\end{tabular}
\end{table}

To retain the benefits of centralised training while ensuring decentralised execution, there are two methods for factorising the joint action space:

\begin{itemize}
    \item \textbf{Policy Factorisation:} $\pi(a|s) := \prod_{i=1}^N \pi_i(a_i|s)$ \\
    Assumes that policies are independent, generally requiring a centralised critic network to coordinate behaviour for stable learning.

    \item \textbf{Value Function Factorisation:} $Q_{\text{tot}}(s, a) = f (Q_1(s, a_1), Q_2(s, a_2), \dots, Q_N(s, a_N))$ \\
    Factorises the joint value function into independent evaluations, each focused on the actions of a single agent, \( f \) represents the factorisation function.
\end{itemize}

\subsubsection{Policy Factorisation}

In policy factorisation, a central critic network oversees training by aggregating local observations from each actor to generate a central value estimate. This estimate guides the updates to the actor's policies. In the shared reward setting, a single centralised critic is sufficient. However, in the reward sum setting, each agent requires its own fully observable critic.

\vspace{12pt}

Leading single-agent actor-critic algorithms, such as \textit{Proximal Policy Optimisation (PPO)} \cite{schulman2017proximal}, which uses policy update clipping to enhance stability, and \textit{Deep Deterministic Policy Gradient (DDPG)} \cite{lillicrap2015continuous}—an off-policy algorithm optimised for continuous action spaces through deterministic policy updates—have been successfully adapted for multi-agent settings. These adaptations include \textit{Multi-Agent Proximal Policy Optimisation (MAPPO)} \cite{yu2022surprising} and \textit{Multi-Agent Deep Deterministic Policy Gradient (MADDPG)} \cite{lowe2017multi}, both of which have demonstrated state-of-the-art performance across various environments with a limited number of agents.

\vspace{12pt}

However, even with a central controller, credit assignment remains challenging in shared reward settings. It is standard practice to use the Advantage function \( A(s, a) = Q(s, a) - V(s) \) instead of the Q-function to reduce variance and improve stability. Foerster et al. introduced \textit{Counterfactual Multi-Agent Policy Gradient (COMA)} \cite{foerster2018counterfactual}, improving credit assignment by using a counterfactual baseline for advantage estimation \( A(s, a_i, a_{-i}) \). COMA calculates a counterfactual value for each agent, reflecting the expected return if the agent had taken an average action \( a'_i \) instead of its actual action \( a_i \), while keeping other agents' actions \( a_{-i} \) fixed.

\begin{equation}
A(s, a_i, a_{-i}) = Q(s, a_i, a_{-i}) - \sum_{a'_i} \pi_i(a'_i|s) Q(s, a'_i, a_{-i})
\label{eq:combined-counterfactual}
\end{equation}

\vspace{-12pt}

\subsubsection{Value Function Factorisation}

To factorise the joint policy, the central critic must consider the entire joint-action space. A more efficient approach is to factorise only the value function using localised critic networks. Agents share their local value estimates \( Q_i \) with a central controller, which serves as a mixing network to produce a global value estimate \( Q_{\text{tot}} \). These individual value functions are tailored to guide the policies of decentralised actor networks during execution. The mixing network must satisfy the \textit{Individual-Global Maximisation (IGM)} constraint \cite{rashid2020monotonic} to ensure maximising the global value aligns with independently maximising each agent’s value function. Here, \( \tau \) represents the vector of local observation-action histories, and \( a \) is the joint-action vector.

\vspace{-12pt}

\begin{align}
    \text{argmax}_{a} Q_{\text{tot}}(\tau, a) = \left( \text{argmax}_{a_1} Q_1(\tau_1, a_1), \ldots, \text{argmax}_{a_n} Q_n(\tau_n, a_n) \right)
\label{eq:IGM}
\end{align}

\vspace{3pt}

Value factorisation methods vary in how they decompose the global value function. The simplest approach, \textit{Value Decomposition Network (VDN)} \cite{sunehag2017value}, sums action values from decentralised critics, \( Q_i \), to produce the joint-action value function, \( Q_{\text{tot}} \). While this method satisfies the IGM constraint and is highly scalable, its linear factorisation has limited representational capacity and lacks global convergence guarantee \cite{du2023review}.

\vspace{12pt}

To better represent complex agent interactions, \textit{QMIX} \cite{rashid2020monotonic} relaxes the strict summation condition by requiring the joint-action value function $Q_{\text{tot}}$ to be monotonically increasing with respect to each action value $Q_i$, satisfying the IGM constraint while using a neural network mixing network with non-negative weights. \textit{QTRAN} \cite{son2019qtran} further relaxes the monotonicity constraint by transforming the joint-action value function into an alternative that can be additively decomposed. \textit{QPLEX} \cite{wang2020qplex} reformulates the IGM constraint by separating $Q_i(o_i, a_i)$ into a value function $V(s)$ and an advantage function $A_i(o_i, a_i)$. Finally, \textit{GraphMix} \cite{naderializadeh2020graph} considers agents as a network of nodes within a complete directed graph and employs a single mixing \textit{Graph Neural Network (GNN)} \cite{scarselli2008graph} to decompose the value function.

\subsection{Decentralised Learning}

\begin{figure}[H]
    \centering
    \includegraphics[width=0.5\linewidth]{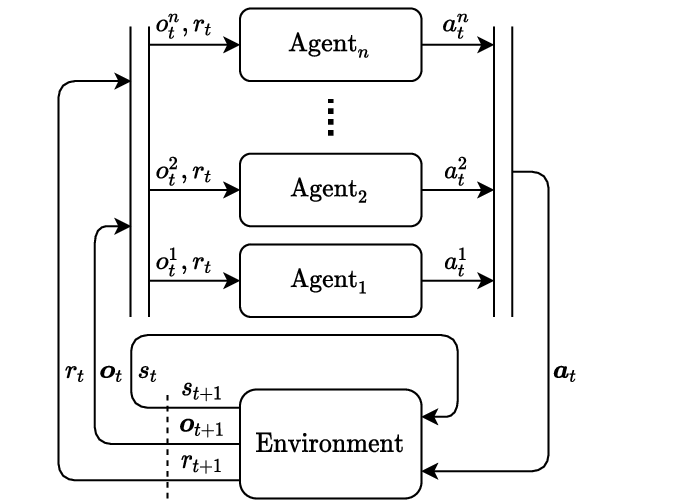}
    \caption{Dec-POMPD: At time-step $t$, the environment is in state $s_t$ and emits a joint observation $o_t$. Each agent $i$ receives observation $o_{i,t}$, takes action $a_{i,t}$, forming joint action $\vec{a}_t$. The environment transitions to state $s_{t+1}$ and emits reward $r_t$. \cite{willemsen2021mambpo}}
    \label{fig:Dec-POMPD}
\end{figure}

The most straightforward approach is to consider each agent as an independent single-agent RL problem. This approach assumes agents make decisions based solely on their local observations without any coordination or data aggregation. Consequently, decentralised learning approaches are infinitely scalable and maximally private, finding broad applications in fields such as autonomous driving \cite{zhang2023high}. With decentralised learning, multiple agents execute solely based on local observations. To formalise this, we extend the single-agent POMDP to the \textit{Decentralised POMDP (Dec-POMDP)} defined by the tuple $(S, \{A_i\}, P, R, \{\Omega_i\}, O, \gamma)$:

\begin{itemize}
    \item \textbf{State Space \((S)\)}: Set of states in the environment.
    \item \textbf{Action Space \((A_i)\)}: Set of actions available to agent \(i\), with the joint action space \(A = \prod_i A_i\).
    \item \textbf{Transition Probability \((P)\)}: The probability \(P(s' | s, a)\) of transitioning from state \(s\) to \(s'\) given joint action \(a\).
    \item \textbf{Reward Function \((R)\)}: Reward function \(R: S \times A \rightarrow \mathbb{R}\) for state-action pairs.
    \item \textbf{Observation Space \((\Omega_i)\)}: Set of observations available to agent \(i\), with the joint observation space \(\Omega = \prod_i \Omega_i\).
    \item \textbf{Observation Probability Function \((O)\)}: Probability \(O(s', a, o) = P(o|s', a)\) of observation \(o\) given state \(s'\) and action \(a\).
    \item \textbf{Discount Factor \((\gamma)\)}: A factor \(\gamma \in [0, 1]\) for weighting future rewards.
\end{itemize}

\subsubsection{Decentralised Value-Based Learning}
\label{sec:independent-value-shared}

In cooperative settings with homogeneous agents, a consistent value function allows them to operate as a unified decision-maker, enabling the use of single-agent RL techniques like Q-learning \cite{watkins1989learning}. Most practical Q-learning implementations use the DQN \cite{mnih2013playing}, however this architecture relies on the assumption of a stationary environment. The \textit{Deep Recurrent Q-Network (DRQN)} \cite{hausknecht2015deep} incorporates an RNN to capture temporal dependencies therefore better mitigating non-stationary environments. 

\vspace{12pt}

Similarly, since Q-learning is off-policy, the relevance of experiences generated with non-stationary policies and transition probabilities is limited and can undermine the replay mechanism. To alleviate this, Foerster et al. \cite{foerster2016ddrqn} initially removed the replay buffer but later introduced two selective sampling strategies: Importance Sampling, which prioritises recent experiences by decaying outdated data, and Fingerprinting, a Meta RL approach that tracks iteration number and exploration rates to gauge the relevance of past experiences \cite{foerster2017stabilising}.

\vspace{12pt}

Without access to joint actions, each agent estimates its Q-value based on individual actions,  making value estimates vulnerable to other agents' exploration in shared reward settings. To address this, \textit{Distributed Q-learning} \cite{lauer2000algorithm} updates the value function only when there's a guaranteed improvement (positive TD error), though it's limited to deterministic environments. \textit{Hysteretic Q-learning} \cite{matignon2007hysteretic} extends this to stochastic environments by applying a higher learning rate for positive TD errors.

\subsubsection{Decentralised Policy-Based Learning}
\label{sec:independent-policy-shared}

In highly non-stationary environments, policy-based methods theoretically excel due to direct policy parameterisation, enabling dynamic adaptation to other agents' changing policies. 
However, the first approach, \textit{Independent Actor-Critic (IAC)} \cite{foerster2018counterfactual}, underperformed due to instability from frequent, large policy updates by other agents. De Witt et al. introduced \textit{Independent Proximal Policy Optimisation (IPPO)} \cite{de2020independent}, a decentralised variant of \textit{PPO}. IPPO outperformed decentralised value-based methods and even matched the performance of leading CTDE methods in some environments. Building on IPPO, \textit{Decentralised Policy Optimisation (DPO)} \cite{su2022decentralized} extends \textit{Trust Region Policy Optimization (TRPO)} \cite{schulman2015trust} to multi-agent settings by constraining policy updates within a trust region, limiting the Kullback-Leibler (KL) divergence. The latest approach, \textit{Total Variation Policy Optimisation (TVPO)} \cite{su2024general}, introduces \( f \)-divergence for measuring distributional distance, with KL divergence as a special case.

\subsection{Decentralised Learning with Networked Agents}

Dec-POMDPs offer a powerful framework but are provably intractable with NEXP-complete complexity \cite{amato2013decentralized}. Consequently, much research focuses on restricted Dec-POMDP variants that are easier to solve yet represent many practical applications. One such variant involves agents sharing information over a time-varying communication network. When communication is free, instantaneous, and lossless, the system becomes effectively centralised, making it solvable as a POMDP since all agents share all observations at each step \cite{oliehoek2012tree}. In practical applications, the challenge lies in creating a communication network that minimises delay, cost, and information loss, thereby approximating centralised performance while maintaining the advantages of a decentralised system. 

\begin{figure}[H]
    \centering
    \includegraphics[width=0.41\linewidth]{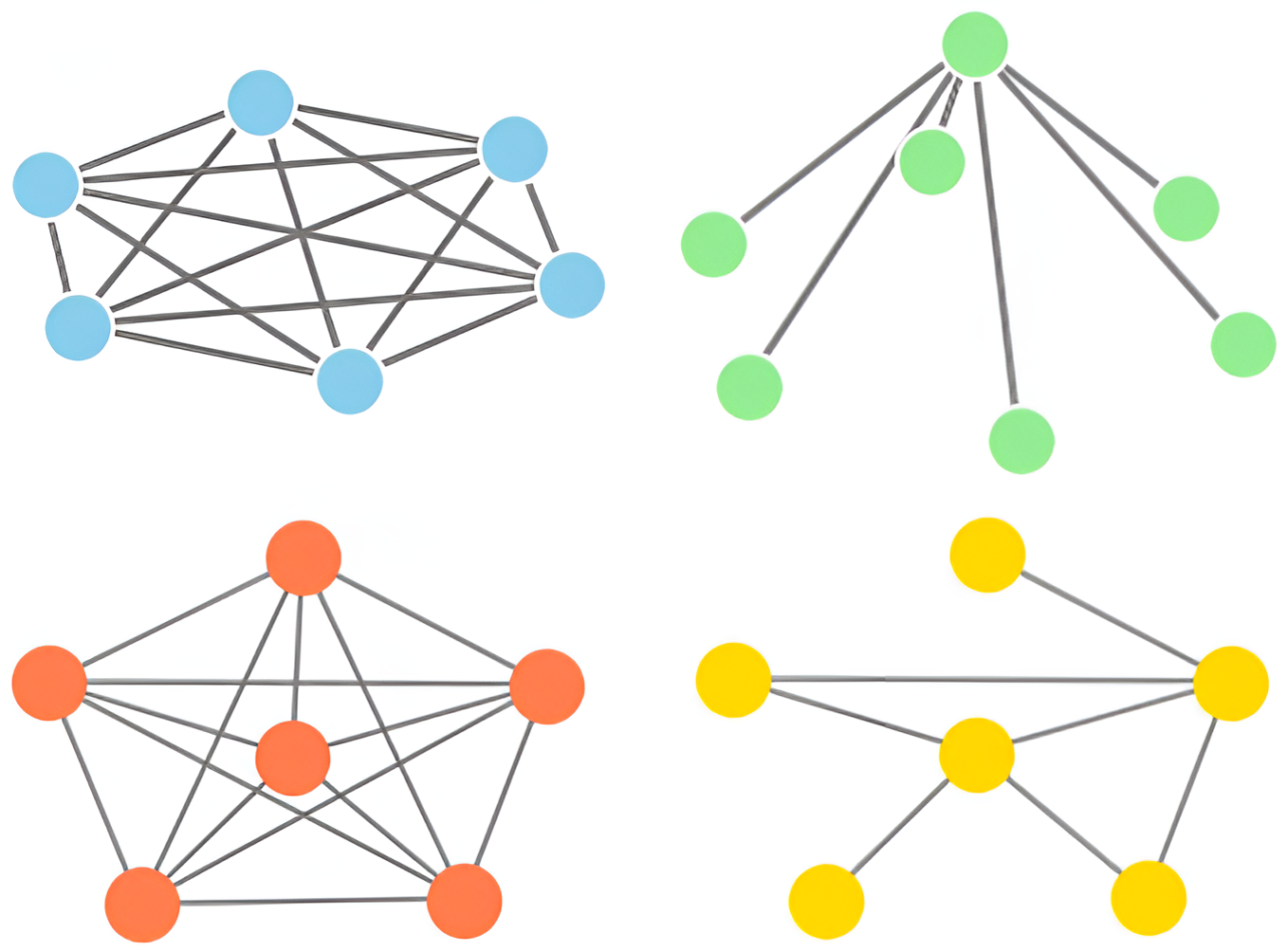}
    \caption{Network Topologies: Peer-to-Peer (blue), Hierarchical (green), Dense (red), and Sparse (yellow)}
    \label{fig:network-topology}
\end{figure}

The topology of the communication network is the first consideration, as it governs the network's efficiency, scalability, and resilience by defining how agents are connected. Networks can be \textit{peer-to-peer}, where agents communicate directly and equally, or \textit{hierarchical}, where communication flows through structured levels of importance. Topologies are also categorised as \textit{static}, with fixed connections determined by factors such as proximity, or \textit{dynamic}, where connections evolve over time due to adaptive mechanisms or changing external conditions. Finally, networks can be \textit{dense}, with all-to-all inter-agent connections scaling quadratically with the number of agents, or \textit{sparse}, where only a subset of agents are directly connected.

\begin{table}[H]
\centering
\caption{Categories of Decentralised Learning with Networked Agents Approaches}
\label{tab:marl-methods}
\begin{tabularx}{\textwidth}{>{\raggedright}p{3cm}>{\raggedright\arraybackslash}X}
\hline
\textbf{Category} & \textbf{Description} \\ \hline

Mean Field Theory &  Applies Mean-Field Theory \cite{stanley1971phase} to approximate single-agent interactions with the averaged effect of the entire population \\ \hline 

Consensus Mechanism & Agents iteratively share information with neighbours to reach agreement on certain variables, enabling coordinated actions. \\ \hline

Communication Protocol & Define or learn strategies for information sharing, including rules on with whom, when, what, and how to communicate.  \\ \hline
     
\end{tabularx}
\end{table}

\subsubsection{Mean Field Theory}

The next key aspect is agent interaction. Mean Field Theory (MFT), which originates from physics \cite{stanley1971phase}, simplifies multi-agent interactions by approximating them as two-agent interactions. Yang et al. \cite{yang2018mean} applied MFT to MARL with MF-Q and MF-AC, which decompose the $Q$-function into pairwise local interactions, $Q^i(s, a^i, a^j)$, between agent $i$ and its neighbours $j$. The MFT approximation reduces these interactions to those between agent $i$ and a virtual "mean agent," represented by $Q^i(s, a^i, \bar{a}^i)$, where $\bar{a}^i$ is the average action of $i$'s neighbours. While this approach allows large systems to be optimised via a centralised agent, it relies on the assumption that agents' influences can be approximated by an average effect, constraining it to environments with homogeneous agents and near-homogeneous policies.

\begin{figure}[H]
\centering
\includegraphics[width=0.4\linewidth]{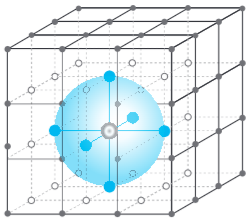}
\caption{Mean field approximation: each agent (node) is influenced by the mean effect of its neighbours (blue region) \cite{yang2018mean}.}
\label{fig:mean-field-approx}
\end{figure}

\subsubsection{Consensus Mechanism}

The second approach involves agents coordinating actions with neighbours via a consensus mechanism. Constraining information transfer to neighbouring agents is a typical approach in one-stage distributed optimisation \cite{xiao2007distributed}, based on the reasonable assumption of weak coupling between distant agents in large-scale systems. However, for sequential decision-making tasks where actions have long-term effects, this approach becomes more challenging. Consensus approaches include \textit{Policy Learning}, where agents share parameters to form a consensus policy, and \textit{Policy Evaluation}, where they share local value functions to optimise a shared value function for a fixed, possibly sub-optimal policy.

\vspace{12pt}

Policy learning approaches centre on the algorithms proposed by Zhang et al. \cite{zhang2018fully}, where each agent maintains its parameter \( \omega^i \) and uses \( Q(\cdot, \cdot; \omega^i) \) as a local estimate of \( Q_\theta \). Agents update their parameters using the TD error, followed by a weighted combination with neighbour estimates \(\tilde{\omega}^j_t\) based on network edge weights \( c_t(i, j) \), as shown by the update rule \( \omega^i_{t+1} = \sum_{j \in \mathcal{N}} c_t(i, j)  \cdot \tilde{\omega}_t^j \). Essentially this becomes a distributed COMA\cite{foerster2018counterfactual} variant, where each agent estimates \( Q(\cdot, \cdot; \omega^i) \) for the counterfactual baseline \( Q_\theta \), applying a consensus constraint to minimise discrepancies. Zhang et al. proved convergence but under the condition of full observability of global states and joint actions, therefore limiting scalability. However, the parameter-sharing mechanism enables network-wide collaboration while preserving privacy, as individual rewards and policies are not shared \cite{zhang2021decentralized}.

\begin{figure}[H]
    \centering
    \includegraphics[width=0.35\linewidth]{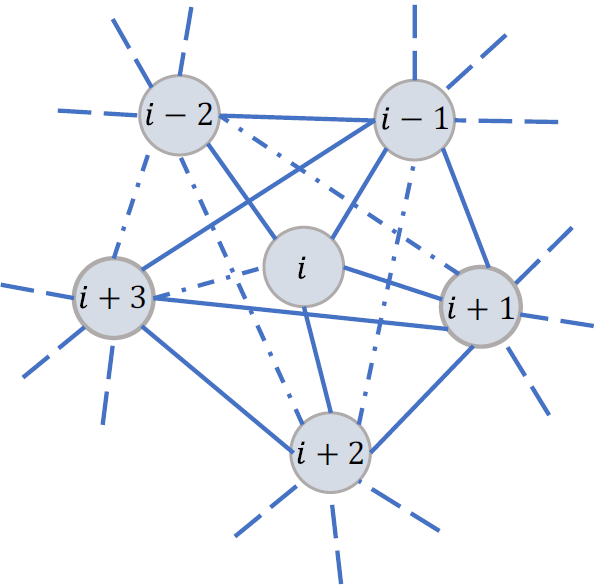}
    \caption{Consensus Network \cite{oroojlooy2023review}: A network where each node \( i \) exchanges information with its immediate neighbors to achieve consensus on a shared variable.}
    \label{fig:consensus-network}
\end{figure}

\vspace{-9pt}

In policy evaluation, agents minimise the Mean Square Projected Bellman Error \textit{(MSPBE)} to learn the value function for a fixed joint policy. Although the Bellman equation recursively computes the value function, exact computation is impractical with function approximators, instead the value function is projected onto a smaller space defined by the approximator, and the Projected Bellman Error quantifies the projection error. In multi-agent settings, MSPBE measures the average squared Mahalanobis distance between each agent’s projected value function and its Bellman update. While minimising MSPBE is a standard goal in policy evaluation, most research prioritises proving convergence \cite{macua2014distributed, stankovi2016multi} over developing practical algorithms. These methods assume a fixed policy, making frequent MSPBE computation impractical in dynamic systems and joint policy evaluation infeasible in large-scale systems.

\subsection{Communication Protocols}
\label{sec:communication-protocols}

Mean Field and Consensus Mechanism approaches are fundamentally restricted special cases of communication protocols. Applications such as network packet routing, which involve distributed agents with partial observability and heterogeneous policies, require more general and robust communication protocols. 

\vspace{12pt}

Effective communication depends on a shared language—an understanding of symbols between speaker and listener—either predefined or learned, with recent work favouring the latter for greater flexibility. Communication can be either \textit{discrete}, involving finite symbols, or \textit{continuous}, using variable values. Agents typically use a shared communication channel, allowing messages to be exchanged before actions \cite{sukhbaatar2016learning}, concurrently \cite{foerster2016learning}, or over multiple rounds \cite{weil2024towards}. Learnable communication protocols are characterised by four key factors: message content, recipients, aggregation, and the learning mechanism.

\subsubsection{Message Content}

Encoding local observations usually involves using a learnable neural network, such as a MLP, RNN, or CNN. It's crucial that all agents share a mutual understanding of these observations to ensure message consistency and interpretability. In centralised training, this is straightforward as parameters can be shared among agents. However, in decentralised training, the challenge is greater, as protocols map action-observation histories to message sequences, leading to a high-dimensional space of possible protocols \cite{foerster2016learning}. While there is extensive research on maintaining consistent encoding across decentralised agents \cite{lin2021learning}, this falls outside the scope of this report.

\subsubsection{Message Recipients}

The next step is determining which agents should receive transmitted information. Early methods, such as \textit{CommNet} \cite{sukhbaatar2016learning}, broadcast observations to all agents, forming a dense static network. However, this results in quadratic communication overhead and floods agents with potentially irrelevant information. To reduce information overload, \textit{NeurComm} \cite{chu2020multi} applies a spatial discount to weight communications, prioritising nearby agents. Externally constrained approaches like \textit{DGN} \cite{jiang2018graph} select recipients based on factors such as proximity and cost, but time-varying external factors can easily disrupt collaboration.

\vspace{12pt}

\textit{Targeting Mechanisms}

\vspace{12pt}

Targeting mechanisms create sparse networks by identifying the most relevant message recipients, thereby maximising efficiency. Effective selection of which agents should communicate typically requires a global overview, with current targeting mechanisms being limited to CTDE. 

\begin{figure}[H]
\centering
\includegraphics[width=0.7\linewidth]{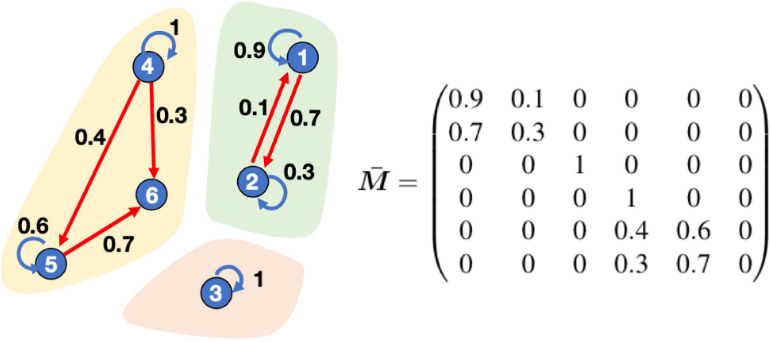}
\caption{When2Comm Adjacency Matrix: Rows with a self-attention score of 1 are replaced with an identity row, indicating no further communication is needed. Other rows are derived from scaled, pruned, and activated matching scores.}
\label{fig:when2comm}
\end{figure}

Conventionally, targeting mechanisms use a key-query matching system. In \textit{TarMAC} \cite{das2019tarmac}, a learnable key is broadcast, encoding the intended recipient's properties. Recipients use soft-attention to assess message relevance, reducing information overload. \textit{ATOC} \cite{jiang2018learning} reduces TarMAC’s communication overhead by broadcasting only to agents within its observable range. It forms dynamic communication groups with hard-attention, restricting broadcasts to these groups for a set duration. \textit{When2Comm} \cite{liu2020when2com} extends this system by introducing \textit{asynchronous} communication. It uses self-attention between an agent’s own key and query vectors, using their correlation as a signal that the agent has enough information to make a decision. 

\vspace{12pt}

\textit{I2C} \cite{ding2020learning} takes a different approach, using a MLP to map local observations to a belief about which individual agents to communicate with, eliminating broadcasting altogether. Using a similar idea, \textit{Model-Based Communication (MBC)} \cite{han2023model} trains agents to predict incoming message content using supervised learning based on past communications. If an agent's confidence in its prediction drops below a threshold, it initiates a broadcast, ensuring messages are sent only when necessary.

\vspace{12pt}

For scalability, \textit{ATOC} and \textit{I2C} limit communication to agents within the sender's observable range, which can hinder convergence if distant agents are relevant. \textit{G2ANet} \cite{liu2020multi} addresses this by starting with a dense network and pruning irrelevant connections using hard-attention gating. Conversely, \textit{AC2C} \cite{wang2023ac2c} begins with local 1-hop connections, training a controller to expand relevant 2-hop connections. Taking a completely different approach, \textit{MAGIC} \cite{niu2021magic} uses a graph attention encoder to directly learn adaptive, directed graphs that specify optimal communication links.

\subsubsection{Message Aggregation}

Agents typically receive multiple messages, and the challenge lies in extracting the most relevant information for policy decisions. For stability and effectiveness, the protocol's output must be invariant to message order. Basic methods include concatenating \cite{foerster2016learning}, averaging \cite{sukhbaatar2016learning}, or summing \cite{du2021learning} message feature vectors, but equally weighting messages causes information loss at large scales. To address this, some methods apply handcrafted rules for unequal weighting \cite{kim2019message}, while others use order-invariant, learnable mechanisms like soft-attention \cite{das2019tarmac} or bi-directional RNNs \cite{peng2017multiagent}, which process messages in both directions.

\vspace{12pt}

\textit{Graph Convolutions}

\vspace{12pt}

For sparse graphs, GNNs can leverage the graph structure to aggregate information. Graph Convolutional Networks (GCNs) \cite{kipf2016semi} encode node or edge features into low-dimensional representations, which are iteratively updated by aggregating information from neighbouring nodes, typically through a weighted sum or average. This process captures both local features and the overall graph structure. The adjacency matrix defines node connectivity and guides aggregation, often normalised to stabilise training by balancing the influence of neighbouring nodes. Algorithm \ref{alg:gcn} outlines a basic graph convolutional layer.

\addvspace{12pt}

\begin{algorithm}[H]
\caption{Graph Convolutional Layer}
\label{alg:gcn}
\KwIn{Adjacency matrix $A$, Input Feature matrix $X$, Graph Weight matrix $W$, Activation function $\sigma$}
$D_{ii} \gets \sum_{j} A_{ij}$ \tcp*{Compute degree matrix}
$\tilde{A} \gets D^{-\frac{1}{2}} A D^{-\frac{1}{2}}$ \tcp*{Normalize adjacency matrix}
$H \gets \sigma(\tilde{A} X W)$ \tcp*{Apply transformation and activation}
\textbf{return:} Updated feature matrix $H$\;
\end{algorithm}

\addvspace{12pt}

\begin{figure}[H]
    \centering
    \includegraphics[width=0.9\linewidth]{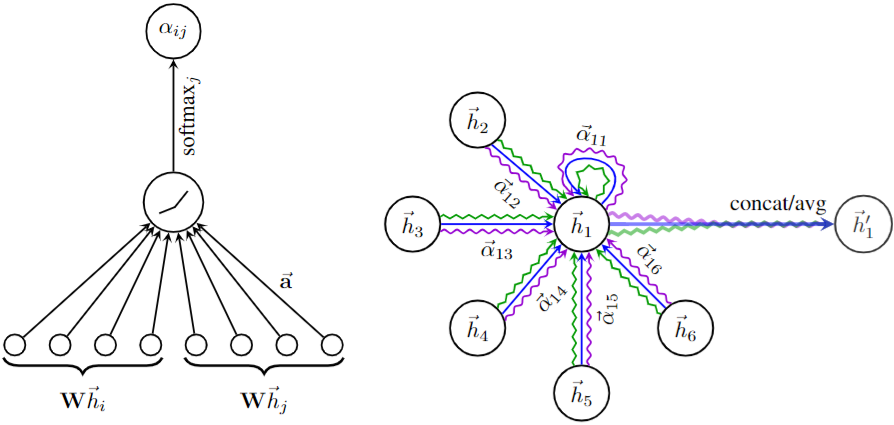}
    \caption{GAT: Computing attention coefficients using a shared mechanism across node pairs (left). Using these coefficients to compute the weighted sum of neighboring node features, updating each node's feature representation (right) \cite{velivckovic2017graph}.}
    \label{fig:gat_diagram}
\end{figure}

The main limitation of GCNs is their static weighting of messages from neighbouring nodes, failing to capture the dynamic importance of information over time. Graph Attention Networks (GATs) \cite{velivckovic2017graph} address this by using an attention mechanism to dynamically weight messages from neighbours, as shown in Figure \ref{fig:gat_diagram}. Algorithm \ref{alg:gat} outlines a single GAT layer.

\addvspace{12pt}

\begin{algorithm}[H]
\caption{Graph Attention Layer}
\label{alg:gat}
\KwIn{Adjacency matrix $A$, Input feature matrix $X$, Graph weight matrix $W$, Attention function $\alpha$, Activation function $\sigma$}
$H' \gets XW$ \tcp*{Apply linear transformation}
\ForEach{edge $(i, j) \in A$}{
    $e_{ij} \gets \alpha(H'_i, H'_j)$ \tcp*{Compute attention score}
    $\alpha_{ij} \gets \frac{\exp(e_{ij})}{\sum_{k \in \mathcal{N}(i)} \exp(e_{ik})}$ \tcp*{Normalize attention scores}
}
\ForEach{node $i \in A$}{
    $H_i \gets \sigma\left(\sum_{j \in \mathcal\{N(i)\}} \alpha_{ij} H'_j\right)$ \tcp*{Apply weighted aggregation and activation}
}
\textbf{return} Updated feature matrix $H$\;
\end{algorithm}

\addvspace{12pt}

Graph convolutional layers are commonly used in message-passing systems for aggregating messages. For example, MAGIC \cite{niu2021magic} and DGN \cite{jiang2018graph} use GATs. A key advantage of these layers is their ability to be stacked. Stacking increases the receptive field, allowing agents to propagate long-range information through multiple hops while communicating only with direct neighbours, efficiently expanding the receptive field without significant communication overhead.

\subsubsection{Learning Mechanism}

The final aspect to consider for learnable communication-based protocols is how they utilise feedback to improve the protocol. Learnable protocols can be characterised as either \textit{Reinforced} or \textit{Differentiable}, as detailed in Table \ref{tab:learning_mechanisms}.

\begin{table}[H]
\centering
\caption{Comparison of Learning Mechanisms for Communication Protocols}
\label{tab:learning_mechanisms}
\begin{tabular}{ p{2.5cm} p{9cm} >{\raggedright\arraybackslash}p{2.7cm} }
\hline
\textbf{Mechanism} & \textbf{Description} & \textbf{Examples} \\ \hline
Reinforced & Assumes successful communication leads to environmental rewards, using environmental feedback to reinforce communication strategies. & \textit{RIAL} \cite{foerster2016learning}, \textit{ATOC} \cite{jiang2018learning} \\ \hline
Differentiable & Uses continuous feedback from other agents to refine communication through backpropagation, allowing real-time improvement. Gradients are computed based on the difference in Q-values after receiving the message. & \textit{DIAL} \cite{foerster2016learning}, \textit{CommNet} \cite{sukhbaatar2016learning} \\ \hline
\end{tabular}
\end{table}

\section{Network Packet Routing}

Network packet routing is a fundamental aspect of computer networking that enables the transmission of data across different networks. This process involves directing data packets from a source node to a destination node through a series of intermediate nodes, such as routers and switches. The objectives of packet routing are to minimise \textit{latency} (the delay from source to destination) and ensure reliable data delivery by minimising \textit{packet loss}. Routers, the primary devices responsible for packet routing, use routing tables to keep track of network paths and make forwarding decisions. Upon receiving a packet, the router examines its destination address and consults the routing table to determine the next hop. This process repeats until the packet reaches its destination.

\begin{figure}[H]
    \centering
    \includegraphics[width=0.7\linewidth]{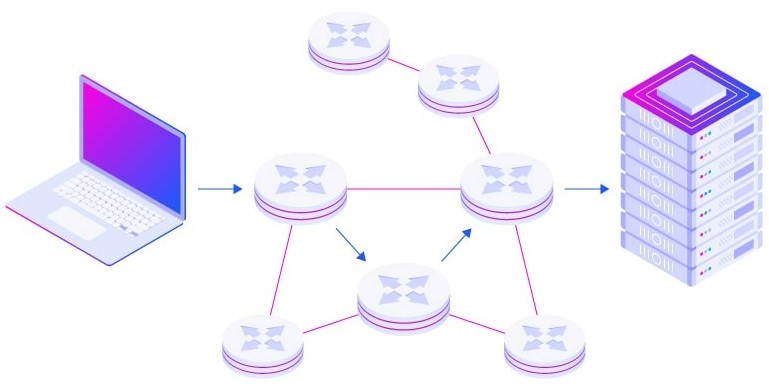}
    \caption{Network packet routing from source to destination via intermediary routers, illustrating the selection of efficient paths based on routing protocols \cite{Anfalovas2024network}.}
    \label{fig:network-packet-routing}
\end{figure}

\subsubsection{Routing Protocols}

Routing decisions rely on protocols that determine the optimal path based on network topology, traffic, and link cost. Common protocols include \textit{Routing Information Protocol (RIP)} \cite{hedrick1988routing}, \textit{Open Shortest Path First (OSPF)} \cite{sidhu1993open}, and \textit{Border Gateway Protocol (BGP)} \cite{rfc1105}. Table \ref{tab:routing_protocols} compares these protocols.

\begin{table}[H]
\centering
\caption{Comparison of RIP, OSPF, and BGP Routing Protocols}
\begin{tabular}{p{2.2cm} p{2.2cm} p{9.8cm}}
\hline
\textbf{Protocol} & \textbf{Type} & \textbf{Description} \\ \hline
RIP & Distance-vector & Uses hop count with a maximum of 15 hops. Regularly exchanges entire routing tables, leading to higher overhead and slower convergence. \\ \hline
OSPF & Link-state & Uses Dijkstra's algorithm to compute the shortest paths. Maintains a map of the topology and exchanges link state advertisements (LSAs) for faster convergence, supporting larger, complex networks. \\ \hline
BGP & Path-vector & Used for routing between autonomous systems (AS). Tracks the full AS path to the destination, preventing routing loops and enabling policy-based routing decisions, essential for reliable internet routing. \\ \hline
\end{tabular}
\label{tab:routing_protocols}
\end{table}

\subsection{MARL in Network Packet Routing}

As networks grow in size and complexity, traditional routing algorithms are becoming increasingly inadequate. A promising solution is using RL for network packet routing, allowing routers to learn and optimise routing policies through interaction with the network. In large, dynamic networks where centralised control is impractical, decentralised execution becomes crucial. MARL enables intelligent routers to cooperate whilst operating independently.  This approach optimises network performance, enhances adaptability to changing conditions, and ensures scalability as the network grows.

\vspace{12pt}

Early approaches like \textit{Q-routing} \cite{boyan1993packet} treated each router as an individual agent, applying Q-learning with a shared set of parameters. Routers maintain Q-tables to estimate route quality, updating the Q-tables based on observed packet delivery times. \textit{DQN-routing} \cite{mukhutdinov2019multi} builds on this by using DQNs \cite{mnih2013playing}, leveraging deep learning to improve routing decisions in more complex network environments.

\vspace{12pt}

However, with the emergence of key technologies such as Software Defined Networking (SDN) \cite{shin2012software}, there has been a shift in how networks are managed. SDN decouples the control plane from the data plane, allowing flexible and dynamic network management. Programmable switches handle data forwarding, while a centralised controller manages routing decisions. This differs from traditional networks, where routers integrate both planes. 

\begin{figure}[H]
    \centering
    \includegraphics[width=0.8\linewidth]{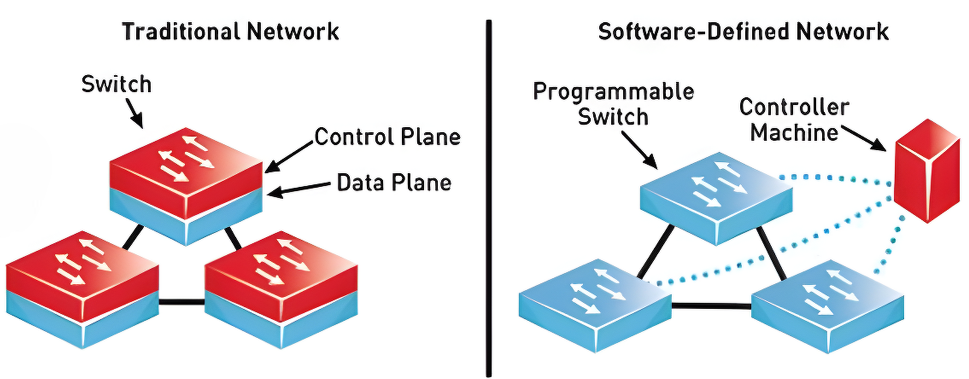}
    \caption{Comparison of Traditional Network and SDN Architectures \cite{maleki2017sdn}.}
    \label{fig:sdn}
\end{figure}

SDN's flexibility supports networked MARL approaches by enabling direct communication between routers and switches for control decisions. \textit{Deep Q-routing with Communication (DQRC)} \cite{you2020toward} builds on \textit{DQN-routing}, adding a communication step between routers before action selection. Drawing from the success of \textit{DRQN} \cite{hausknecht2015deep} in non-stationary settings, DQRC also uses an LSTM layer to aggregate local and neighbouring observations, thereby improving performance in dynamic networks.

\vspace{12pt}

Conventional neural networks (feed-forward, recurrent, and convolutional) can specialise in specific networks but struggle to generalise.  In contrast, \textit{DRL+GNN} \cite{almasan2022deep} leverages GNNs to generalise across diverse graph sizes and structures, providing a robust solution for dynamic graphs. However, like many leading routing approaches \cite{casas2020intelligent}, the \textit{DRL+GNN} functions as a fully observable single agent as the GNN's effectiveness depends on the graph representation quality, limiting scalability and reactivity compared to decentralised approaches. 

\vspace{12pt}

Therefore to leverage GNNs in a decentralised setting, the \textit{Graph-Query Neural Network} \cite{geyer2018learning} uses multi-round communication with a recurrent aggregation function to form high-quality representations of the network. However, the non-stationary environment created by decentralised agents required the use of supervised learning on labelled shortest-path data for stable training, limiting adaptability to new graphs without retraining.

\subsection{NetMon}

NetMon, recently introduced by Weil et al. \cite{weil2024towards}, is the leading framework for decentralised multi-agent reinforcement learning in network packet routing.  Weil et al. made several novel contributions, which are discussed in the following section.

\fancypagestyle{netmon-page}{
    \fancyfoot[LE]{\sffamily\textbf{\thepage}} 
    \fancyfoot[RO]{\sffamily\parbox[t]{\dimexpr\textwidth-1cm}{\scriptsize While not explicitly mentioned in the paper, the name "NetMon" is used internally within the authors' code repository. For details, see \href{https://github.com/jw3il/graph-marl}{https://github.com/jw3il/graph-marl}.}} 
}

\thispagestyle{netmon-page} 

\subsubsection{Decoupling Node and Agent Observations}

An intelligent router has two responsibilities: first, to build an internal representation of the network by storing experiences and communicating with neighbouring routers; second, to use this representation to make routing decisions. These tasks require different types of information and processing methods. For example, while packet-level data may be irrelevant for understanding the overall network, it is crucial for routing decisions. Similarly, GNNs are essential for learning network representations, but a simpler DQN may suffice for making routing decisions.

\begin{figure}[H]
    \centering
    \includegraphics[width=0.7\linewidth]{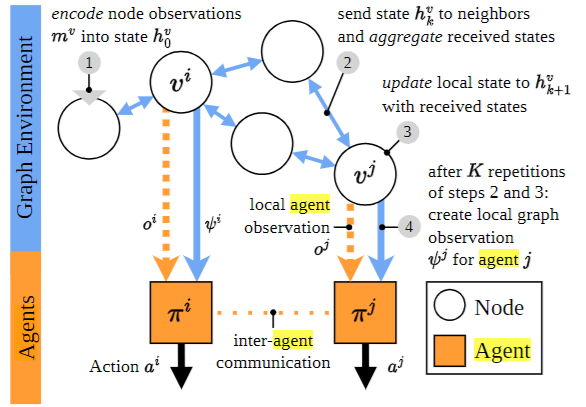}
    \caption{NetMon Process \cite{weil2024towards}: Nodes encode and share observations to update local states which are transferred to the agent for routing decisions.}
    \label{fig:netmon-process}
\end{figure}

Previous MARL approaches treat this as a single objective, requiring complex models and often limiting methods to centralised control or supervised learning. NetMon introduces a novel approach by decoupling graph representation learning from routing decisions. By removing the main source of non-stationarity—packet movement—graph representation learning becomes nearly stationary, therefore enabling effective graph representation learning without relying on centralisation or supervised learning.

\vspace{12pt}

The key advantage is that the system can be trained end-to-end using RL. When a packet arrives, the node model provides its learned graph representation to the agent, which combines it with packet-level observations to make routing decisions. The resulting loss is backpropagated through both models, refining the decision-making process and its understanding of the network.

\subsubsection{Multi-Round Recurrent Message Passing}

In large distributed systems, effective graph representation learning relies on efficient information propagation among agents. Stacking graph convolutional layers centralises execution and creates a rigid structure, while multi-round recurrent communication \cite{geyer2018learning} offers a more flexible decentralised approach for dynamic networks. NetMon uses two RNNs for recurrent message passing: one encodes local observations, and the other aggregates messages from neighbouring agents. Each node maintains a local hidden state \(h^v\), reset after each environment step, and a cell state \(c^v\), which persists. Despite its simplicity, Weil et al. demonstrated that NetMon vastly outperformed state-of-the-art message-passing models.

\begin{figure}[H]
\centering
\includegraphics[width=1\linewidth]{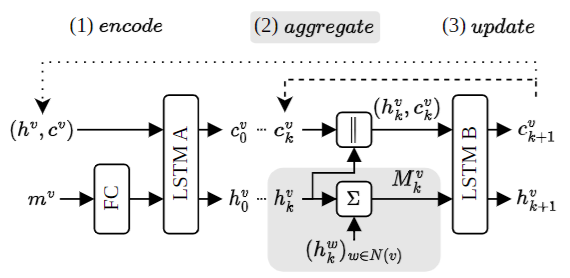}
\caption{\textit{NetMon} uses a recurrent message-passing model with two LSTM networks to encode and aggregate information across the network \cite{weil2024towards}.}
\label{fig:netmon-reccurent-message}
\end{figure}

\vspace{-6pt}

\begin{algorithm}[H]
\caption{NetMon Distributed Node State Update}
\label{alg:node_state_update}
\KwIn{Node $v$ with direct neighbors $N(v)$, state $s$, previous node state $h^v$, node observation $m^v$, and $k$ communication rounds}
$h^v_0 \gets \text{encode}(h^v, m^v)$ \tcp*{$\triangleright$ Encode node observation}
\For{$k \gets 0$ to $K-1$} {
    Send $h^v_k$ to all neighbors $w \in N(v)$\;
    Receive $h^w_k$ from all neighbors $w \in N(v)$\;
    $M^v_k \gets \text{aggregate}^k(\{h^w_k\}_{w \in N(v)})$\;
    $h^v_{k+1} \gets \text{update}^k(h^v_k, M^v_k)$\tcp*{$\triangleright$ Update with message passing}
}
\textbf{return:} Updated node state $h^v_K$;
\end{algorithm}

\subsubsection{Generalisable Pre-Training}

The final contribution is the strategy for training the model, Weil et al. pre-train their model across a diverse set of networks to ensure it is both generalisable and adaptable. This strategy mitigates the limitations of specialised models, such as susceptibility to local optima and the necessity for retraining when network conditions evolve. The pre-trained model can then be fine-tuned online during deployment, allowing for rapid adaptation to specific network conditions without the prohibitive costs and performance issues of training from scratch in a live environment. 

\chapter{System Design}
\label{cha:system-design}
\section{Design Objectives}

This design builds upon the existing NetMon framework \cite{weil2024towards}, focusing on dynamic communication to address key practical limitations of the current system. 

\vspace{12pt}

The key design objectives for the dynamic communication system are:

\begin{enumerate}
    \item \textbf{Optimise Routing Decisions}: Optimise routing decisions by dynamically weighting received information from neighbouring nodes, leading to more accurate graph representations.
    
    \item \textbf{Reduce Communication Overhead}: Minimise communication by selectively determining which nodes require updated information, reducing unnecessary data exchange and improving scalability.
    
    \item \textbf{Adapt to Dynamic Networks}: Implement a flexible aggregation mechanism and dynamic communication control to enhance generalisation and adaptability in variable, real-world network conditions.
\end{enumerate}


\section{Design Assumptions and Constraints}

The system's design is shaped by critical constraints that define the operational boundaries for effective performance in large-scale network packet routing.

\begin{longtable}{>{\raggedright}p{0.22\textwidth}>{\raggedright\arraybackslash}p{0.73\textwidth}}
\caption{Design Constraints: The following constraints define the operational bounds for an effective large-scale packet routing system} \\
\hline
\textbf{Constraint} & \textbf{Description} \\
\hline
\endfirsthead

\hline
\textbf{Constraint} & \textbf{Description} \\
\hline
\endhead

\hline
\endfoot

Large-Scale System & The system must efficiently scale to manage a large number of routers without performance degradation. \\
\hline
Decentralised Control & Each agent makes real-time decisions using local information, without a central controller. \\
\hline
Localised Communication & Each router communicates only with immediate neighbours, requiring local decision-making and efficient information propagation across the network. \\
\hline
\end{longtable}

While extensive literature exists on multi-agent reinforcement learning involving heterogeneous agents, multi-task objectives, and edge computing. This is not the primary objective of the design, therefore a series of simplifying assumptions were made for the proposed design and the simulations used to evaluate performance. 

\begin{longtable}{>{\raggedright}p{0.22\textwidth}>{\raggedright\arraybackslash}p{0.73\textwidth}}
\caption{Design Assumptions: The following assumptions outline the key simplifications made during the system's design process.} \\
\hline
\textbf{Assumption} & \textbf{Description} \\
\hline
\endfirsthead

\hline
\textbf{Assumption} & \textbf{Description} \\
\hline
\endhead

\hline
\endfoot

\hline
\endlastfoot

Homogeneous Agents & All routers are assumed to have identical hardware and software capabilities, with the same action and observation spaces, allowing parameter sharing. \\
\hline
Uniform Task Priority & All network packets are treated with equal priority, simplifying the routing algorithm to a single optimisation objective. \\
\hline
Idealised Control Plane & The system assumes instantaneous, unlimited, and lossless communication between routers on the control plane during environment steps, within the context of SDN. \\
\hline
Synchronous Communication & It is assumed that all agents can synchronise their communication rounds, ensuring consistent and predictable information propagation across the network. \\
\hline
Sufficient Computational Resources & Each router is assumed to have adequate memory and processing power to handle deep reinforcement learning computations. \\
\hline
\end{longtable}

\clearpage 

\section{Design Foundation}

The core architecture of the design closely follows the original implementation by Weil et al., with several minor adaptations to improve efficiency, as detailed in Section \ref{sec:design-config}. Each node in the network starts by encoding its local observations using an MLP. The encoded representation is first processed by RNN-A to capture local temporal information, after which the updated hidden state is transmitted to neighbouring nodes. Each node then aggregates the received hidden states to update its hidden state representation. This aggregated hidden state is passed through RNN-B, which processes the aggregated temporal information. The output hidden state of RNN-B therefore represents a temporal representation of the surrounding network. 

\begin{figure}[H]
    \centering
    \includegraphics[width=1\linewidth]{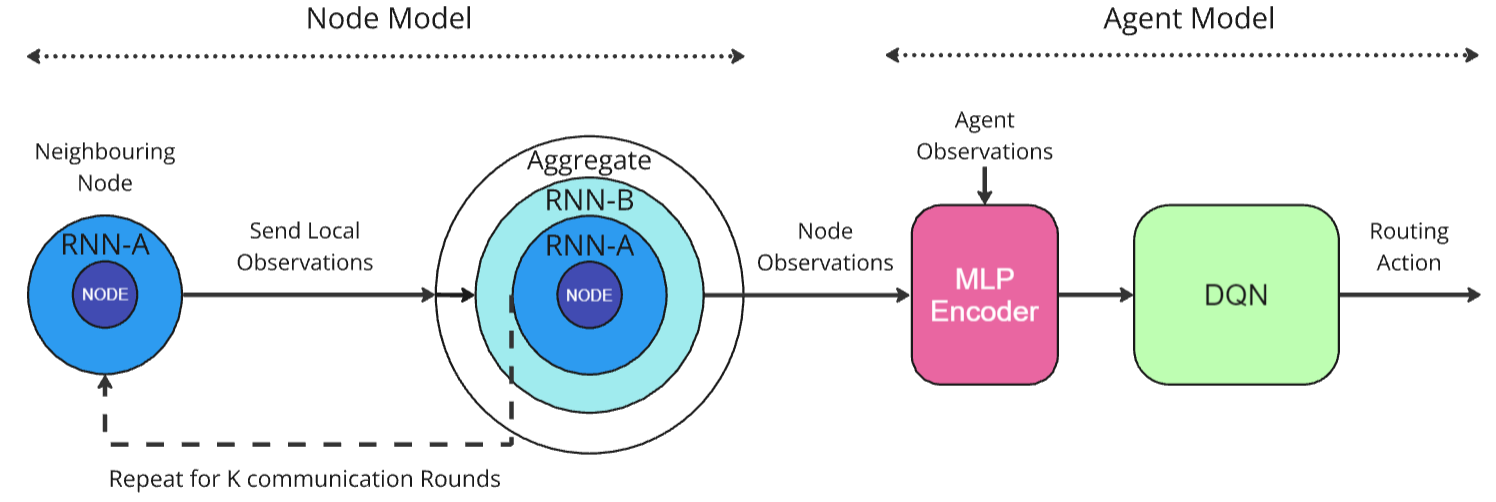}
    \caption{Foundational Architecture: Showing the left-to-right flow from the node model, where local observations are processed and exchanged between nodes, to the agent model, which integrates these observations and makes routing decisions.}
    \label{fig:baseline-architecture}
\end{figure} 

For the next communication round the updated hidden states are exchanged, aggregated and updated using RNN-B. This iterative process continues until the maximum number of communication rounds. At that point, the node model concatenates its updated local hidden state with the latest received neighbouring hidden states and transfers this to the agent model. The agent model then concatenates the node graph representation with any local packet-specific observations to feed into an MLP encoder. The encoded node and agent representation is then passed into a DQN to make routing decisions.

\section{Design Overview}

The design centres around dynamic node communication and introduces a novel, dynamic multi-round communication mechanism that improves the effectiveness and efficiency of information propagation. The mechanism dynamically aggregates information from received messages and intelligently selects which neighbouring nodes should receive the updated hidden state in subsequent communication rounds.

\begin{figure}[H]
\centering
\includegraphics[width=1\linewidth]{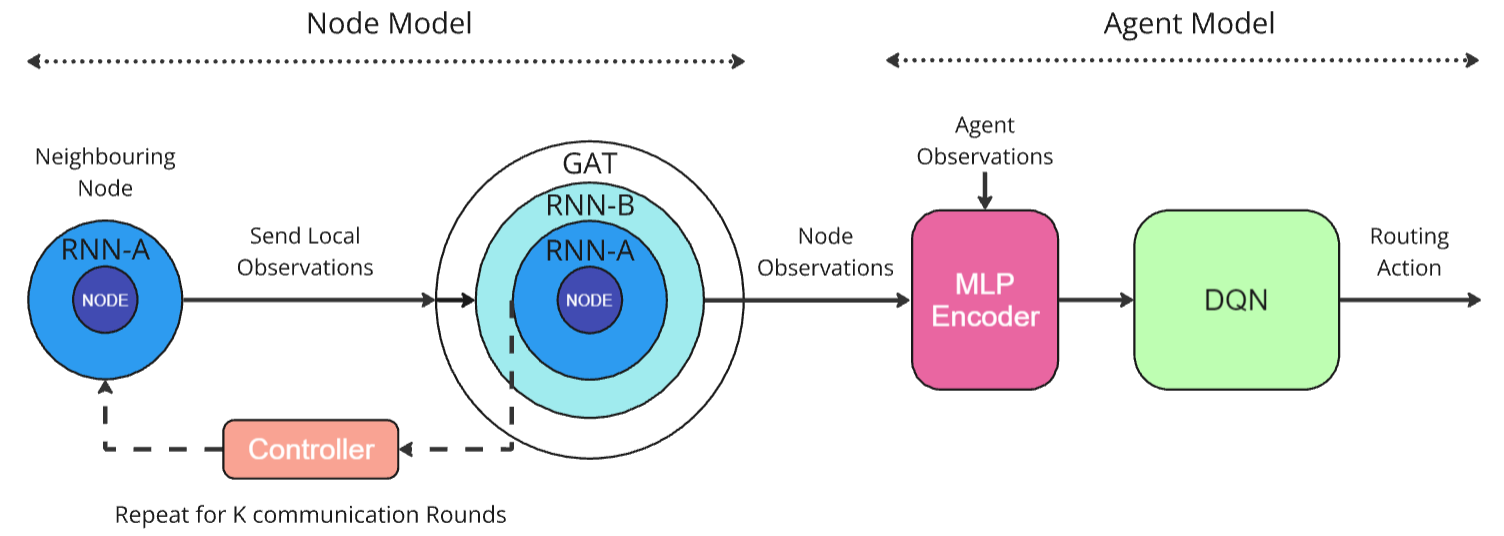}
\caption{Proposed Architecture: Flow through a GAT layer (white) for aggregation and an Iteration Controller (red) for dynamic communication control.}
\label{fig:controller-architecture}
\end{figure}

The design exclusively focuses on optimising the performance of the node model, further improvements to the agent model, beyond the existing implementation using a DQN, are left for future work. The design consists of two primary components, which are briefly summarised below. For a more detailed explanation, please refer to Chapters  \ref{cha:aggregation-mechanism} and \ref{cha:iteration-controller}:

\begin{enumerate}
    \item \textbf{Aggregation Mechanism}: This mechanism utilises a Graph Attention Network (GAT) \cite{velivckovic2017graph} to dynamically aggregate information from received messages. The GAT ensures that the most relevant information is collected and used to update the node's graph representation.

    \item \textbf{Iteration Controller}: This component is a dynamic multi-round communication controller responsible for determining which neighbouring nodes should receive the updated hidden state in each communication round. It optimises the communication process by selectively propagating the most relevant information to neighbouring nodes.
    
\end{enumerate}

\section{Training Process}

This section describes the data flow during training in a network packet routing environment, aiming to develop a model that performs robustly across varying network conditions. In practice, this would be followed by online fine-tuning within the target network. The system is trained end-to-end using reinforcement learning within a CTDE framework, where a central controller updates shared agent parameters without aggregating observations. Off-policy training allows agents to learn from replayed past experiences instead of only recent interactions.

\subsubsection{Example Implementation}

Adapted from Weil et al. \cite{weil2024towards}, Algorithms \ref{alg:data_collection_dqn} and \ref{alg:training_dqn} describe the off-policy, end-to-end training process for an independent DQN agent model within a network packet routing environment. Specifically, Algorithm \ref{alg:data_collection_dqn} focuses on the data collection phase, while Algorithm \ref{alg:training_dqn} outlines the off-policy training phase.

\vspace{12pt}

Node observations are denoted as $o_t$, node states as $h_t$, and the distributed node state update function from Algorithm \ref{alg:node_state_update_with_gat} as $U(h_t, m_t, s_t; \theta_U)$, where $\theta_U$ parameterises the function. This update function includes GAT aggregation and the Iteration Controller, returning the next state of all nodes, $h_{t+1} \doteq (h_{t+1}^v)_{v \in \mathcal{V}}$, along with the overall graph representation for each node $\psi_t \doteq (\psi_t^i)_{i \in \mathcal{I}}$ which corresponds to the concatenated neighbouring hidden states $h_{\text{concat}}$.

\addvspace{12pt}

Transition data is collected using an $\epsilon$-greedy exploration strategy. Given the use of RNNs, it is crucial to ensure that the length of transition sequences (sequence length = $J$) matches the RNN's unroll depth—the number of consecutive timesteps the RNN processes before making a prediction. Therefore, for each sample, $J$ consecutive timesteps are stored to capture the necessary temporal dependencies. This transition data is then stored in replay memory for future use.

\addvspace{12pt}

\begin{algorithm}[H]
\caption{Data Collection with Independent DQN}
\label{alg:data_collection_dqn}
\KwIn{Replay Memory $D$, Action-Value Function $Q$ with weights $\theta_Q$, Node State Update Function $U$ with weights $\theta_U$}
\For{episode $\gets 0$ \KwTo $\dots$}{
    $h_0 \gets 0$ \tcp{Initialize node states}
    Obtain $s_0$, $o_0$ by resetting the environment\;
    $m_0 \gets \text{MLP}(o_0)$ \tcp{Encode observation}
    \For{$t \gets 0$ \KwTo $T-1$}{
        $h_{t+1}, \psi_t \gets U(h_t, m_t, s_t; \theta_U)$ \tcp{Node state update}
        \For{each agent $i$}{
            Select random action $a_t^i$ with probability $\epsilon$\;
            \Else{select action $a_t^i \gets \mathbf{argmax}_{a} \, Q(o_t^i \mid \psi_t^i, a; \theta_Q)$\;}
        }
        Perform environment step with actions $a_t$ and get reward $r_t$, state $s_{t+1}$, and observation $o_{t+1}$\;
        $m_{t+1} \gets \text{MLP}(o_{t+1})$ \tcp*{Encode new observation}
        Store $(h_t, h_{t+1}, m_t, m_{t+1}, s_t, s_{t+1}, o_t, a_t, r_t, o_{t+1})$ in $D$\;
    }
}
\end{algorithm}

\addvspace{12pt}

After an initial exploration period, the model is trained off-policy at a specified frequency of environment steps.  During training, batches of transition data are sampled from the replay memory. For the first step in each sequence, the corresponding node state is retrieved from the replay memory, and the node state update is then simulated for the subsequent timesteps in the sequence.

\addvspace{12pt}

An example reward function, $Z_j$, assigns a scalar reward of 1 if a packet successfully reaches its destination node. The mean squared TD error is then calculated for each sample based on this reward. Losses are accumulated over the entire batch, and gradient descent is used to update the parameters of the behaviour network. The parameters of the target network are subsequently updated at the desired frequency.

\addvspace{12pt}

\begin{algorithm}[H]
\caption{Offline Training with Independent DQN}
\label{alg:training_dqn}
\KwIn{Replay Memory $D$, Action-Value Function $Q$ with weights $\theta_Q$, Target Weights $\hat{\theta}_Q$, Node State Update Function $U$ with weights $\theta_U$}
\For{batch sequence indices in $D_j \gets j_0$ \KwTo $j_0 + (J-1)$}{
    \If{$j = j_0$}{
        $h_{j'} \gets h_j$ \tcp{Load node state from replay memory}
    }
    $h_{j'+1}, \psi_{j'} \gets U(h_{j'}, m_j, s_j; \theta_U)$ \tcp{Train node state update}
    $h_{j'+2}, \psi_{j'+1} \gets U(h_{j'+1}, m_{j+1}, s_{j+1}; \theta_U)$ \tcp{Target input}
    $y_j \gets r_j + Z_j \gamma \mathbf{argmax}_{a} \, Q(o_{j+1} \mid \psi_{j+1}, a; \hat{\theta}_Q)$\;
    $Z_j \gets \begin{cases}
    0 & \text{if agent $i$ is done at step $j+1$}\\
    1 & \text{otherwise}
    \end{cases}$\;
    $L \gets L + \left(y_j - Q(o_j \mid \psi_j, a_j; \theta_Q)\right)^2$\;
}
Perform gradient descent on $L$ with respect to parameters $\theta_Q$ and $\theta_U$\;
Update target weights $\hat{\theta}_Q$\;
\end{algorithm}

.

\section{Design Configuration}
\label{sec:design-config}

This section outlines the key configuration settings for both the node and agent models within the system. These configurations were carefully chosen to optimise model performance and efficiency while maintaining consistency with the original implementation. Full configurations are provided in the appendix.

\vspace{12pt}

The node model is based on the configuration proposed by Weil et al., utilising two RNN cells with shared hidden and cell states. A \textit{Gated Recurrent Unit (GRU)} \cite{cho2014learning} was chosen over an LSTM for better computational efficiency. The RNN's unroll depth is set to 8, consistent with Weil et al.'s implementation, based on their finding that while greater unroll depth can enhance long-term stability, it also increases training time. In our experiments, the base implementation of NetMon consistently outperformed the environment's benchmarks, leading us to simplify the models. For example, we reduced the hidden dimension from 128 to 64 to better evaluate performance improvements.

\vspace{12pt}

Although any RL algorithm that operates using only an input feature vector could theoretically be used, this design opts for a DQN as the agent model, chosen for its simplicity and consistency with Weil et al.'s approach. The DQN model includes both behaviour and target networks. The agent model encodes its own observations along with shared observations from the local node. These encoded observations are then processed through the Q-network using a single linear layer of input and output sizes  $(128, D + 1)$ to generate value estimates for each potential routing action.

\chapter{Environment Setup}
\label{cha:environment-implementation}
This chapter details the testing environments used to evaluate the proposed design. It covers the graph generation process for creating datasets, the components of the observation space for both node and agent models, the setup of the testing environments, and the configuration settings for the experiments.

\vspace{12pt}

Due to time constraints and the need for consistent evaluation, the first environment—Shortest Path Regression—was directly adopted from Weil et al. \cite{weil2024towards}. The second environment, Dynamic Network Packet Routing, builds upon the static network packet routing environment from Weil et al. by simulating node failures, providing a more realistic representation of practical network conditions. A summary of each environment is presented below:

\begin{table}[H]
\centering
\caption{Overview of Testing Environments}
\begin{tabular}{>{\raggedright}p{0.2\textwidth}>{\raggedright\arraybackslash}p{0.75\textwidth}}
\toprule
\textbf{Environment} & \textbf{Description} \\
\midrule
Shortest Path Regression & A simplified environment designed for iterative testing of the node model, evaluating the effectiveness of information propagation by assessing each node's collective knowledge of the network. The task is framed as a multi-task supervised regression problem, predicting the shortest path length (delay) to each node. \\
\midrule
Dynamic Network Packet Routing & Simulates a realistic network to assess packet delivery efficiency. It tests the system's ability to manage and route traffic under dynamic conditions, including randomized bandwidth limitations on each edge and node failures. This environment evaluates the model's adaptability to sudden network changes, focusing on robustness and efficiency in unpredictable conditions. \\
\bottomrule
\end{tabular}
\label{tab:testing-environments}
\end{table}

\section{Graph Generation}
\label{sec:graph-generation}

The graph generation process for both tasks follows the method described by Weil et al. The process begins with randomly placing \(L\) nodes on a 2D plane. Each node connects to its nearest neighbors until it reaches a fixed degree \(D\), creating a discrete action space for each node. Disconnected graphs are excluded to ensure network connectivity. Although action masking \cite{schneider2021distributed} could accommodate variable-degree networks, this is not the focus of our experiments. Node delays, calculated as the 2D Euclidean distance, are rounded to the nearest integer to match environment steps, improving simulation efficiency.

\begin{figure}[H]
    \centering
    \includegraphics[width=1\linewidth]{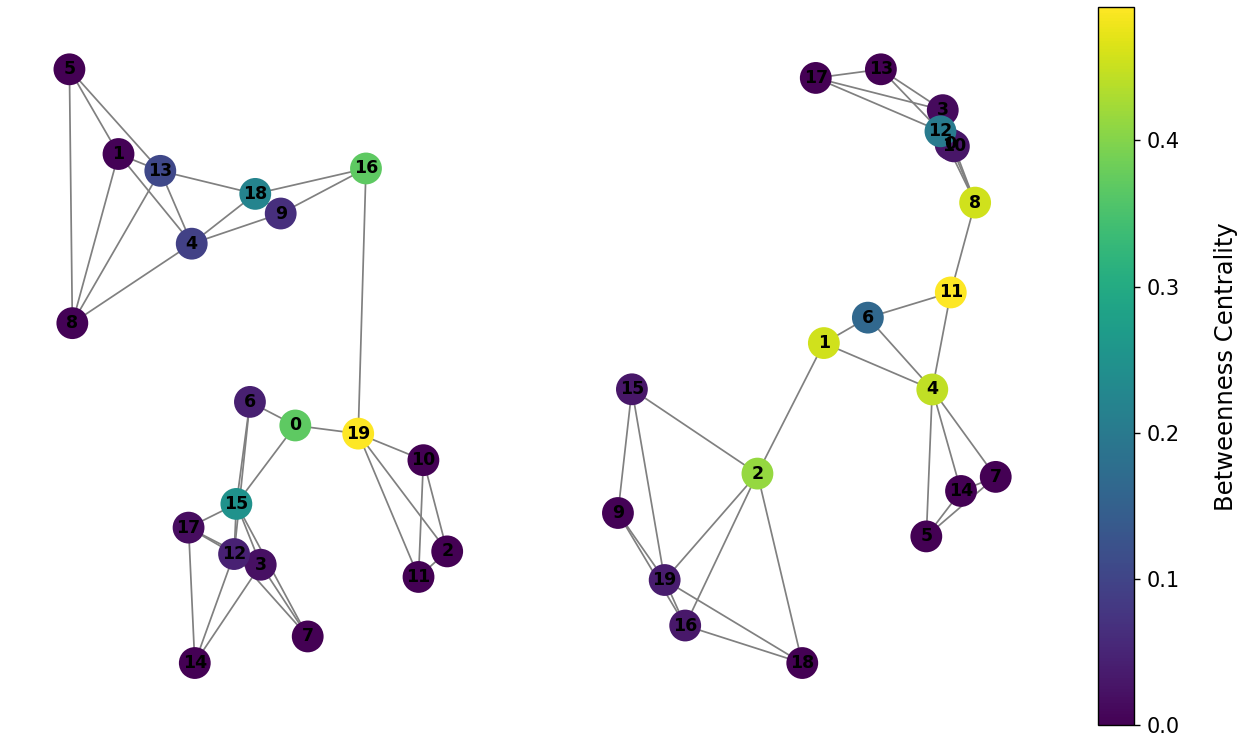}
    \caption{Example generated graphs with \(L = 20\) and \(D = 3\), where nodes are color-coded by their betweenness centrality.}
    \label{fig:graph-visualisation}
\end{figure}

Figure \ref{tab:network_metrics} provides key metrics for 1000 test graphs used in the Dynamic Network Packet Routing environment. These metrics are representative for both tasks, as the same graph generation method is used. Metrics include order (number of nodes), node degree (edges per node), size (total edges), diameter (in hops and delays), APSP (All-Pairs Shortest Paths), and node betweenness centrality.

\vspace{12pt}

APSP measures the shortest path between node pairs, expressed in hops or delays. The diameter, representing the maximum APSP, indicates the minimum hops/delay required to propagate information across the entire network. Betweenness centrality identifies potential network bottlenecks by measuring the proportion of shortest paths passing through each node. For further details on the generated graphs, such as the distribution of edge lengths (delay), refer to the appendix.

\begin{longtable}{p{7.25cm} p{1.5cm}<{\centering} p{1.5cm}<{\centering} p{1.5cm}<{\centering} p{1.5cm}<{\centering}}
\caption{Graph Statistics Summary for the Dynamic Network Packet Routing 1000-Graph Test Dataset} \\
\toprule
\textbf{Metric} & \textbf{Min} & \textbf{Max} & \textbf{Mean} & \textbf{Std} \\
\midrule
\endfirsthead

\toprule
\textbf{Metric} & \textbf{Min} & \textbf{Max} & \textbf{Mean} & \textbf{Std} \\
\midrule
\endhead

\bottomrule
\caption{(Continued)} \\
\endfoot

\bottomrule
\endlastfoot

Order & 20 & 20 & 20 & 0 \\
Node degree & 3 & 3 & 3 & 0 \\
Size & 30 & 30 & 30 & 0 \\
Diameter (hops) & 5.00 & 12.00 & 7.21 & 1.42 \\
Diameter (delays) & 8.00 & 23.00 & 12.84 & 2.72 \\
APSP (hops) & 0.00 & 12.00 & 3.33 & 1.92 \\
APSP (delays) & 0.00 & 23.00 & 5.70 & 3.60 \\
Node betweenness centrality & 0.00 & 0.70 & 0.15 & 0.12 \\
\end{longtable}
\label{tab:network_metrics}

Both tasks focus on generating a diverse set of graphs to create a robust and generalisable model. By exposing the model to a wide range of betweenness centrality values, we challenge it with networks that contain bottlenecks, improving its ability to generalise to more complex and varied topologies.

\begin{figure}[H]
    \centering
    \includegraphics[width=0.9\linewidth]{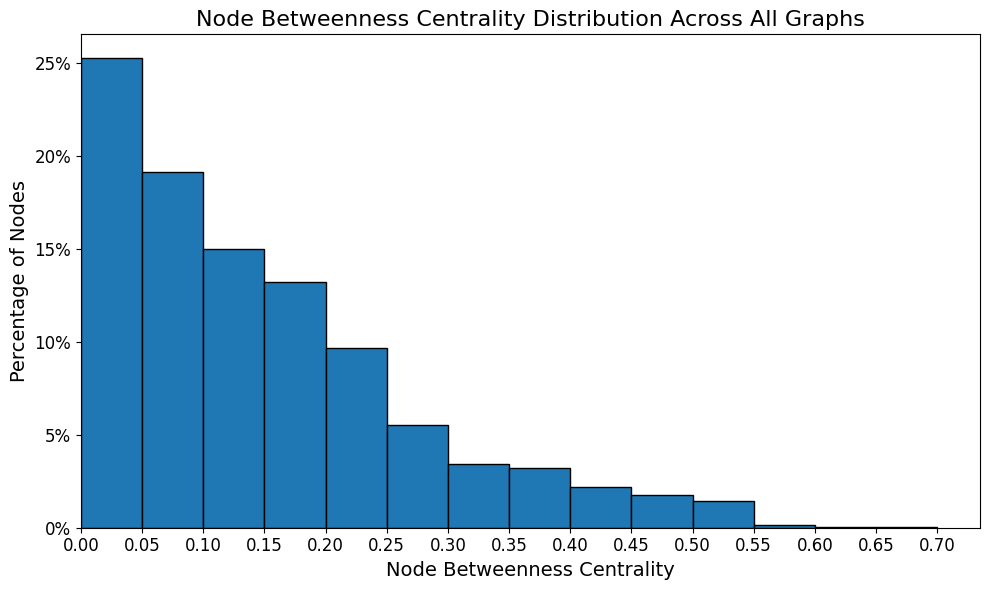}
    \caption{Histogram showing the distribution of node betweenness centrality across the 1000 test graphs for the Dynamic Network Packet Routing Environment.}
    \label{fig:node-betweenness-centrality-distribution}
\end{figure}

\begin{figure}[H]
    \centering
    \includegraphics[width=0.9\linewidth]{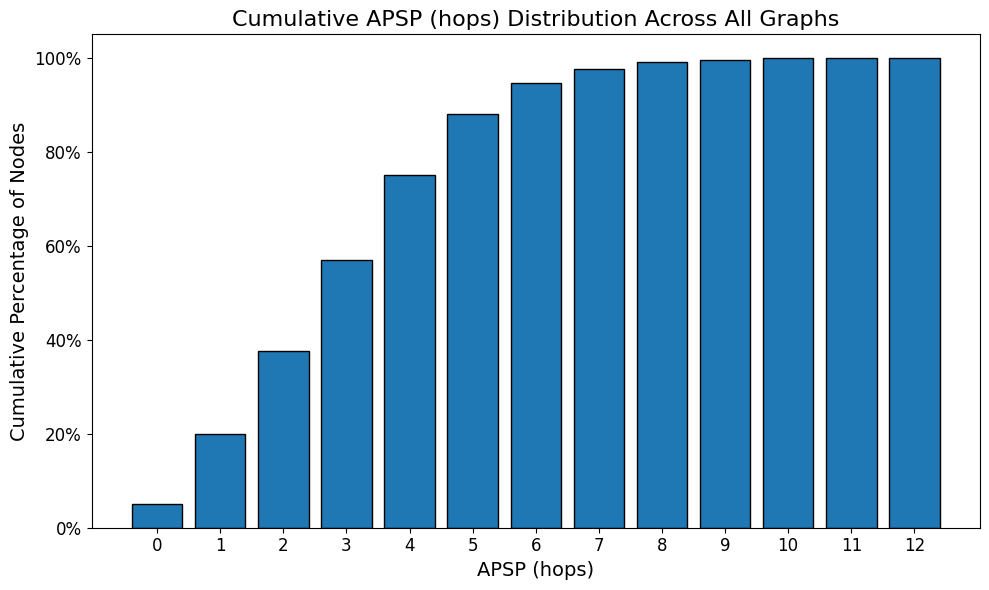}
    \caption{Cumulative distribution of APSP (hops) across the 1000 test graphs for the Dynamic Network Packet Routing Environment.}
    \label{fig:cumulative-apsp-hops-distribution-across_all_graphs}
\end{figure}

The cumulative APSP (hop) distribution in Figure \ref{fig:cumulative-apsp-hops-distribution-across_all_graphs} shows that 99\% of all shortest paths are under 8 hops. Weil et al. observed that for networks of this size, more than four communication rounds per step had a diminishing effect on information sharing, as nodes were already exchanging information in opposing directions.

\section{Observation Space}

Not all information required for routing decisions is necessary for learning graph representations. By removing packet-specific details, we can reduce the feature space for the node model. For instance, instead of knowing the size of each packet, the node model may only need to know the total load per node.

\begin{table}[H]
\centering
\caption{Node-level observations for learning graph representations, focusing on node status and neighbouring connections without packet-specific details}
\begin{tabular}{p{4.5cm} p{10cm}<{\raggedright}}
\toprule
\textbf{Observation} & \textbf{Description} \\
\midrule
Node ID & One-hot encoded unique router identifier. \\
Packet Count & Total number of packets at the router. \\
Total Load & Aggregate size of all packets at the router. \\
Neighbour Node IDs & One-hot encoded IDs of adjacent routers. \\
Neighbour Edge Lengths & Lengths of edges to neighbouring routers. \\
Neighbour Edge Loads & Data loads on edges to neighbouring routers. \\
\bottomrule
\end{tabular}
\label{tab:router-observations}
\end{table}

When a packet arrives at a node, the node shares its learned graph representation with the agent, offering a broader view of the network. For routing decisions, the agent also requires packet-specific information, including packet size, destination, current node, and relevant edge attributes. To prevent routing loops, knowledge of the previous node is crucial. Additionally, understanding the remaining time on the current edge allows the agent to better anticipate the future network state and make informed routing decisions.

\begin{table}[H]
\centering
\caption{Agent-level observation components needed for routing decisions, including packet-specific details and contextual information about the packet's journey.}
\begin{tabular}{p{4.5cm} p{10cm}<{\raggedright}}
\toprule
\textbf{Observation} & \textbf{Description} \\
\midrule
Packet ID & One-hot encoded unique packet identifier. \\
Packet Size & Size of the packet. \\
Destination Node & One-hot encoded target router. \\
Current Node & One-hot encoded current router. \\
Previous Node & One-hot encoded previous router or placeholder. \\
On Edge & Binary value indicating if the packet is on an edge. \\
Remaining Time & Time left to travel on the current edge. \\
Neighbour Node IDs & One-hot encoded IDs of neighbouring routers. \\
Neighbour Edge Lengths & Lengths of edges to neighbouring routers. \\
Neighbour Edge Loads & Loads on edges to neighbouring routers. \\
\bottomrule
\end{tabular}
\label{tab:packet-observations}
\end{table}

\section{Shortest Path Regression}
\label{sec:spr-experiment}

Weil et al. introduced a simplified evaluation environment aimed at quickly iterating on the node model design. This environment evaluates the quality of learned graph representations by predicting shortest path lengths in a multi-target regression task. Accurate predictions reflect a strong understanding of the network’s topology and state, which is expected to lead to better routing decisions in the dynamic network packet routing environment. A summary of the experiment configuration settings for both training and evaluation is provided below, with full configuration details available in the appendix.

\vspace{12pt}

The node model is trained using supervised learning on a labelled dataset that includes node observations and shortest path lengths from each node to every other node. This dataset is generated by resetting the routing environment (see Section \ref{sec:dynamic-experiment}) and collecting labelled initial observations. Training data is derived from 99,000 diverse static graphs, with an additional 1,000 graphs reserved for validation. While Weil et al.'s original implementation trained the model for 50,000 iterations, preliminary tests indicated continued improvement beyond this point, prompting an extension to 100,000 iterations. Model performance is evaluated using Mean Squared Error (MSE) between predicted and actual shortest path lengths, which also serves as the loss function during training.

\vspace{12pt}

Model performance is evaluated on 1,000 held-out test graphs, with results averaged across five different seeds to ensure statistical robustness. To assess the model's adaptability, sequence lengths—representing the number of prior timesteps used for predictions—are varied from 1 to 256. This variation tests the model's effectiveness in both static environments, where historical data may be more critical, and dynamic environments, where fewer preceding timesteps are relevant.

\section{Dynamic Network Packet Routing}
\label{sec:dynamic-experiment}

This environment simulates a small-scale network packet routing scenario, where the goal is to efficiently route packets from source to destination node within a dynamic network. Node failures and bandwidth congestion are simulated during both training and evaluation, mimicking the challenges of online learning within a live network. This setup aims to develop a model capable of forming robust, generalisable graph representations in a highly dynamic environment, providing a realistic test of the model's ability to handle unexpected network changes.

\begin{figure}[H]
    \centering
    \includegraphics[width=1\linewidth]{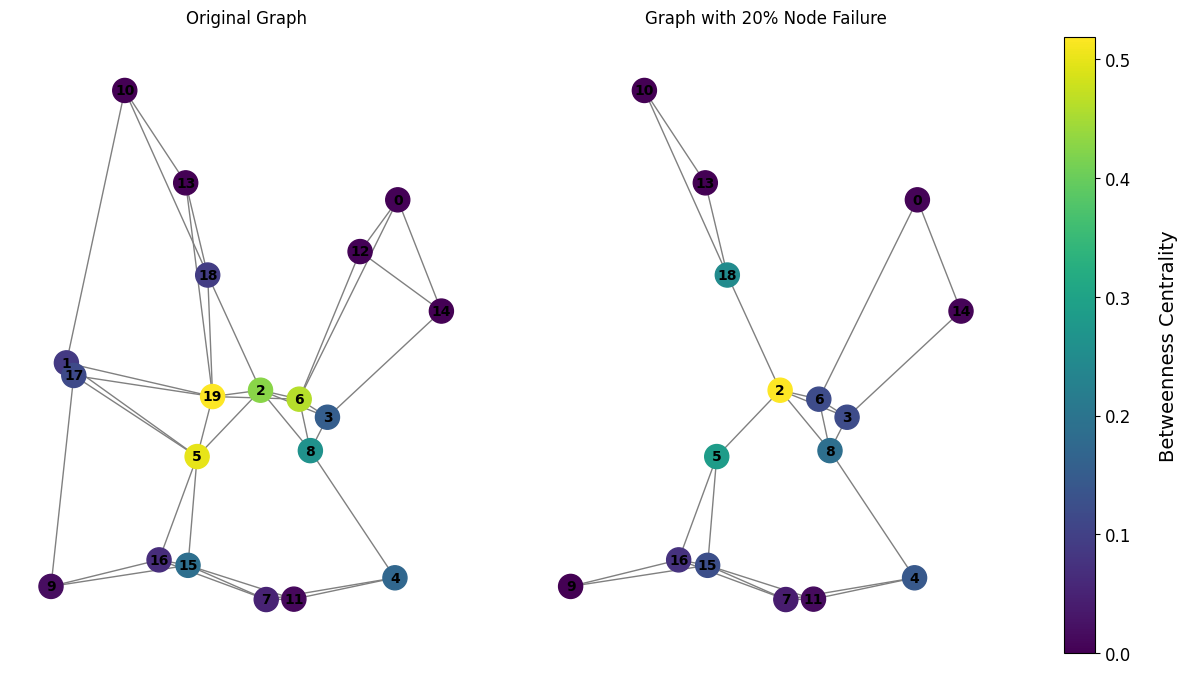}
    \caption{Comparison of original (left) and 20\% node failure (right) networks, with nodes colour-coded by betweenness centrality to show shifts in importance.}
    \label{fig:dynamic-dropout-graph}
\end{figure}

\clearpage 

To make the network dynamic, each node has a 20\% probability of failure at each environment step. When a node fails, it cannot share local observations or receive packets. Failure duration is randomly set between 5 and 10 steps, and no more than 40\% of nodes can be inactive at any given time to maintain network functionality.

\vspace{12pt}

The model is trained end-to-end using reinforcement learning, with rewards based on successful packet delivery. A reward of +10 is given for each successful delivery, incentivising efficient routing. All edges have a fixed bandwidth limit of 1, to encourage cooperative behaviour and reduce congestion, a penalty of -0.2 is applied whenever packets are "blocked" due to bandwidth limitations. The same -0.2 penalty is also applied for routing to inactive nodes, helping to minimise packet loss.

\vspace{12pt}

The model is trained over 1 million steps, with each episode capped at 50 steps. Although this is fewer steps than those used by Weil et al., their results suggest that this is sufficient for convergence. In each episode, a new random graph is generated, with 20 packets of randomised sizes [0,1] routed across 20 nodes. Off-policy training is performed every 10 steps, using a batch size of 32 and a sequence length of 8 steps. An $\epsilon$-greedy strategy is employed to balance exploration and exploitation during training, with $\epsilon$ set to zero during evaluation to fully utilise the learned policy. The training process minimises the mean squared TD error, measuring the model’s accuracy in predicting the effectiveness of its routing decisions.

\vspace{12pt}

The trained models are evaluated on 1,000 unseen test graphs to ensure robustness and generalisability, with results averaged across five different seeds for statistical reliability. Evaluation episodes are extended to 300 steps, allowing more packets to reach their destinations, thereby offering a more comprehensive assessment of the model's capabilities. Performance is measured using key network metrics such as Throughput and Delay, as well as indicators of cooperative behaviour, including the frequency of "Blocked" and "Looped" packets.

\begin{table}[H]
\centering
\caption{Performance Metrics for Dynamic Network Packet Routing Evaluation}
\begin{tabular}{>{\raggedright}p{0.25\textwidth} p{0.7\textwidth}<{\raggedright}}
\toprule
\textbf{Metric} & \textbf{Description} \\
\midrule
Throughput & Packets delivered to their destination node per step. \\
Delay & Steps taken for a packet to reach its destination node, used as a proxy for latency. \\
Blocked Packets & Number of packets unable to travel to their desired node due to bandwidth limitations and node inactivity. \\
Looped Packets & Number of packets revisiting previously visited nodes per step. \\
Shortest Path Ratio (SPR) & Ratio of actual steps taken by a packet to the minimum possible steps to reach the destination node. \\
\bottomrule
\end{tabular}
\label{tab:network-performance-metrics}
\end{table}

\chapter{Aggregation Mechanism}
\label{cha:aggregation-mechanism}
\section{Motivation and Objectives}

Weil et al.'s existing implementation, designed for maximum efficiency, uses summation to aggregate received node hidden states. While this approach is effective in static, fixed-degree networks, it has significant limitations.  The method overlooks the structure of the network, and the simplicity of summation may fail to capture complex inter-agent relationships and dependencies.

\vspace{12pt}

For practical network applications, the design must handle a variable number of input messages to withstand node failures and be capable of adapting to sudden network changes through dynamically prioritising message importance. Introducing a GAT layer within the recurrent message passing model, therefore, presents an opportunity to not only learn higher-quality graph representations through capturing more intricate agent interactions but also to make the overall system more robust and adaptable for practical network applications.

\section{Related Work}
\label{sec:related-work}

Leveraging graph convolutional layers within a message-passing framework is a common technique for improving feature representations. Models like the DGN \cite{jiang2018graph} stack these layers to propagate information, with each node using its own non-recurrent GNN.

\vspace{12pt}

However, stacking more layers tends to centralise the network's execution. The \textit{Anti-Symmetric DGN (A-DGN)} \cite{gravina2022anti} addresses this by instead using multiple decentralised communication rounds per step and adding a diffusion term to better capture long-range agent dependencies. To factor in temporal dependencies, the \textit{GCRN-LSTM} \cite{seo2018structured} uses an LSTM layer after the encoder, then aggregates the intermediate hidden states across agents using Chebyshev spectral graph convolutions \cite{defferrard2016convolutional}.

\vspace{12pt}

Despite relying on simple summation for message aggregation, NetMon's dual RNN message-passing framework outperformed the above methods in the Shortest Path Regression task \cite{weil2024towards}, indicating that incorporating graph convolutional layers could further improve performance.

\vspace{12pt}

\textit{MAGIC} \cite{niu2021magic} demonstrates the successful integration of GAT as a message aggregator, showing its ability to capture more complex relationships and dependencies between agents. However, among related works, only the Graph-Query Neural Network \cite{geyer2018learning} has effectively implemented an attention-based aggregation mechanism within a message-passing framework in a network packet routing environment. This approach, however, relied on supervised learning to learn graph representations, raising the question of whether end-to-end training using reinforcement learning could achieve similar or improved results.

\section{Design Concept}

The implementation of the GAT follows the standard approach outlined in Algorithm \ref{alg:gat}. Since the GAT is to be used within a decentralised multi-round communication system, more than one layer is unnecessary. Similarly, to maximise computational efficiency, only a single attention head is used.

\begin{figure}[H]
    \centering
    \includegraphics[width=1\linewidth]{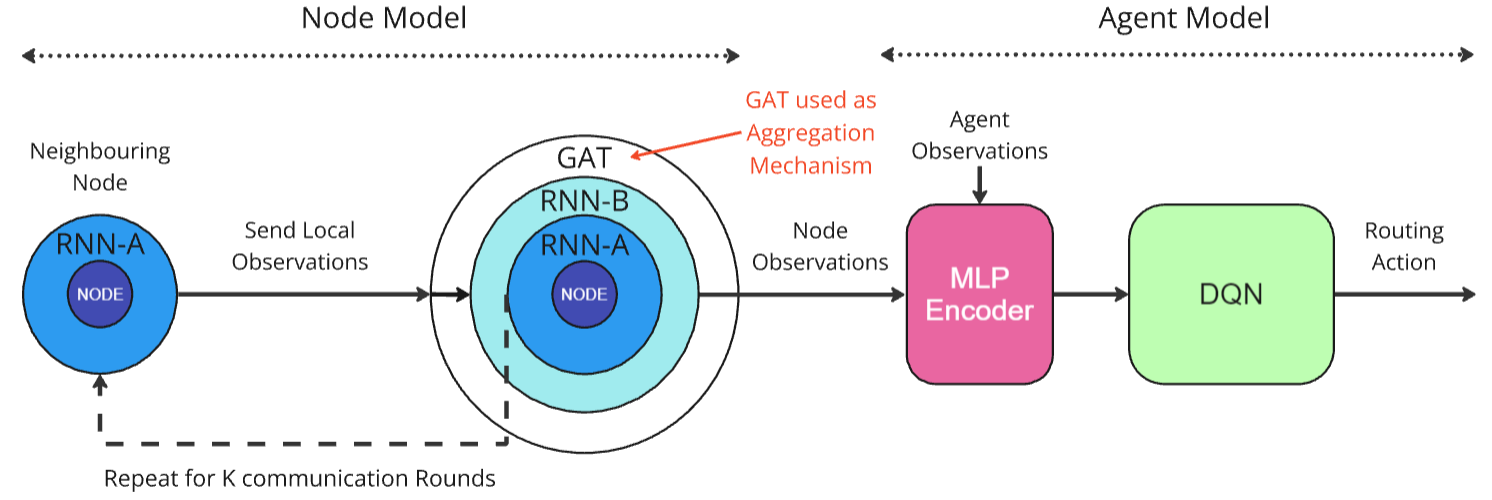}
    \caption{GAT aggregation: Neighbouring nodes process local observations with RNN-A, exchange states, and then the GAT aggregates these states before passing them to RNN-B. The output from RNN-B is then shared with the agent model.}
    \label{fig:gat-architecture}
\end{figure}

\addvspace{12pt}

The adapted distributed node state update with the GAT layer is outlined in Algorithm \ref{alg:node_state_update_with_gat}. The process begins with encoding node observations using an MLP. Next, the initial hidden and cell states of the RNN are initialised. The encoded observations are then fed into RNN-A to generate the initial node hidden state. In each communication round, the algorithm collects the hidden states of neighbouring nodes along with the node's own hidden state, then applies a linear transformation via a learnable weight matrix to dynamically adjust these representations.

\vspace{12pt}

The GAT layer then computes unnormalised attention scores \(e_{ij}\) through pairwise attention between the transformed states, applying a Leaky ReLU \cite{maas2013rectifier} activation for non-linearity. These scores are normalised using a softmax function to produce attention coefficients, reflecting the relative importance of each neighbouring node. These coefficients are used to weight the aggregation of the neighbouring nodes' transformed feature representations. The aggregated features, along with the current hidden and cell state pair, are passed through RNN-B to update each node’s temporal representation of the surrounding network. Finally, the hidden states from the latest communication round are concatenated with those of the neighbouring nodes to create a broader network representation. 

\addvspace{12pt}

\begin{algorithm}[H]
\caption{Distributed Node State Update with GAT Layer}
\label{alg:node_state_update_with_gat}
\KwIn{Node $v$ with direct neighbours $N(v)$, initial hidden state $h^v$ and cell state $c^v$, node observation $o^v$, adjacency matrix $A$, weight matrix $W$, communication rounds $K$, and activation function $\alpha$}

$m^v \gets \text{MLP}(o^v)$ \tcp*{Encode observation}
$(h^v_0, c^v_0) \gets \text{RNN-A}(m^v, c^v)$ \tcp*{Initial RNN-A pass}

\For{$k \gets 0$ to $K-1$} {
    $H^v_k \gets \{h^u_k \mid u \in N(v)\} \cup \{h^v_k\}$ \tcp*{Collect states}
    $H^v_k \gets H^v_k W$ \tcp*{Linear transform}
    
    \ForEach{edge $(i, j) \in A$}{
        $e_{ij} \gets \alpha(H^v_{k}[i], H^v_{k}[j])$ \tcp*{Pairwise attention}
    }
    \ForEach{node $i \in N(v) \cup \{v\}$}{
        $\alpha_{ij} \gets \frac{\exp(e_{ij})}{\sum_{k \in N(i) \cup \{i\}} \exp(e_{ik})}$ \tcp*{Normalise scores}
        $M^v_k \gets \sum_{j \in N(i)} \alpha_{ij} H^v_k[j]$ \tcp*{Aggregate messages}
        $(h^v_{k+1}, c^v_{k+1}) \gets \text{RNN-B}(M^v_k, c^v_k)$ \tcp*{Update with RNN-B}
    }
}
$h^{\text{concat}} \gets \text{concatenate}(h^v_K, \{h^u_K \mid u \in N(v)\})$ \tcp*{Concatenate final states}

\textbf{return:} Concatenated hidden states $h^{\text{concat}}$;
\end{algorithm}

\section{Methodology}

This research evaluates the GAT as an aggregation mechanism within a distributed recurrent message-passing system, focusing on three key aspects: the quality of graph representations from supervised learning, the effectiveness of end-to-end reinforcement learning, and the GAT's adaptability in dynamic networks.

\vspace{12pt}

The standard experiment configurations were used for both tasks. To magnify the impact of the aggregation mechanism, four communication rounds were used for the Shortest Path Regression task. However, due to computational limitations, only one communication round was used for the Dynamic Network Packet Routing task. 

\subsubsection{Baselines}

The GAT is benchmarked against three established aggregation methods: Summation, Mean, and a single Graph Convolutional Network (GCN) layer. Summation and Mean are straightforward and computationally efficient but do not differentiate based on the importance of messages. The GCN layer, on the other hand, leverages network structure to weight messages, providing a more sophisticated baseline that evaluates the importance of structural information within the aggregation mechanism. 

\section{Results and Discussion}

\subsection{Shortest Path Regression}

The objective of the Shortest Path Regression task is to evaluate the quality of the learned graph representations. This section discusses the challenges encountered, the proposed solutions, the behaviour of each aggregation mechanism during training, and their performance on an unseen test dataset.

\subsubsection{Challenges and Solutions}

\vspace{12pt}

\textit{Exploding Gradient Problem}

\vspace{12pt}

During preliminary testing, we observed sudden spikes in validation loss in the later stages of training, especially as the number of communication rounds increased. The root cause was traced to backpropagating losses after every environment step. In a multi-round communication system, this approach causes gradients to accumulate over multiple rounds, ultimately destabilising the learning process.

\begin{figure}[H]
    \centering
    \includegraphics[width=0.9\linewidth]{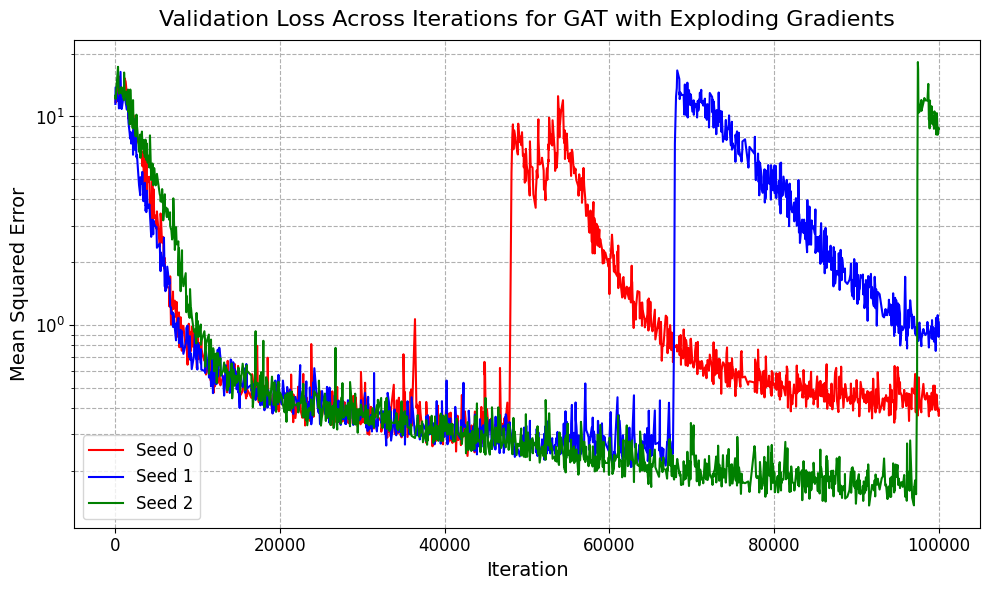}
    \caption{Sudden spikes in validation loss (log scale) during the later stages of training on the Shortest Path Regression task, observed across multiple seeds.}
    \label{fig:exploding-gradients-gat}
\end{figure}

To mitigate this, we incorporated gradient clipping, limiting gradient norms to a maximum of 1.0. This approach stabilised learning dynamics, improving the overall performance and reliability of the GAT. The integration of gradient clipping within the training process is shown in Algorithm \ref{alg:training_dqn_gradient_clipping}.

\addvspace{12pt}

\begin{algorithm}[H]
\caption{Offline Training with DQN and Gradient Clipping}
\label{alg:training_dqn_gradient_clipping}
\KwIn{Replay Memory $D$, Action-Value Function $Q$ with weights $\theta_Q$, Target Weights $\hat{\theta}_Q$, Node State Update Function $U$ with weights $\theta_U$, Clipping Norm $\lambda$, Sequence Length $J$, Communication Rounds $K$}
\For{batch sequence indices in $D_j \gets j_0$ \KwTo $j_0 + (J-1)$}{
    \If{$j = j_0$}{
        $h_{j'} \gets h_j$ \tcp{Load node state from replay memory}
    }
    \For{$k \gets 0$ to $K-1$}{
        $h_{j'+1}, \psi_{j'} \gets U(h_{j'}, m_j, s_j; \theta_U)$ 
        
        $h_{j'+2}, \psi_{j'+1} \gets U(h_{j'+1}, m_{j+1}, s_{j+1}; \theta_U)$ \tcp{Target input}
    }
    $y_j \gets r_j + Z_j \gamma \mathbf{argmax}_{a} \, Q(o_{j+1} \mid \psi_{j+1}, a; \hat{\theta}_Q)$\;
    $Z_j \gets \begin{cases}
    0 & \text{if agent $i$ is done at step $j+1$}\\
    1 & \text{otherwise}
    \end{cases}$\;
    $L \gets L + \left(y_j - Q(o_j \mid \psi_j, a_j; \theta_Q)\right)^2$\;
}
$\nabla_{\theta_Q} L$, $\nabla_{\theta_U} L \gets \text{compute gradients of } L \text{ w.r.t. } \theta_Q \text{ and } \theta_U$\;
$\nabla_{\theta_Q} L, \nabla_{\theta_U} L \gets \text{clip}(\nabla_{\theta_Q} L, \lambda), \text{clip}(\nabla_{\theta_U} L, \lambda)$\;
Perform gradient descent on $L$ for parameters $\theta_Q$ and $\theta_U$\;
Update target weights $\hat{\theta}_Q$\;
\end{algorithm}

\subsubsection{Training Behaviour and Trends}

All four aggregation mechanisms tested—GAT, GCN, Mean, and Summation—proved capable of learning high-quality graph representations in the supervised task, continuing to improve well beyond the 100,000 training iterations.

\vspace{12pt}

Mean aggregation proved effective in the early stages of training, providing a simple and consistent approach that allowed the model to focus on optimising the encoder and RNN components. However, as training progressed, treating all messages equally began to limit the model's ability to generate nuanced feature representations. Similarly, the limitations of Summation became apparent through the high variance observed in its validation loss curves. Unlike Mean aggregation, Summation lacks a normalisation mechanism, allowing certain node state feature vectors to dominate disproportionately. This leads to instability in the early iterations and ultimately hampers the model's ability to produce quality graph representations.

\addvspace{12pt}

GAT, on the other hand, although requiring extra iterations to jointly optimise the parameters of the encoder, RNN, and attention mechanism, proved capable of capturing more complex relationships between nodes and eventually converged to the lowest validation loss of all methods tested. Notably, however, GCN hindered learning relative to Mean and Summation. One hypothesis is that weighting messages using the graph structure biases predictions towards closer nodes, making it detrimental for predicting shortest paths across the network. 

\begin{figure}[H]
    \centering
    \includegraphics[width=0.9\linewidth]{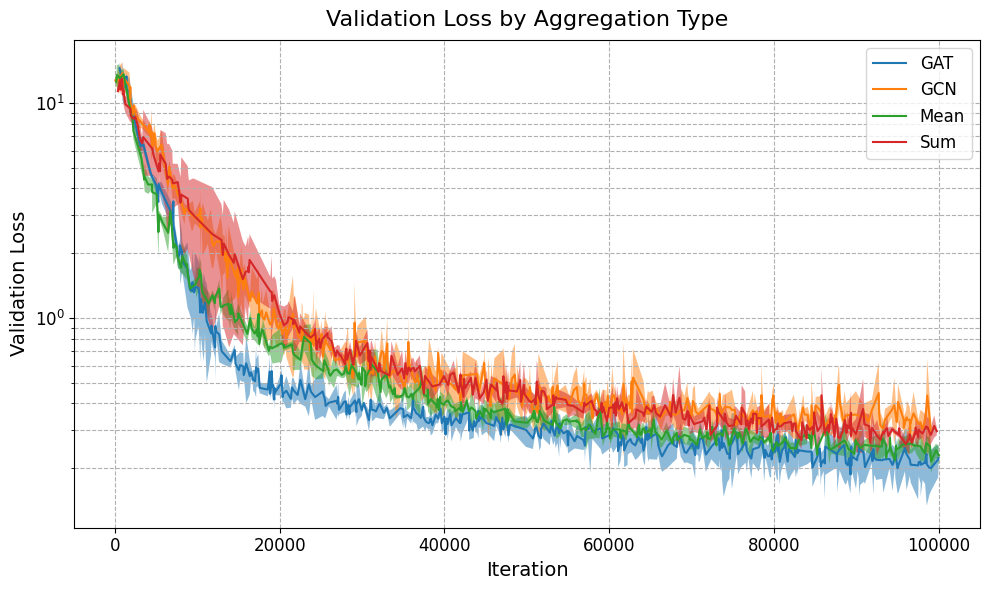}
    \caption{Validation loss over 100,000 iterations for the aggregation mechanisms. Logarithmic scale for the y-axis. Shaded areas show standard deviation bounds.}
    \label{fig:training_loss_curve}
\end{figure}

\subsubsection{Generalisation to Unseen Graphs}

\begin{figure}[H]
    \centering
    \includegraphics[width=0.9\linewidth]{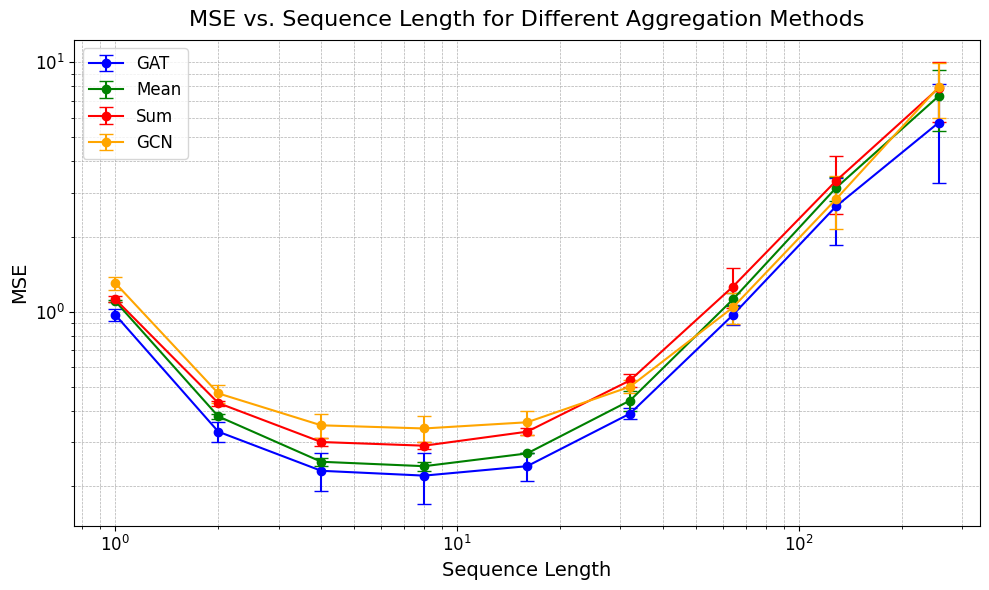}
    \caption{Test loss across sequence lengths for the aggregation mechanisms. Logarithmic scale for both axes. Error bars represent the standard deviation}
    \label{fig:mse_vs_sequence_length}
\end{figure}

GAT consistently outperformed other aggregation methods across all sequence lengths on the held-out test graphs, confirming its ability to learn richer, more generalisable, and robust graph representations. On average, GAT improved performance by 23.23\% over Summation, 12.73\% over Mean, and 24.59\% over GCN, for a detailed performance breakdown see the Appendix.

\vspace{12pt}

Consistent with Weil et al., all methods performed best when the observation window matched the RNN unroll depth of 8, with performance declining as the observation window length deviated from this value. GAT's consistent performance across all sequence lengths highlights its robust ability to generate high-quality graph representations, making it effective in both static and dynamic networks.

\subsection{Dynamic Network Packet Routing}

While improved graph representation learning is promising, the true potential of a dynamic aggregation system lies in its ability to handle variable inputs, reflecting real-world challenges such as hardware failures. This section compares GAT's performance in a dynamic network with baseline methods. The primary objectives are to evaluate GAT's effectiveness in online training within a dynamic, sparse-reward environment using end-to-end reinforcement learning and to determine whether the dynamic aggregation mechanism leads to improved routing metrics. 

\subsubsection{Training Behaviour and Trends}

\begin{figure}[H]
    \centering
    \includegraphics[width=0.9\linewidth]{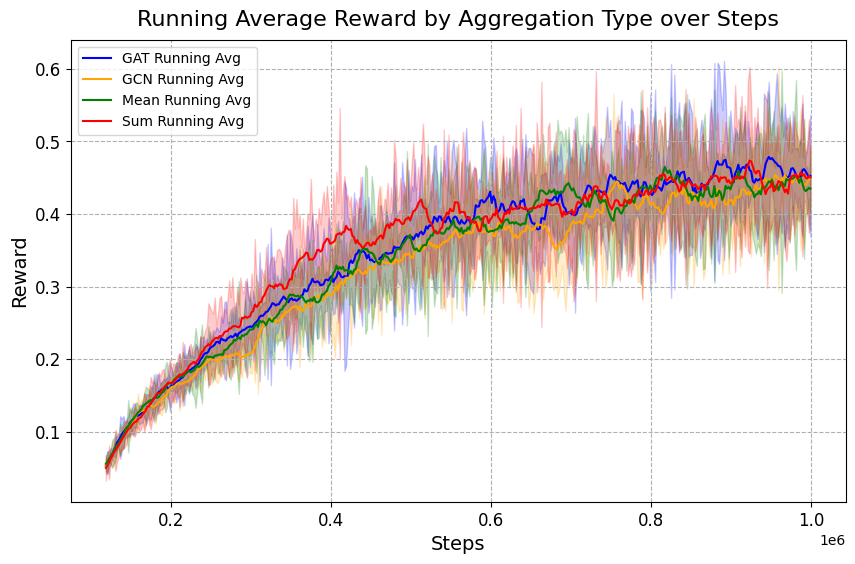}
    \caption{Running average (500-step) of Rewards in the Dynamic Network Packet Routing Environment. Shaded areas show standard deviation.}
    \label{fig:reward-dynamic-aggregation}
\end{figure}

The first notable observation is the significant volatility in rewards during training, which is expected in a sparse reward environment where major rewards are only given when a packet reaches its destination. Unlike in the Shortest Path Regression task, GAT neither converges faster nor achieves higher performance. This supports Geyer et al.'s \cite{geyer2018learning} conclusion that sparse reward networks lack the detailed feedback necessary to effectively train an attention mechanism solely through RL.

\vspace{12pt}

Although all methods converge to similar rewards, it's surprising that the Summation initially outperforms both Mean and GCN. Its fast convergence is expected due to its simplicity, but outperforming the others without a normalisation mechanism is unexpected. This can be attributed to two factors: centrally trained agents with shared parameters reduce the impact of singular disruptive messages, and the lack of regularisation may introduce exploration noise, accelerating learning.

\begin{figure}[H]
    \centering
    \includegraphics[width=0.9\linewidth]{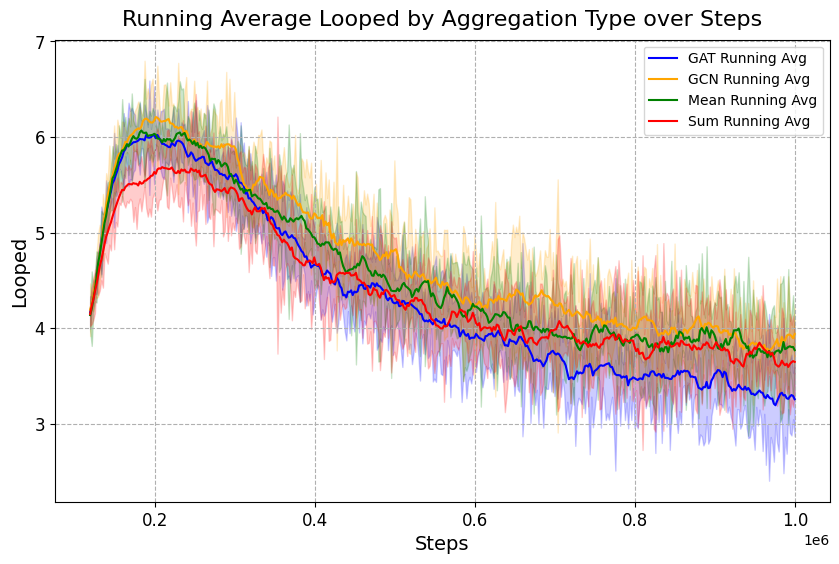}
    \caption{Running average (500-step) of Looped Packets in the Dynamic Network Packet Routing Environment. Shaded areas show standard deviation.}
    \label{fig:looped-dynamic-aggregation}
\end{figure}

However, since rewards are primarily driven by throughput, they don’t capture the full picture. The reduction in looped packets suggests that nodes are showing greater collective awareness of the network, avoiding unnecessary revisits to previously encountered nodes. This behaviour is critical in dynamic networks, where nodes must constantly update and reassess the importance of their neighbours.

\subsubsection{Routing Performance}

While all aggregation methods converged to similar rewards during training, testing over the longer 300-step window revealed GAT's clear advantages. GAT outperformed all baselines, achieving a 4.8\% increase in rewards, 4.2\% higher throughput, and 3.4\% lower delay compared to the next best method. These results suggest that in a highly dynamic, sparse reward network of this size, a 50-step evaluation may not capture enough successful packet deliveries to fully represent performance. However, this must be balanced against the greater diversity of training graphs when using shorter episodes, enhancing the model's generalisability.

\begin{figure}[H]
    \centering
    \includegraphics[width=1\linewidth]{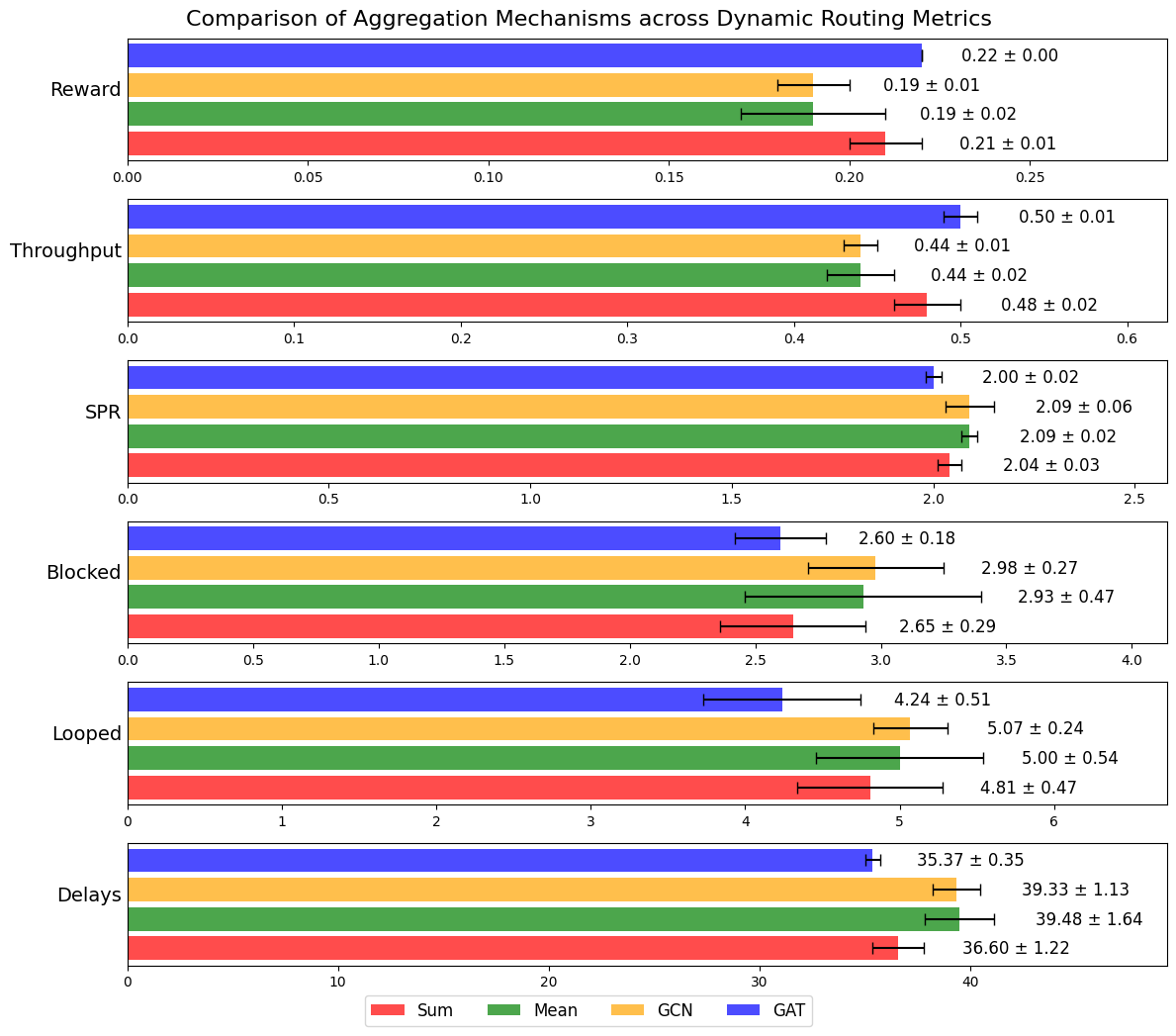}
    \caption{Routing metrics in the Dynamic Network Packet Routing environment, with each subplot having its own x-axis. Error bars represent standard deviations.}
    \label{fig:dynamic-aggregation-results}
\end{figure}

GAT reduced looped packets by 11.9\% compared to the next best method. This result supports the hypothesis that GAT enhances information propagation, providing a broader and more accurate view of the network, especially in dynamic environments. However, this improvement comes with increased variability in looped packets, indicating that the dynamic aggregation mechanism can lead to less consistent outcomes, with occasional lapses in critical information flow. Despite this, GAT demonstrated greater stability across other network metrics, with relatively low standard deviations compared to other methods. This highlights the importance of dynamically weighting node importance for maintaining stability in environments with frequent node failures or shifting topologies.

\vspace{12pt}

Interestingly, GCN exhibited the lowest performance among all aggregation mechanisms. This suggests that, in network packet routing applications, the network structure may not accurately reflect the relative importance of nodes, and relying on it could hinder performance compared to simpler averaging techniques.

\section{Summary of Results}

This study evaluated the effectiveness of the GAT as an aggregation mechanism in a distributed message-passing system, comparing its performance against three baseline methods—Summation, Mean, and a Graph Convolutional Network layer. The results consistently demonstrated that GAT outperformed these baselines, particularly in dynamic environments where its ability to assign varying importance to messages allowed it to capture more nuanced relationships between nodes, promoting cooperative behaviour and leading to performance improvements over both tasks. 

\vspace{12pt}

A key challenge identified during the study was integrating a learnable aggregation mechanism within a multi-round communication system, which initially caused issues like exploding gradients. This was successfully mitigated by implementing gradient clipping, ensuring stable training and convergence.

\vspace{12pt}

In the Shortest-Path Regression supervised learning task, GAT achieved the lowest MSE on the test dataset across all sequence lengths, underscoring its ability to efficiently propagate critical information throughout the network and learn higher-quality graph representations in both static and dynamic environments.

\vspace{12pt}

This capability extended to end-to-end reinforcement learning, where GAT proved to be the superior aggregation mechanism during the longer 300-step evaluation on the test dataset. Its effectiveness in training an attention-based aggregation mechanism in a sparse reward, dynamic environment can be largely attributed to the decoupling of node and agent observation spaces, creating a more stationary node environment. GAT achieved an 11.9\% reduction in looped packets—a key indicator of cooperative behaviour—compared to other methods. This improvement led to significant gains in network performance, including a 4.8\% increase in rewards, a 4.2\% increase in throughput, and a 3.4\% reduction in delay. Additionally, GAT demonstrated greater stability in dynamic environments relative to the other baselines, likely due to its ability to dynamically adjust node importance in response to changing network conditions.

\vspace{12pt}

In conclusion, GAT's advanced aggregation capabilities make it a powerful tool for enhancing network performance in complex and dynamic environments. Its ability to learn and adapt to varying node importance not only fosters more effective cooperative behaviour but also ensures more stable and efficient network operations.

\chapter{Iteration Controller}
\label{cha:iteration-controller}
\section{Motivation and Objectives}

The inherent challenge with multi-round communication systems is the significant communication overhead, which can burden bandwidth and escalate operational costs. While these systems typically improve performance relative to single-round communication and are crucial for propagating information across large distributed networks, they often lead to inefficiencies when every agent is required to communicate with all neighbours in each round.

\begin{figure}[H]
\centering
\includegraphics[width=0.5\linewidth]{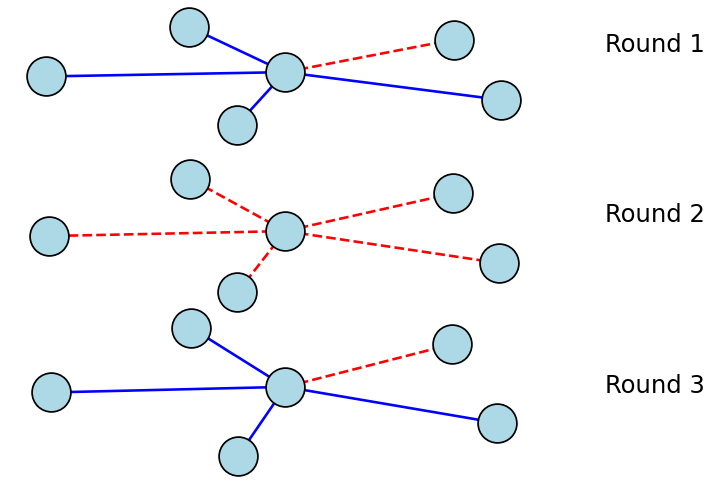}
\caption{Nodes evaluate neighbours' hidden states in each round, dynamically deciding whether to transmit (blue) or not transmit (red)}
\label{fig:controller-concept}
\end{figure}

The inclusion of GAT, a flexible and dynamic aggregation mechanism capable of handling a variable number of inputs, paves the way for developing a novel targeting system designed specifically to leverage the multi-round communication process to reduce communication overhead.

\section{Related Work}

Traditional targeting mechanisms in multi-agent systems typically fall into two categories. The first category relies on a central controller with greater observability to determine the relative importance of agent-to-agent interactions. This can be achieved through various methods, such as using a key-query matching system \cite{das2019tarmac, liu2020when2com}, inferring agent beliefs \cite{ding2020learning}, or pruning irrelevant connections \cite{liu2020multi}.

\vspace{12pt}

The second category, while capable of forming decentralised communication groups, leverages a centralised controller during the training phase to improve performance. Notable examples of this approach include MBC \cite{han2023model} and AC2C \cite{wang2023ac2c}, which incorporate supervised and self-supervised learning objectives, respectively. The proposed approach falls within this category but introduces a novel method that uniquely employs multi-round communication for decentralised targeting, utilising only end-to-end reinforcement learning without the need for additional supervised learning objectives.

\section{Design Concept}

This work introduces a decentralised targeting mechanism for multi-round communication. Each node maintains hidden states from neighbouring agents and uses an attention mechanism to decide whether to keep communication links active in the next round. This approach optimises communication by reducing unnecessary transmissions and creates a more adaptive system that assesses the importance of communication partners, enhancing performance in dynamic networks.

\begin{figure}[H]
    \centering
    \includegraphics[width=0.5\linewidth]{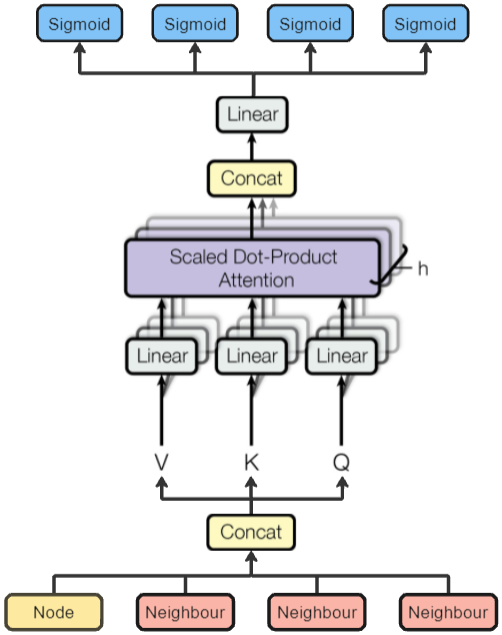}
    \caption{Iteration Controller Forward Pass: Node and neighbouring hidden states are concatenated, processed through an MHA mechanism, and projected to a sigmoid binary classifier to determine whether to transmit the updated state.}
    \label{fig:mha-iteration-controller}
\end{figure}

Each node processes the combined local and neighbouring states using Multi-Head Attention \cite{vaswani2017attention}. The output passes through a sigmoid function to produce a binary decision for each neighbour, indicating whether further communication is needed. If so, the node transmits its updated hidden state to the selected neighbours. This decision is based on the confidence in the current graph representation or the need for additional information. The process repeats until all nodes decide no further communication is necessary or the maximum number of rounds is reached. Finally, each node concatenates its own hidden state with the latest received states to form a collective network representation.

\addvspace{12pt}

\begin{algorithm}[H]
\caption{Iteration Controller Forward Pass}
\label{alg:attention_network_initial}
\KwIn{Input $\mathbf{x} \in \mathbb{R}^{M \times H}$, weights $W_Q, W_K, W_V \in \mathbb{R}^{H \times H}$, $W_{FC} \in \mathbb{R}^{H \times 1}$, bias $b_{FC}$, heads $h$, dimension per head $d_k = H / h$}
$\mathbf{Q}, \mathbf{K}, \mathbf{V} \gets \mathbf{x} W_Q, \mathbf{x} W_K, \mathbf{x} W_V$ \tcp*{Project to q, k, v}
\For{$i \gets 1$ to $h$}{
    $\mathbf{Q}_i \gets \mathbf{Q}[:, :, i \times d_k : (i+1) \times d_k]$ \tcp*{Slice query for head $i$}
    $\mathbf{K}_i \gets \mathbf{K}[:, :, i \times d_k : (i+1) \times d_k]$ \tcp*{Slice key for head $i$}
    $\mathbf{V}_i \gets \mathbf{V}[:, :, i \times d_k : (i+1) \times d_k]$ \tcp*{Slice value for head $i$}
    $\mathbf{A}_i \gets \text{softmax}\left(\frac{\mathbf{Q}_i \mathbf{K}_i^\top}{\sqrt{d_k}}\right) \mathbf{V}_i$ \tcp*{Scaled dot-product attention}
}
$\mathbf{A} \gets \text{Concatenate}(\mathbf{A}_1, \mathbf{A}_2, \ldots, \mathbf{A}_h)$ \tcp*{Concatenate attention heads}
$\mathbf{z} \gets \mathbf{A} W_{FC} + b_{FC}$ \tcp*{Project to number of neighbours}
$\mathbf{y} \gets \sigma(\mathbf{z})$ \tcp*{Apply sigmoid function}

\textbf{return:} $\mathbf{y}$
\end{algorithm}

\section{Methodology}

This research evaluates the impact of an Iteration Controller within a distributed message-passing system, focusing on three aspects: the quality of learned graph representations, communication overhead, and routing performance in dynamic network environments.

\vspace{12pt}

Standard experimental configurations are used throughout. The Iteration Controller is first assessed on the Shortest Path Regression task using 1 to 4 communication rounds to evaluate its effectiveness across different overhead levels. It is then tested in the Dynamic Network Packet Routing environment with 4 communication rounds to maximise the Iteration Controller's impact. Summation is used as the aggregation mechanism to isolate the Iteration Controller's effects.

\subsubsection{Baselines}

Two baselines are used to evaluate the Iteration Controller. First, the Iteration Controller is compared against the Maximum Communication baseline. Then, its average communication overhead is used to set the communication probability for the Matched Communication baseline.

\begin{table}[H]
\caption{Baseline communication strategies tested against the Iteration Controller.}
\centering
\begin{tabular}{>{\raggedright}p{0.22\textwidth}>{\raggedright\arraybackslash}p{0.73\textwidth}}
\toprule
\textbf{Baseline} & \textbf{Description} \\ \midrule
\textbf{Maximum Communication} & Assesses the Iteration Controller's impact on performance and overhead by comparing it to a fixed system with the maximum communication rounds. \\ \addlinespace
\textbf{Matched Communication} & Matches the communication overhead of the Iteration Controller using a set communication probability for each communication round to isolate the performance impact. \\
\bottomrule
\end{tabular}
\end{table}

\section{Results and Discussion}

\subsection{Shortest Path Regression}

This section evaluates the Iteration Controller's graph representation quality across different overhead levels, covering the challenges encountered and their respective solutions, along with analysis of the performance during training and testing.  

\subsubsection{Challenges and Solutions}

\vspace{12pt}

\textit{Bias to Communicate}

\vspace{12pt}

\begin{figure}[H]
    \centering
    \includegraphics[width=0.9\linewidth]{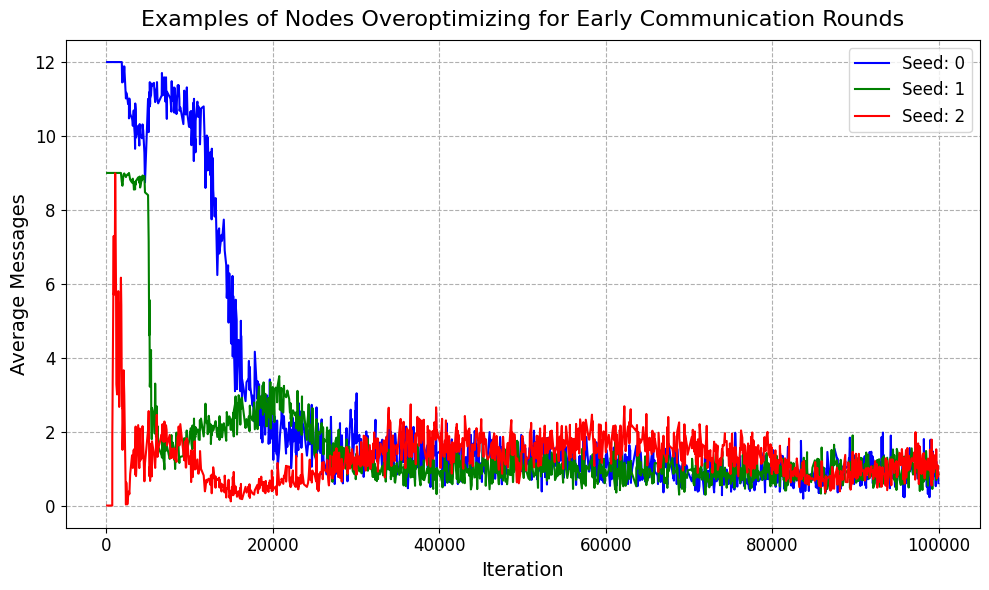}
    \caption{Average messages per node per step (out of 12) on the Shortest Path Regression task, showing the initial neglect of later communication rounds.}
    \label{fig:dropping-avg-messages}
\end{figure}

The initial design exhibited instability, especially during the early stages of training. Agents overly focused on early communication rounds, prioritising basic initial information while neglecting later rounds, which they perceived as noise. This led to rapid convergence but ultimately hindered long-term performance.

\begin{figure}[H]
    \centering
    \includegraphics[width=0.5\linewidth]{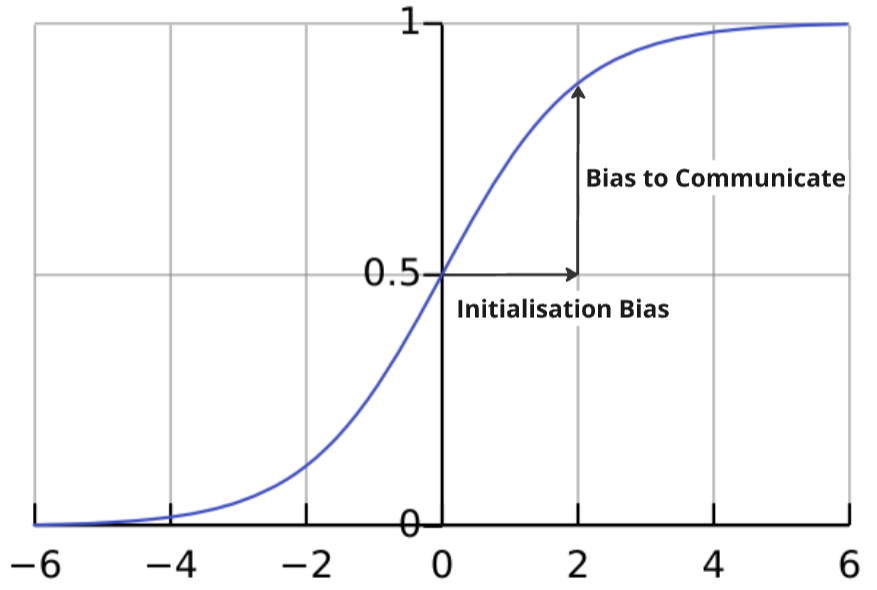}
    \caption{Increasing the bias weight initialisation shifts sigmoid input, biasing the system towards outputting 1 (communicate).}
    \label{fig:sigmoid-bias}
\end{figure}

To address this issue, two measures were introduced. First, the bias weights of the final feed-forward layer before the sigmoid function were increased using a "Communication Bias" hyperparameter. The default zero-initialisation of the bias weights, combined with Xavier initialisation \cite{glorot2010understanding} of the weight matrix, resulted in near-zero input to the sigmoid function. By increasing the bias initialization, agents were encouraged to maintain communication in later rounds.

\begin{figure}[H]
    \centering
    \includegraphics[width=0.9\linewidth]{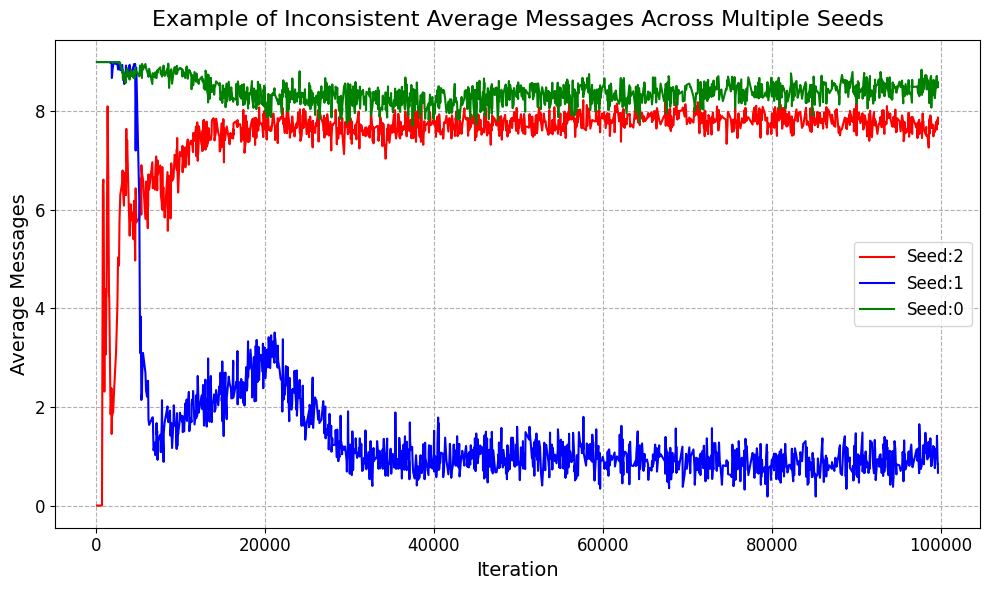}
    \caption{Average messages per node per step (out of 12) on the Shortest Path Regression task, highlighting volatility from insufficient exploration.}
    \label{fig:inconsistant-avg-messages}
\end{figure}

\addvspace{12pt}

\textit{Introducing Exploration Noise}

\addvspace{12pt}

Despite these adjustments, Figure \ref{fig:inconsistant-avg-messages} shows how the results remained inconsistent across different seeds, with agents either maximising communication or avoiding non-critical communication altogether. To counter sub-optimal policy adoption due to insufficient exploration, random noise, scaled by the "noise scale" hyperparameter, was added to the output logits before applying the sigmoid function. This noise encouraged exploration by introducing variability to the agents' actions, preventing them from getting stuck in sub-optimal policies and promoting more robust learning. The adjustment stabilised behaviour and led to consistent outcomes across different seeds. The final design of the Iteration Controller is outlined in Algorithm \ref{alg:attention_network_final}. 

\addvspace{12pt}

\begin{algorithm}[H]
\caption{Iteration Controller Modified Forward Pass}
\label{alg:attention_network_final}
\KwIn{Input $\mathbf{x} \in \mathbb{R}^{M \times H}$, weight matrices $W_Q, W_K, W_V \in \mathbb{R}^{H \times H}$, fully connected weight $W_{FC} \in \mathbb{R}^{H \times 1}$, bias $b_{FC}$, number of heads $h$, dimension per head $d_k = H / h$, noise scale $\epsilon$, communication bias $\beta$}
$\mathbf{Q} \gets \mathbf{x} W_Q, \; \mathbf{K} \gets \mathbf{x} W_K, \; \mathbf{V} \gets \mathbf{x} W_V$ \tcp*{Project to q, k, v}
\For{$i \gets 1$ to $h$} {
    $\mathbf{Q}_i \gets \mathbf{Q}[:, :, i \times d_k : (i+1) \times d_k]$\;
    $\mathbf{K}_i \gets \mathbf{K}[:, :, i \times d_k : (i+1) \times d_k]$\;
    $\mathbf{V}_i \gets \mathbf{V}[:, :, i \times d_k : (i+1) \times d_k]$\;
    $\mathbf{A}_i \gets \text{softmax}\left(\frac{\mathbf{Q}_i \mathbf{K}_i^\top}{\sqrt{d_k}}\right) \mathbf{V}_i$ \tcp*{Scaled dot-product attention}
}

$\mathbf{A} \gets \text{Concatenate}(\mathbf{A}_1, \mathbf{A}_2, \ldots, \mathbf{A}_h)$ \tcp*{Concatenate attention heads}
$\mathbf{z} \gets \mathbf{A} W_{FC} + b_{FC} + \beta$ \tcp*{Include communication bias in projection}
$\mathbf{\tilde{z}} \gets \mathbf{z} + \mathcal{N}(0, \epsilon^2)$ \tcp*{Add exploration noise}
$\mathbf{y} \gets \sigma(\mathbf{\tilde{z}})$ \tcp*{Apply sigmoid function}

\textbf{return:} $\mathbf{y}$\;
\end{algorithm}

\addvspace{12pt}

\textit{Hyperparameter Optimisation}

\addvspace{12pt}

While these additional hyperparameters improve performance, they also add complexity, requiring further optimisation. Due to limited computational resources and time, the remaining hyperparameters were adopted from Weil et al. \cite{weil2024towards}.

\addvspace{12pt}

Performance degrades significantly when sequence length deviates too much from the RNN unroll depth, leading to high variability. To show a consistent trend, total MSE for sequence lengths (2, 4, 8, 16, 32) is used as a performance proxy, excluding extreme deviations. A grid search tested various hyperparameter combinations on the validation dataset, as shown in the scatter plot below. Refer to the appendix for figures on the isolated impact of communication bias and noise scaling on validation loss.

\vspace{12pt}

To balance performance with reduced communication overhead, we chose a noise scaling of 0.3 and a communication bias of 0.5. A high communication bias was crucial for maintaining performance by ensuring later communication rounds weren't ignored, while moderate noise scaling offered sufficient exploration to ensure consistency, whilst providing the stability required to learn an effective policy.

\begin{figure}[H]
\centering
\includegraphics[width=0.9\linewidth]{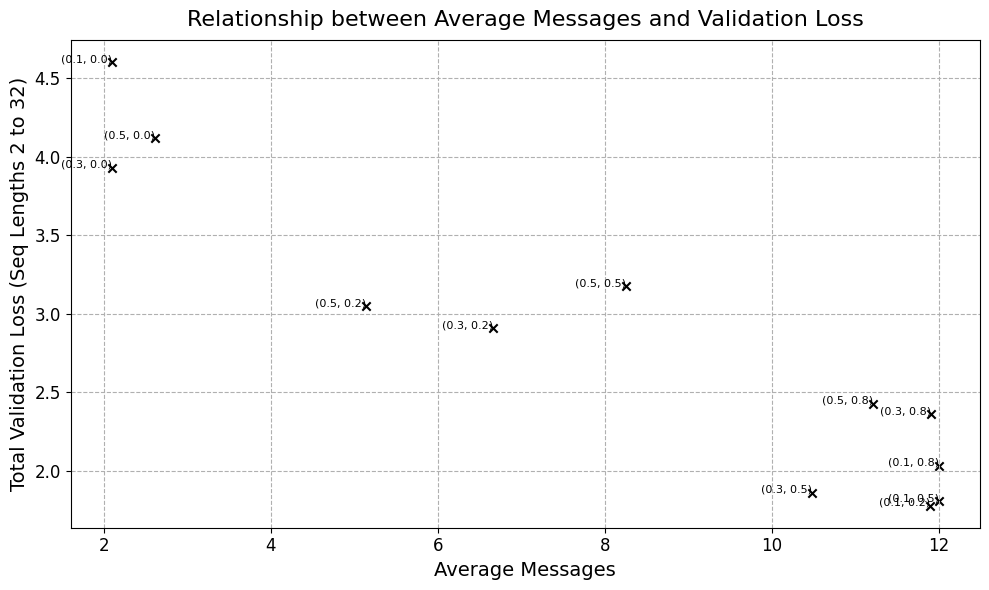}
\caption{Scatter plot of Average Messages vs. Total Validation Loss (summed over sequence lengths 2 to 32). Point labels (Noise Scaling, Communication Bias).}
\label{fig:iteration-controller-hparam}
\end{figure}

\subsubsection{Training Behaviour and Trends}

\begin{figure}[H]
    \centering
    \includegraphics[width=0.9\linewidth]{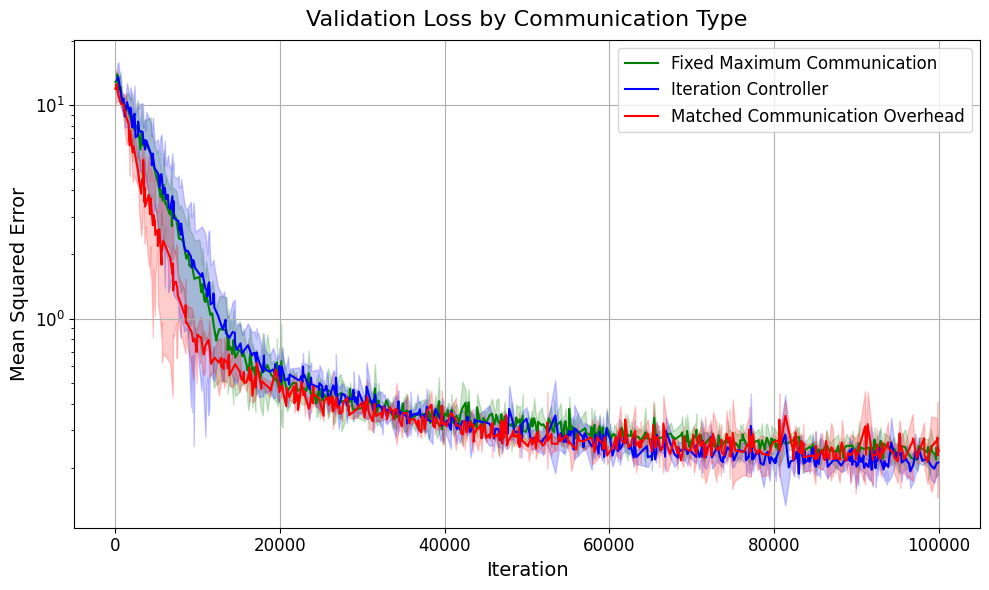}
    \caption{Validation loss (log scale) per Communication Type (4 communication rounds) on Shortest Path Regression task. Shaded areas show standard deviation.}
    \label{fig:controller-validation-training}
\end{figure}

In the early stages of training, the Matched Communication baseline learned the fastest, benefiting from fewer learnable parameters compared to the Iteration Controller and lower overhead than the Maximum Communication baseline which created a less noisy environment for identifying critical information. The Iteration Controller exhibited greater volatility early on as it optimised its additional parameters. However, all three methods eventually converged to similar validation losses, with minimal differences in performance.

\begin{figure}[H]
    \centering
    \includegraphics[width=0.9\linewidth]{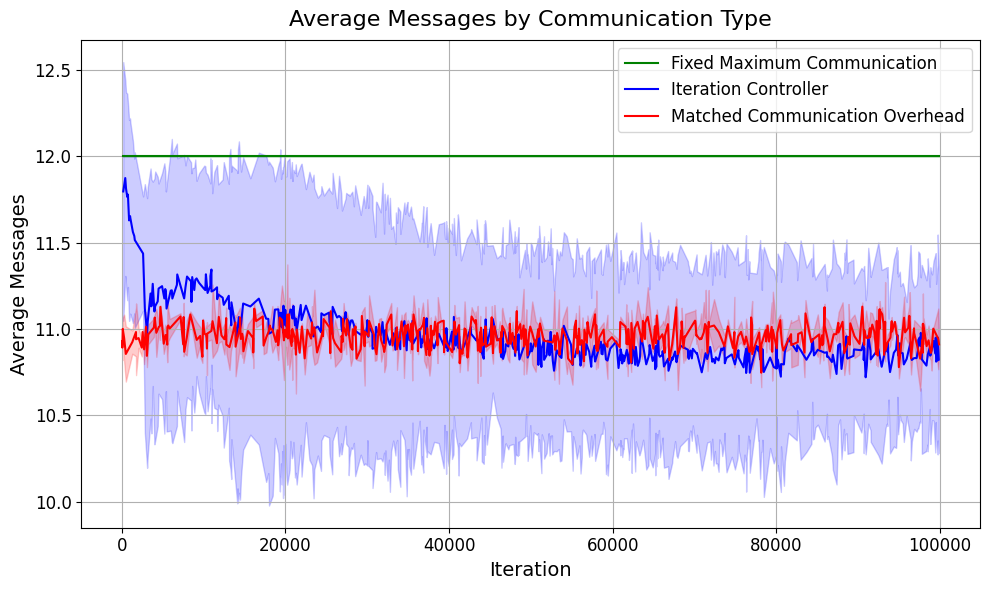}
    \caption{Averages Messages per Node (max = 12) per Communication Type (4 communication rounds). Shaded areas show standard deviation.}
    \label{fig:training-controller-messages}
\end{figure}

The advantages of the Iteration Controller become evident when analysing communication overhead. Initially, the average number of messages sent per node per iteration is close to the maximum, demonstrating the effectiveness of the Communication Bias parameter. However, this number rapidly decreases, eventually converging to around 11 messages, resulting in an 8.3\% overhead reduction.

\vspace{12pt}

As expected, the exploration noise is more pronounced in the early stages of training, causing greater variability in communication overhead. Over time, however, the Iteration Controller becomes more confident in its predictions, with outputs gravitating toward the extremes of the sigmoid function. This behaviour effectively reduces the impact of exploration noise, similar to an $\epsilon$-greedy exploration policy.

\vspace{12pt}

Breaking down the impact of individual communication rounds on overhead reveals notable trends. The Iteration Controller causes a sharper initial reduction in communication for later rounds, quickly identifying that these rounds are less critical for establishing an early, preliminary policy. As training progresses, the Iteration Controller continues to reduce communication in later rounds, while early rounds stabilise more quickly, highlighting the importance of early-round information. Eventually, all rounds converge to stable values, although later rounds exhibit greater volatility, reflecting the less frequent need for long-distance communication..

\begin{figure}[H]
    \centering
    \includegraphics[width=0.9\linewidth]{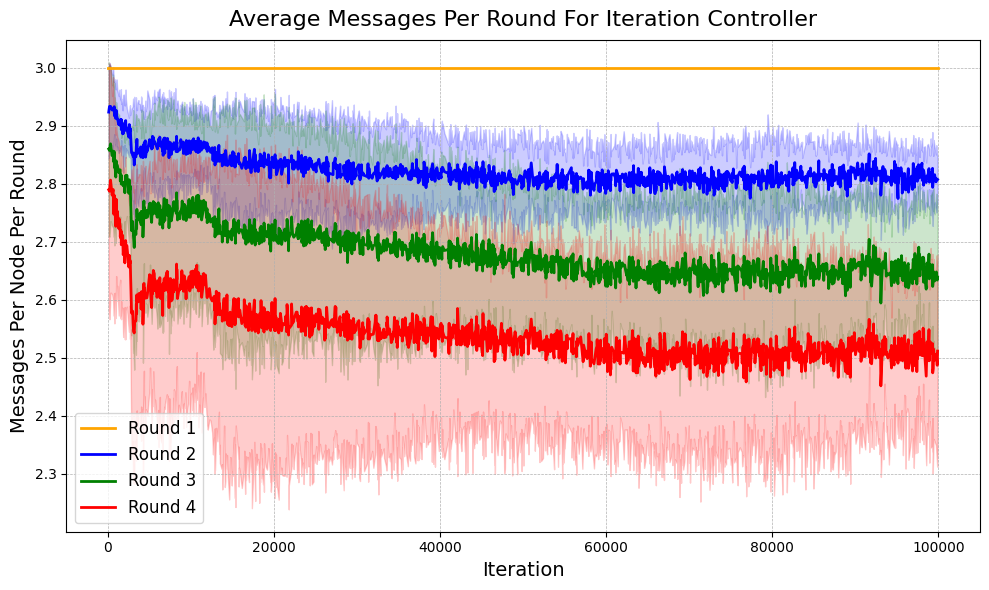}
   \caption{Averages Messages per Round (max = 3) for the Iteration Controller on the Shortest Path Regression task. Shaded areas show standard deviation.}
    \label{fig:messages-per-round-training}
\end{figure}

\subsubsection{Generalisation to Unseen Graphs}

The first step in evaluating performance impact is to assess the quality of learned graph representations on unseen graphs. While one might expect that dynamically limiting communication could lead to missed critical information and hinder performance, the test results reveal a more nuanced outcome.

\begin{figure}[H]
    \centering
    \includegraphics[width=0.97\linewidth]{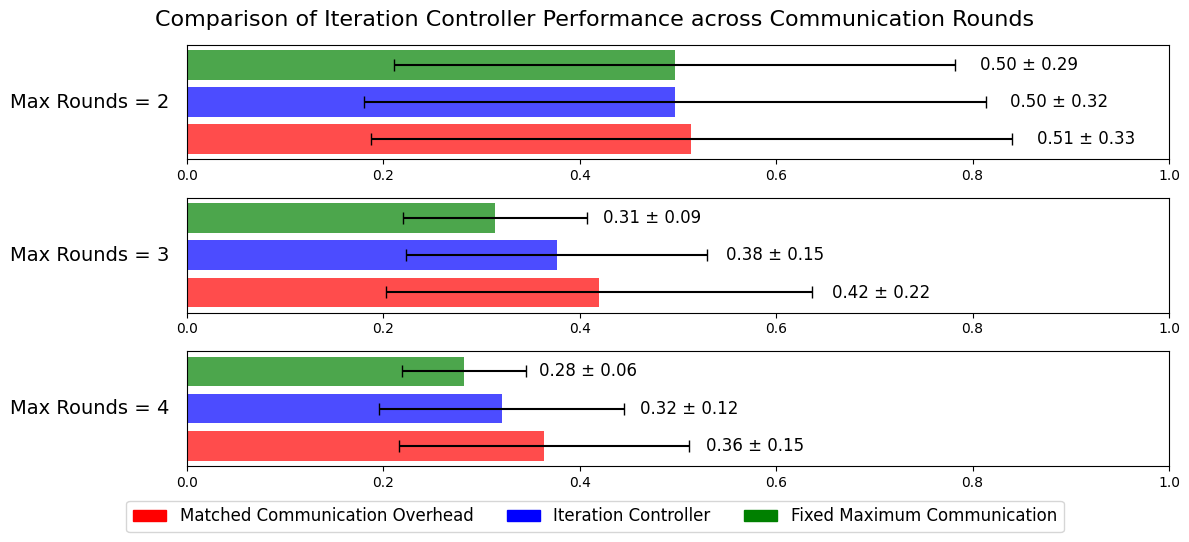}
    \caption{Total MSE across sequence lengths (2,4,8,16,32) on the test graphs. Error bars represent the standard deviation}
    \label{fig:sl-variable-comm-controller}
\end{figure}

As expected, Figure \ref{fig:sl-variable-comm-controller} shows that increasing the number of communication rounds improves prediction accuracy and enhances graph representation quality across all methods tested. Consequently, it is not surprising that the Fixed Maximum Communication baseline outperforms the Iteration Controller due to the higher communication overhead. However, the benefits of the Iteration Controller become evident when compared to the Matched Communication baseline, delivering an 11.1\% improvement in performance at 4 maximum communication rounds.

\vspace{12pt}

Deep diving into the performance of each communication strategy at 4 maximum communication rounds reveals some interesting trends. The Maximum Communication baseline performs best at shorter sequence lengths, likely because, with fewer preceding time steps, maximizing information exchange becomes more critical. However, as sequence lengths increase, excessive information exchange becomes detrimental, potentially overloading nodes with irrelevant information.

\begin{figure}[H]
    \centering
    \includegraphics[width=0.9\linewidth]{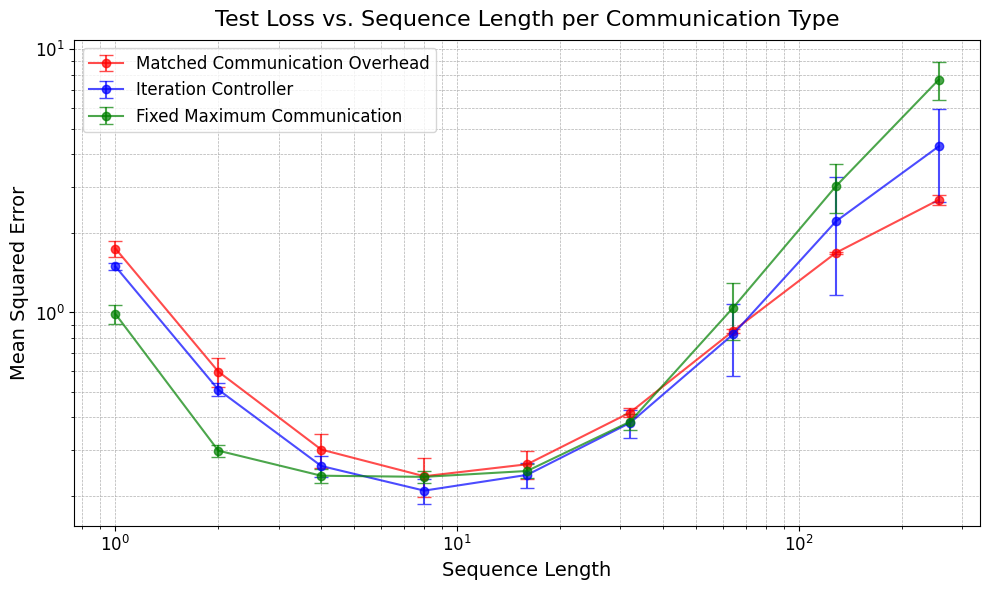}
    \caption{Test MSE across sequence lengths per Communication Type (4 rounds). Logarithmic scale for both axes. Error bars represent the standard deviation.}
    \label{fig:test-loss-sl-controller}
\end{figure}

This leads to the Iteration Controller outperforming both baselines when the sequence length matches the RNN unroll depth at 8. This suggests that dynamic communication control improves information propagation, resulting in higher-quality graph representation learning at the optimal sequence length. However, as sequence lengths increase, dynamic communication control leads to more volatile outcomes. In these cases, the Matched Communication baseline often outperforms, as longer sequences combined with higher overhead can overwhelm nodes with excessive information and reduce performance. Refer to Table \ref{tab:mse_supervised_controller_comparison} in the Appendix for detailed mean and standard deviation values.

\vspace{12pt}

\clearpage %

The second key aspect to assess is the impact on communication overhead. With up to four communication rounds, the Iteration Controller reduced total overhead by 8.3\% compared to the Maximum Communication baseline. This reduction is even more significant in the later rounds, with a 15.7\% decrease in average message count by the fourth round. These results indicate that the Iteration Controller becomes increasingly effective as the number of communication rounds grows, further optimising efficiency in systems with higher communication demands.

\begin{figure}[H]
    \centering
    \includegraphics[width=1\linewidth]{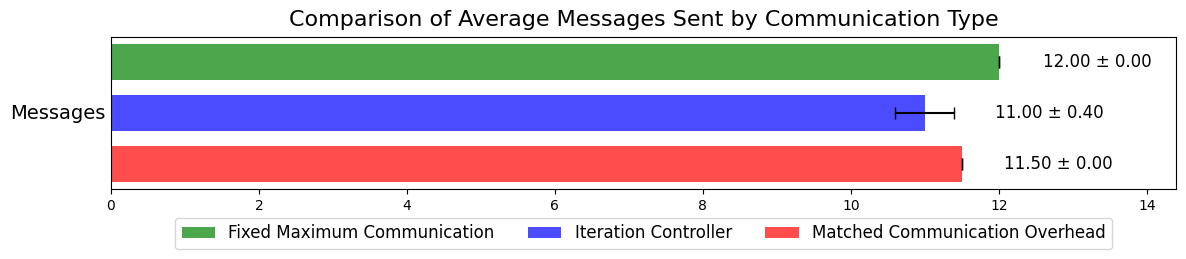}
    \caption{Average messages per Communication Type (4 communication rounds) on the Shortest Path Regression task. Error bars represent the standard deviation}
    \label{fig:communication_efficiency_comparison}
\end{figure}

\begin{figure}[H]
    \centering
    \includegraphics[width=0.7\linewidth]{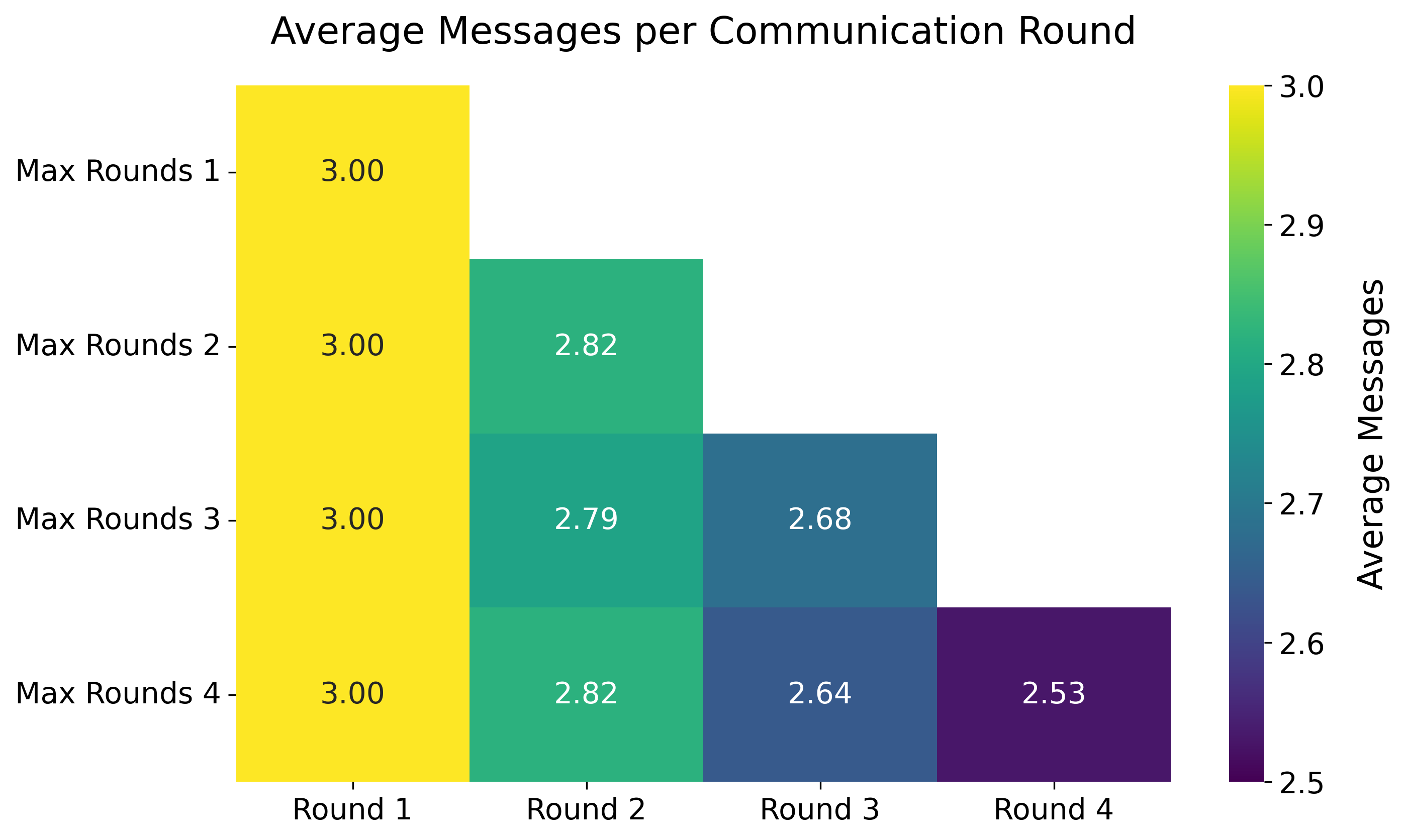}
    \caption{Heatmap showing the Average Messages per Node per Communication Round when using the Iteration Controller.}
    \label{fig:messages-per-round-heatmap}
\end{figure}

\subsection{Dynamic Network Packet Routing}

Building on the promising results from the Shortest Path Regression task, this section evaluates the Iteration Controller's performance in the more challenging Dynamic Network Packet Routing environment. We compare its performance against the two baselines at four maximum communication rounds.

\subsubsection{Training Behaviour and Trends}

As anticipated, training was highly volatile across all methods due to reward sparsity and the challenges of online training in a dynamic network. Despite this volatility, the Iteration Controller appeared to converge faster and achieve higher reward values compared to the Matched Communication baseline, performing similarly to the Maximum Communication baseline. This trend is also evident in other metrics like throughput and the Q-values of the behaviour network, detailed in the Appendix. 

\begin{figure}[H]
    \centering
    \includegraphics[width=0.9\linewidth]{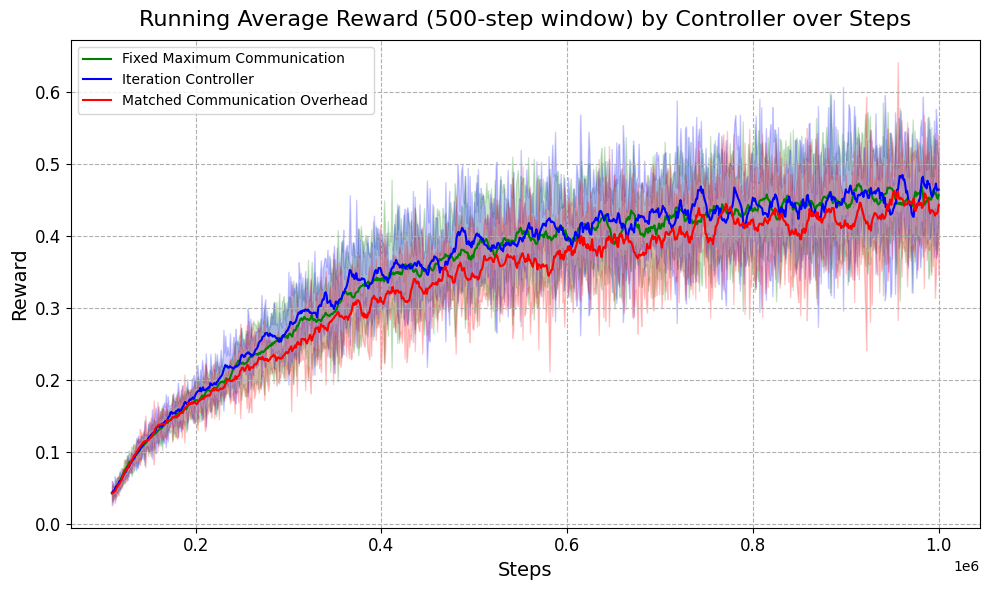}
    \caption{Running average (500-step) of Rewards in the Dynamic Network Packet Routing Environment. Shaded areas show standard deviation.}
    \label{fig:dynamic-controller-rewards}
\end{figure}

\begin{figure}[H]
    \centering
    \includegraphics[width=0.9\linewidth]{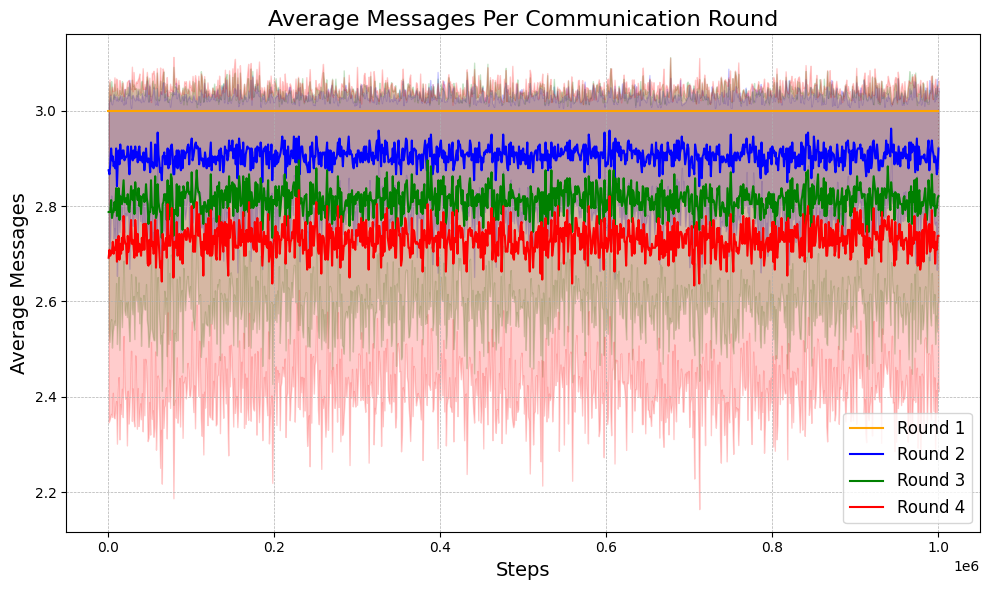}
    \caption{Average Messages per Round (4 communication rounds) for the Iteration Controller. Shaded areas represent the standard deviation.}
    \label{fig:messages-dynamic-controller}
\end{figure}

Moreover, the volatility in average messages per round is notably higher in the dynamic packet routing environment. Each round tends to stabilise around its starting point, suggesting that the environment's instability hampers the Iteration Controller's ability to effectively prioritise the importance of shared neighbouring node states, as illustrated in Figure \ref{fig:messages-dynamic-controller}.

\subsubsection{Routing Performance}

\begin{figure}[H]
    \centering
    \includegraphics[width=1\linewidth]{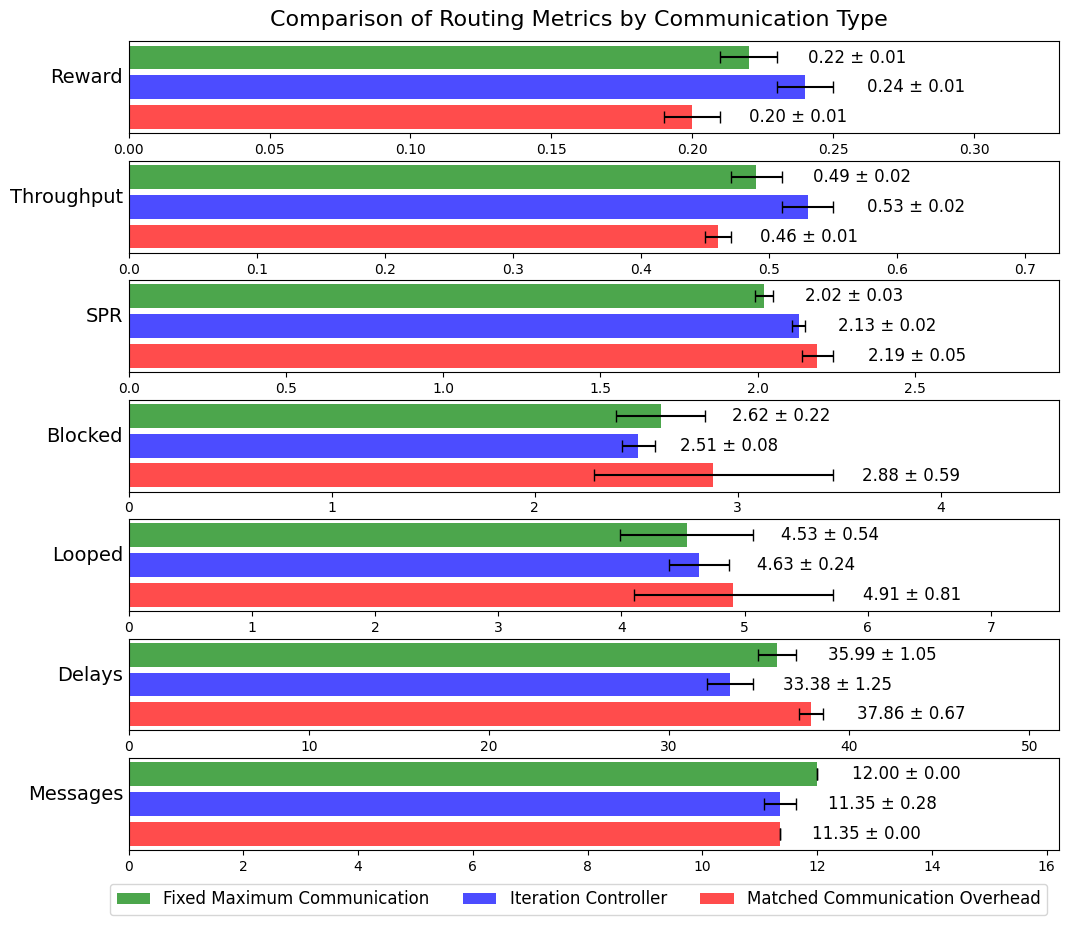}
    \caption{Routing metrics in the Dynamic Network Packet Routing environment, with each subplot having its own x-axis. Error bars represent standard deviations.}
    \label{fig:dynamic-controller-test-metrics}
\end{figure}

Despite a 5.4\% reduction in Messages (overhead), the Iteration Controller outperformed the Maximum Communication baseline, achieving 9.1\% higher rewards on the test dataset. This trend is consistent across key network metrics, with the Iteration Controller resulting in 8.2\% higher throughput and 7.3\% lower delays. The likely reason is that the Iteration Controller, trained without a secondary objective to reduce communication, achieves these reductions based solely on performance considerations, thereby enhancing the model's robustness and generalisability.

\vspace{12pt}

Interestingly, when examining cooperative metrics like Blocked and Looped packets, the Iteration Controller results in 4.2\% fewer blocked packets but 2.2\% more looped packets. Reducing blocked packets requires efficient bandwidth management and cooperation between nodes, achievable through effective information propagation. However, dynamic communication control might occasionally miss critical information, leading to looped packets, which can adversely affect the shortest path ratio.

\vspace{12pt}

A noteworthy finding is the impact on variability. The variation in routing metrics for the Iteration Controller in the dynamic environment is significantly lower than that of the Maximum Communication and Matched Communication baselines, indicating that dynamic communication control in a non-stationary environment results in a more stable learning process.

\section{Summary of Results}

This research has demonstrated the effectiveness of the Iteration Controller in improving the quality of learned graph representations and routing performance in dynamic environments while reducing communication overhead in a distributed multi-round message-passing system. The Iteration Controller was evaluated against two baselines: one using the maximum allowable communication volume to assess communication reduction, and another matching the Iteration Controller's communication volume to isolate performance improvements.

\vspace{12pt}

Key challenges in implementing the Iteration Controller were identified and successfully addressed. Over-optimising for early-round communication was mitigated by adjusting the bias of the Iteration Controller's final feed-forward layer, and inconsistencies due to insufficient exploration were resolved by introducing exploration noise to the sigmoid input.

\vspace{12pt}

In the Shortest Path Regression task, the Iteration Controller outperformed the Matched Communication baseline at all communication volumes and even surpassed the Maximum Communication baseline at the trained sequence length while reducing overhead by up to 8.3\%. This highlights the Iteration Controller's ability to learn higher-quality graph representations through effective information propagation. Notably, it significantly reduced communication in later rounds, demonstrating its effectiveness in systems with higher communication volumes.

\vspace{12pt}

In the Dynamic Network Packet Routing environment, the Iteration Controller achieved a 9.1\% increase in rewards while reducing communication by 5.4\% compared to the maximum communication baseline. It also showed less variability across routing metrics, suggesting a more stable learning process in dynamic settings, making it suitable for real-world online learning applications. However, dynamic communication control occasionally missed critical information, leading to 2.2\% more looped packets, negatively affecting the shortest path ratio.

\chapter{Dynamic Communication System}
\label{cha:integrated-dynamic-communication}
Chapters \ref{cha:aggregation-mechanism} and \ref{cha:iteration-controller} demonstrated significant performance improvements when each component was evaluated in isolation. The final objective is to assess how the overall system's performance compares with other leading communication-based MARL approaches when tested in the dynamic network packet routing environment.

\section{Methodology}

We use the standard dynamic network packet routing configuration from Chapter \ref{cha:environment-implementation}, allowing up to four communication rounds per step to highlight the potential of the multi-round communication system. Routing performance is evaluated using key network metrics and its impact on communication overhead.

\subsubsection{Baselines}

The selected baselines represent leading decentralised communication-based MARL approaches, all centred on the DQN architecture. Except for NetMon, these methods use a single model without decoupling node and agent observations. We initially considered a centralised agent for comparison, but the RAM requirements for a 20-node network exceeded our resources. Baseline configurations were carefully chosen for fairness; for example, DGN uses four stacked convolutional layers to match the four communication rounds used by NetMon. Full configuration details are provided in the appendix.

\begin{longtable}{p{0.15\textwidth} p{0.8\textwidth}}
\caption{Baseline MARL approaches tested against the communication system.} \\
\toprule
\textbf{Baseline} & \textbf{Description} \\ 
\midrule
\endfirsthead

\toprule
\textbf{Baseline} & \textbf{Description} \\ 
\midrule
\endhead

\bottomrule
\endfoot

\textbf{DQN} \cite{mnih2013playing} & A fully decentralised Deep Q-Network where agents independently learn policies using Q-learning based on local observations.\\

\textbf{DRQN} \cite{hausknecht2015deep} & A fully decentralised Deep Recurrent Q-Network (DRQN) extending DQN with an LSTM layer for memory.\\

\textbf{CommNet} \cite{sukhbaatar2016learning} & Extends DRQN with dense inter-agent communication via a centralised, differentiable mechanism. It employs four communication rounds, aggregating messages by adding a node's hidden state to the mean of its neighbours' hidden states.\\

\textbf{DGN} \cite{jiang2018graph} & Decentralised agents communicate through four stacked self-attention layers, with a Q-Network applied to the final layer to guide policy decisions. Consistent with the original implementation, we use 8 attention heads and a key and value size of 16.\\

\textbf{NetMon} \cite{weil2024towards} & The original configuration used as the basis of our design. Uses summation for aggregation and four fixed communication rounds. \\

\end{longtable}

\section{Results and Discussion}

\subsection{Dynamic Network Packet Routing}

\subsubsection{Training Behaviour and Trends}

\begin{figure}[H]
    \centering
    \includegraphics[width=0.9\linewidth]{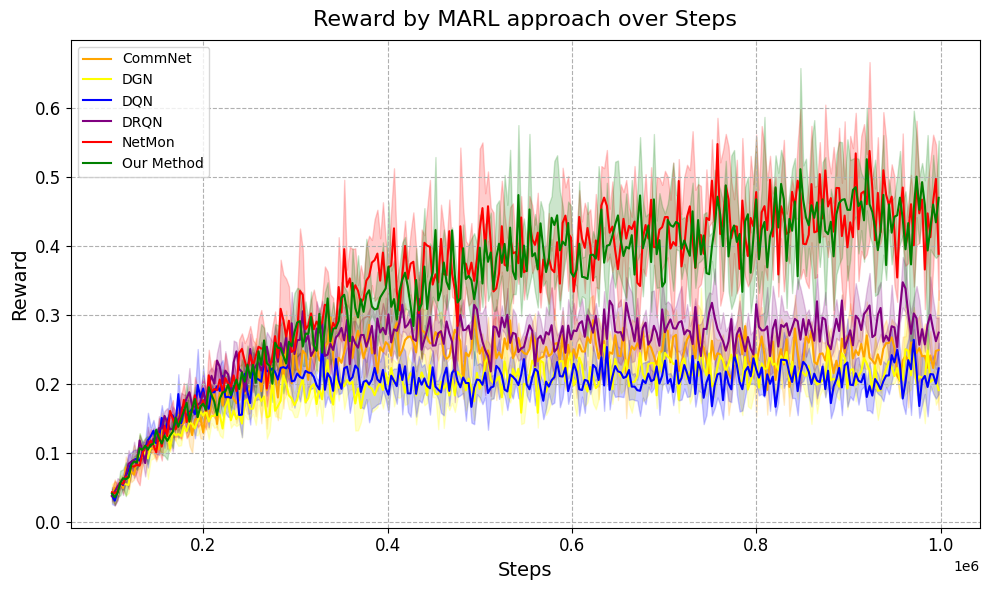}
    \caption{Rewards over 1,000,000 steps in the Dynamic Network Packet Routing Environment. The shaded areas represent the standard deviation.}
    \label{fig:dynamic-final-rewards}
\end{figure}

\vspace{-6pt}

Decoupling node and agent observations shows clear benefits as evidenced by the superior performance of both NetMon and our method. Interestingly, inter-agent communication had minimal impact, with DGN and CommNet converging to similar rewards as the fully decentralised methods. This suggests that a single model struggles to learn effective communication in highly non-stationary environments. Notably, DRQN achieved the highest reward among single-model approaches, underscoring the value of recurrent networks in partially observable environments.

\begin{figure}[H]
    \centering
    \includegraphics[width=0.9\linewidth]{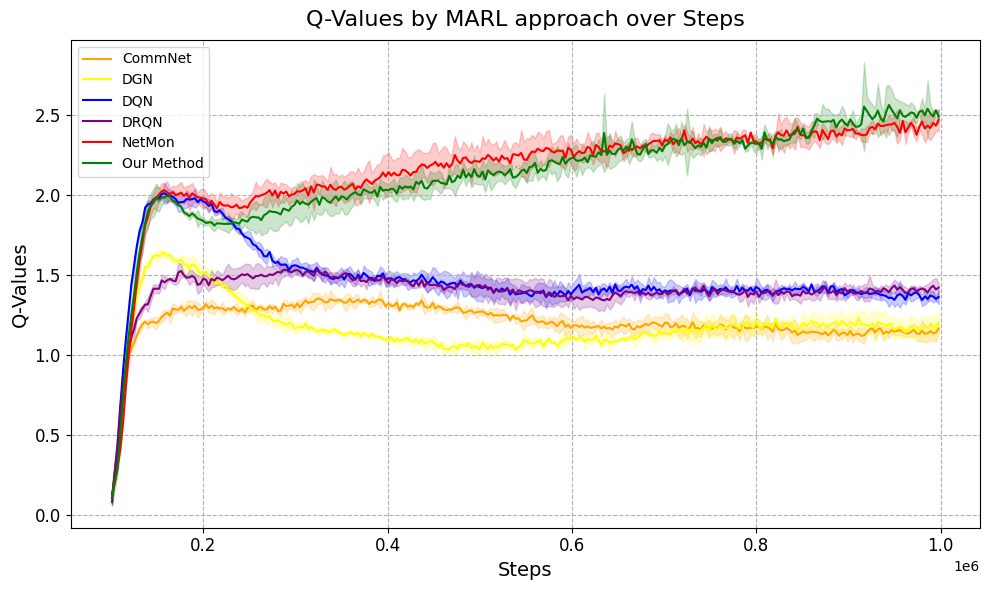}
    \caption{Behaviour Network Q-values over 1,000,000 steps in the Dynamic Network Packet Routing Environment. Shaded areas represent the standard deviation}
    \label{fig:dynamic-final-q-values}
\end{figure}

\vspace{-6pt}

These findings are supported by the Q-values of the behaviour network, which provide a stable leading indicator of performance through the agent model's assessment. Both NetMon and our proposed system continue to learn beyond the 1,000,000-step mark, clearly outperforming other approaches. Additionally, our method achieves slightly higher Q-values than the base NetMon configuration, suggesting the potential of our system.

\subsubsection{Routing Performance}

These results are confirmed on the test dataset, where the proposed method and NetMon consistently outperform other baselines across nearly all routing metrics. Notably, even the base NetMon configuration achieves 129\% higher throughput, 133\% higher rewards, and a 49\% reduction in delay compared to DRQN, the next best-performing baseline.

\addvspace{12pt}

\begin{figure}
    \centering
    \includegraphics[width=0.86\linewidth]{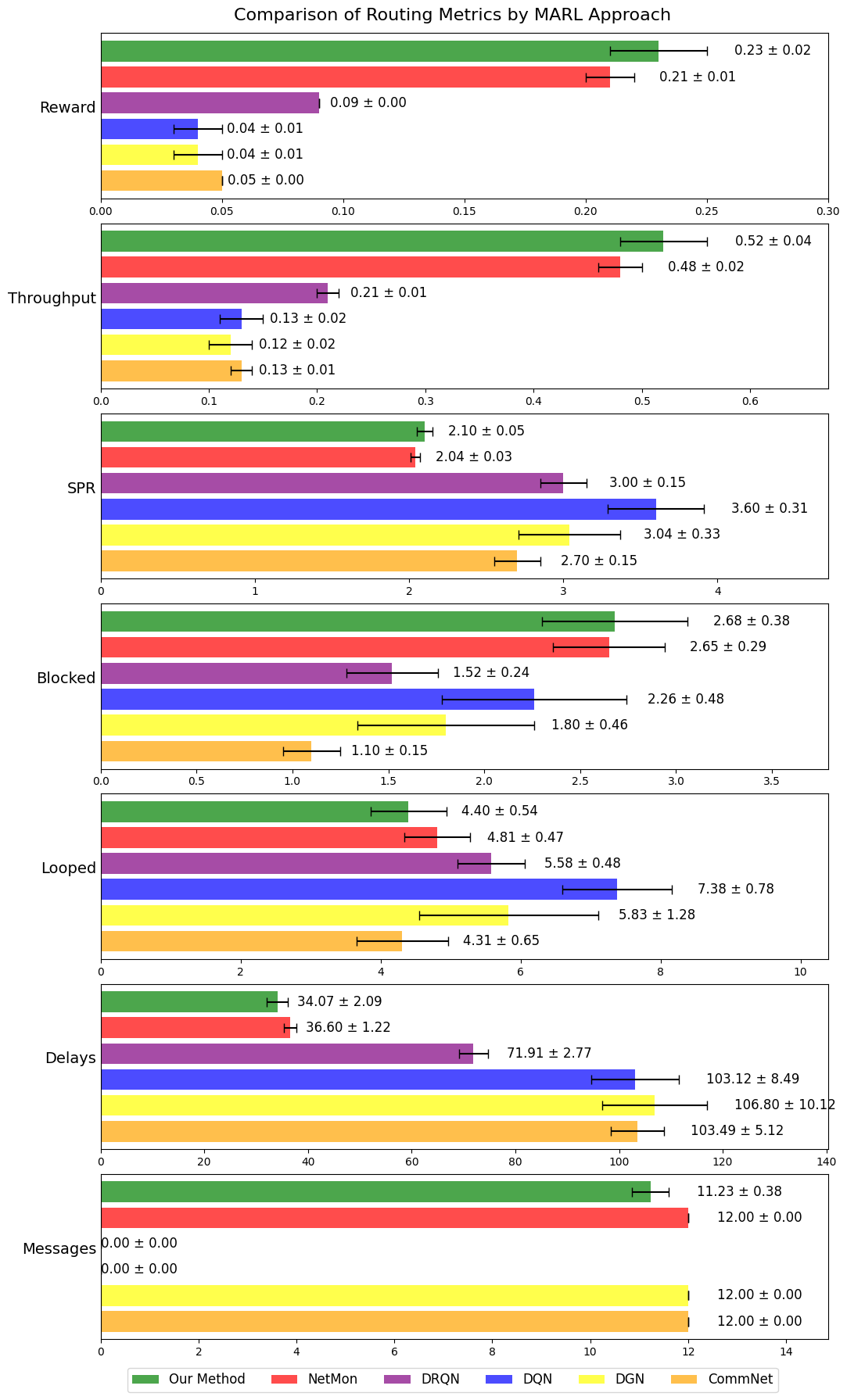}
    \caption{Routing metrics in the Dynamic Network Packet Routing environment, with each subplot having its own x-axis. Error bars represent standard deviations.}
    \label{fig:dynamic-final-test-metrics}
\end{figure}

An exception to this trend is the number of blocked packets. Due to their limited visibility over the network, the DQN, DRQN, DGN, and CommNet models focused on the more straightforward, reward-dense task of reducing blocked packets which is achievable with only local knowledge. CommNet demonstrated the strongest inter-agent communication, resulting in the lowest number of blocked and looped packets. However, this advantage came at the cost of reduced throughput and increased delay.

\vspace{12pt}

The DGN model significantly underperformed in this environment, even falling behind the fully decentralised DRQN. This under-performance may be attributed to two factors: the rigidity and centralisation imposed by stacked convolutional layers, which are ill-suited for a dynamic environment with node failures, and the lack of an RNN, which is essential for capturing temporal dependencies.

\vspace{12pt}

The proposed communication system significantly outperformed the base NetMon configuration, achieving a 9.5\% higher reward, 8.3\% greater throughput, a 6.9\% reduction in delays, and an 8.5\% decrease in looped packets, all while using 6.4\% less communication. Although the base NetMon had a slightly lower shortest path ratio (2.9\%) and fewer blocked packets (1.1\%), these differences are relatively minor. These results highlight the effectiveness of integrating the GAT aggregation mechanism and the Iteration Controller, which individually increased rewards by 4.8\% and 9.1\%, respectively, and demonstrated a powerful synergy when combined, leading to a 9.5\% higher reward.

\section{Summary of Results}

This research corroborated the findings of Weil et al. \cite{weil2024towards}, who demonstrated that decoupling the node and agent observation space and utilising a recurrent distributed message-passing model significantly stabilises the learning of graph representations, particularly in dynamic environments. Even the base NetMon configuration outperformed all other fully decentralised and inter-agent communication-based MARL approaches in the dynamic network packet routing environment, achieving a 133\% higher reward than the next best-performing baseline.

\vspace{12pt}

The next conclusion was that the proposed communication system designed specifically for dynamic environments which integrates the GAT aggregation mechanism and Iteration Controller, further optimises performance within the Dynamic Network Packet Routing environment.  The individual reward impacts of the GAT (4.8\% increase) and Iteration Controller (9.1\% increase) combined to yield a 9.5\% overall improvement relative to the base NetMon configuration. This translated into significant gains across key routing metrics, including throughput, delay, and looped packets, all while reducing communication overhead by 6.4\%.

\vspace{12pt}

This study highlights the potential of the dynamic communication system for effective and efficient information propagation in large-scale dynamic networks. Ultimately, this enables decentralised agents to gain a greater overview of the current state of the network, thereby allowing for more effective decision-making in applications such as network packet routing.




\chapter{Conclusion}
\label{cha:conclusion}
\section{Summarised Contributions \& Achievements}

This study advances the design of decentralised message-passing systems in dynamic network environments, particularly by enhancing the collective knowledge and communication efficiency within these networks. The research contributes to the development of more robust and high-performing network protocols, with a focus on improving scalability and adaptability within dynamic networks.

\begin{enumerate}

    \item Adapted the static network packet routing environment developed by Weil et al. \cite{weil2024towards} to include node failures with randomized failure probabilities and durations to more accurately represent a real practical network. 
    
    \item Demonstrated that a single GAT layer within the recurrent message passing model improves information propagation across a network, capturing more nuanced agent interactions and forming higher-quality graph representations.

    \item Verified that by decoupling the node and agent observation space, the GAT layer can be successfully trained end-to-end using reinforcement learning in a sparse reward, dynamic network packet routing environment. 

    \item Showed that by forming higher-quality graph representations, the GAT layer fosters greater inter-agent cooperation, resulting in an 11.9\% reduction in looped packets, contributing to a 4.8\% increase in routing performance.
    
    \item Introduced a novel multi-round communication targeting mechanism, called the Iteration Controller, tailored for large-scale dynamic networks, achieving a 9.1\% reward increase and 6.4\% reduction in communication.

    \item Extensively evaluated the Iteration Controllers' performance at varying overhead levels, revealing that later communication rounds are less critical, suggesting greater potential in high overhead systems.
    
    \item Established that the performance advantages of the GAT layer and Iteration Controller have a cumulative effect within the dynamic network packet routing environment, ultimately achieving a 9.5\% reward increase whilst using 6.4\% less communication.

\end{enumerate}

\section{Limitations \& Future Work}

While this study has made significant advancements, several limitations were encountered that may affect the generalisability and applicability of the results. However, these challenges also open up new opportunities for future research, paving the way for enhancing the robustness and practicality of the proposed methods in more complex and realistic scenarios.

\begin{itemize}

    \item \textbf{Network Size:}  Resource constraints limited our experiments to 20 nodes, which may reduce the applicability of our findings to larger, real-world networks with thousands of nodes. Future work should focus on scaling the framework to accommodate larger networks, addressing the challenges associated with increased computational and memory requirements.

    \item \textbf{Centralised Comparison:} Limited RAM prevented evaluation against a centralised model, likely representing peak performance. To fully assess the framework's efficiency, future studies should include comparisons with centralised models, leveraging more powerful hardware to provide a comprehensive evaluation.

    \item \textbf{Synchronous Communication:} The simulations assumed synchronous communication, which does not accurately reflect real-world network conditions that are subject to latency and potential data loss.  Future work should explore integrating an asynchronous system by implementing stop condition mechanisms \cite{liu2020when2com}.
   
    \item \textbf{Computational Overhead:} The GAT and Iteration Controller increase computational demands during training and inference. Future studies should balance performance gains with the added computational costs, especially in time-sensitive applications like network packet routing.

    \item \textbf{Unified Attention Mechanism:} Currently, the GAT and Iteration Controller apply MHA sequentially. Future research could explore a unified approach, where MHA is performed once with different attention heads allocated to each task, potentially reducing overall time complexity through parallel processing.
    
    \item \textbf{Heterogeneous Networks:} Future work should explore GAT applications in heterogeneous networks, where parameter sharing across nodes is not feasible. Investigating GAT performance in these complex environments would improve its relevance and applicability to real-world scenarios.

    \item \textbf{Bandwidth Reduction:}  Previous studies \cite{liu2020when2com} have shown that attention-based aggregation mechanisms enable significant compression of the transmitted key vector without compromising performance. Future work should explore combining these compression techniques within the communication framework. 

    \item \textbf{Convergence Conditions}: Future work should also explore proving that increased collective knowledge of the network will lead to convergence and identify the necessary conditions.  
    
\end{itemize}

\bibliographystyle{plainnat}

{\small 
\setstretch{0.9} 
\bibliography{bibliography}
} 
\chapter*{Appendix} 
\addcontentsline{toc}{chapter}{Appendix} 
\markboth{Appendix}{}  
\renewcommand{\thesection}{\Alph{section}}
\setcounter{section}{0}

\setcounter{table}{0}
\renewcommand{\thetable}{\Alph{section}.\arabic{table}}

\section{Test Graphs}

\begin{figure}[H]
    \centering
    \begin{minipage}[t]{0.48\linewidth}
        \centering
        \includegraphics[width=\linewidth]{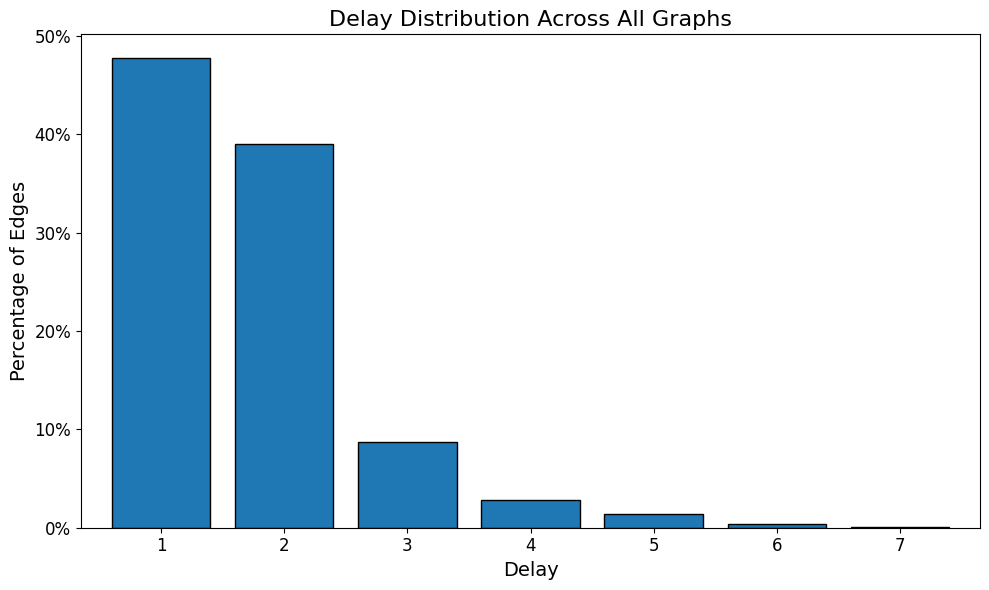}
        \caption{Distribution of edge lengths (delays) across the 1000 test graphs for the Dynamic Network Packet Routing Environment}
        \label{fig:edge-lengths-distribution}
    \end{minipage}
    \hfill
    \begin{minipage}[t]{0.48\linewidth}
        \centering
        \includegraphics[width=\linewidth]{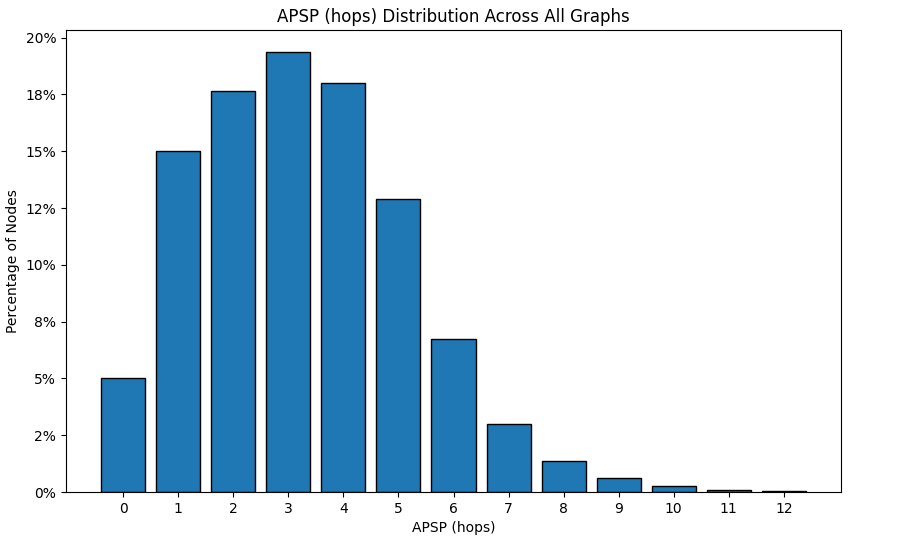}
        \caption{Distribution of APSP (hops) across the 1000 test graphs for the Dynamic Network Packet Routing Environment}
        \label{fig:apsp-hops-distribution} 
    \end{minipage}
\end{figure}

\clearpage 
\vspace*{-12pt}

\section{Environment Configuration}

\subsection{Shortest Path Regression}

\begin{table}[H]
\centering
\caption{Training Configurations for the Shortest Path Regression task}
\begin{tabular}{p{8cm} p{6.5cm}<{\centering}}
\toprule
\textbf{Configuration} & \textbf{Value} \\
\midrule
Training Graphs & 99,000 \\
Training Iterations & 100,000 \\
Validation Graphs & 1,000 \\
Validation Frequency & 1,000 iterations \\
Learning Rate & 0.001 \\
Optimizer & AdamW \cite{loshchilov2017decoupled} \\
Batch Size & 32 \\
Loss Function & MSE \\
\bottomrule
\end{tabular}
\label{tab:training-config-spr}
\end{table}

\begin{table}[H]
\centering
\caption{Evaluation Configurations for Shortest Path Regression}
\begin{tabular}{p{8cm} p{6.5cm}<{\centering}}
\toprule
\textbf{Configuration} & \textbf{Value} \\
\midrule
Evaluation Graphs & 1,000 \\
Evaluation Seeds & 5 \\
Sequence Lengths & [1, 2, 4, 8, 16, 32, 64, 128, 256] \\
\bottomrule
\end{tabular}
\label{tab:evaluation-config-spr}
\end{table}

\clearpage 
\vspace*{-12pt}

\subsection{Dynamic Network Packet Routing}

\begin{table}[H]
\centering
\caption{Dynamic Modifications for Dynamic Network Packet Routing}
\begin{tabular}{p{8cm} p{6.5cm}<{\centering}}
\toprule
\textbf{Configuration} & \textbf{Value} \\
\midrule
Edge Bandwidth Limitation & 1 \\
Node Failure Probability & 20\% \\
Node Failure Duration (Min) & 5 steps \\
Node Failure Duration (Max) & 10 steps \\
Maximum Inactive Nodes & 40\% \\
Routing to Inactive Node Reward & -0.2 \\
\bottomrule
\end{tabular}
\label{tab:evaluation-config-dynamic}
\end{table}

\begin{table}[H]
\centering
\caption{Training Configuration for Dynamic Network Packet Routing}
\begin{tabular}{p{8cm} p{6.5cm}<{\centering}}
\toprule
\textbf{Configuration Parameter} & \textbf{Value} \\
\midrule
Packets Per Episode & 20 \\
Total Steps & 1,000,000 \\
Maximum Steps Per Episode & 50 \\
Replay Memory Size & 200,000 \\
Training Frequency & 10 steps \\
Training Iterations & 1 \\
Batch Size & 32 \\
Sequence Length ($J$) & 8 \\
\bottomrule
\end{tabular}
\label{tab:training-config-npr}
\end{table}

\begin{table}[H]
\centering
\caption{Exploration Strategy Configuration for Dynamic Network Packet Routing}
\begin{tabular}{p{8cm} p{6.5cm}<{\centering}}
\toprule
\textbf{Configuration Parameter} & \textbf{Value} \\
\midrule
Initial Exploration Steps & 100,000 \\
Initial Exploration Rate ($\epsilon$) & 1.00 \\
Epsilon Decay Rate & 0.999 every 100 steps \\
Minimum Exploration Rate ($\epsilon$) & 0.01 \\
\bottomrule
\end{tabular}
\label{tab:exploration-config-npr}
\end{table}

\begin{table}[H]
\centering
\caption{Reward Shaping Configuration for Dynamic Network Packet Routing}
\begin{tabular}{p{8cm} p{6.5cm}<{\centering}}
\toprule
\textbf{Condition} & \textbf{Reward / Penalty} \\
\midrule
Packet reaches destination & +10 \\
Packet blocked due to bandwidth & -0.2 \\
Packet routed to inactive node & -0.2 \\
\bottomrule
\end{tabular}
\label{tab:reward-shaping-config}
\end{table}

\begin{table}[H]
\centering
\caption{Evaluation Configurations for Dynamic Network Packet Routing}
\begin{tabular}{p{8cm} p{6.5cm}<{\centering}}
\toprule
\textbf{Configuration} & \textbf{Value} \\
\midrule
Evaluation Graphs & 1,000 \\
Evaluation Seeds & 5 \\
Packets Per Episode & 20 \\
Maximum Steps Per Episode & 300 \\
\bottomrule
\end{tabular}
\label{tab:evaluation-config-npr}
\end{table}

\clearpage 

\vspace*{-12pt}

\section{Design Configuration}

\begin{table}[H]
\centering
\caption{Full Configuration Parameters for Dynamic Communication System}
\begin{tabular}{p{8cm} p{6.5cm}<{\centering}}
\toprule
\textbf{Configuration} & \textbf{Value} \\
\midrule
Activation Function & Leaky ReLU \\
Aggregation Type & GAT \\
Communication Bias & 0.5 \\
Communication Rounds & 4 \\
Discount Factor ($\gamma$) & 0.9 \\
Encoder Dimensions & 256,128 \\
Epsilon Decay & 0.999 \\
Epsilon Update Frequency & 100 \\
Initial Epsilon & 1 \\
Iteration Controller Attention Heads & 4 \\
Learning Rate & 0.001 \\
Mini Batch Size & 32 \\
Agent Model & DQN \\
Noise Scaling & 0.3 \\
Replay Buffer Capacity & 100000 \\
RNN Hidden State Dimensions & 64 \\
RNN Type & GRU \\
RNN Unroll Depth & 8 \\
Steps Before Training & 100000 \\
Steps Between Training & 10 \\
Target Network Update Frequency ($\tau$) & 0.01 \\
Target Update Steps & 0 \\
\bottomrule
\end{tabular}
\label{tab:configurations-lilac-leaf-11}
\end{table}

\clearpage 

\vspace*{-12pt}

\section{Baseline Configuration}

The following table lists the configuration parameters for the baselines used in Chapter \ref{cha:integrated-dynamic-communication}. $d_o$ represents the number of observation dimensions, $D$ represents the node degree, therefore $ D + 1$ is size of the action space.

\begin{table*}[h]
\centering
\caption{Baseline Agent Architecture Configuration}
\begin{tabular*}{\textwidth}{@{\extracolsep{\fill}}llp{0.48\textwidth}@{}}
\toprule
\textbf{Architecture} & \textbf{Component} & \textbf{Configuration} \\ 
\midrule
\multirow{3}{*}{DQN}   & Input Layer        &  $(d_o, 512, 256)$  \\
                       & Activation Function     & Leaky ReLU \\
                       & Output Layer       &  $(256, D + 1)$ \\ 
\midrule
\multirow{4}{*}{DQNR}  & Input Layer        &  $(d_o, 512, 256)$ \\
                       & Activation Function     & Leaky ReLU \\
                       & Recurrent Layer    & LSTM, hidden size 256, cell state size 256 \\
                       & Output Layer       & $(256, D + 1)$ \\ 
\midrule
\multirow{4}{*}{CommNet} & Input Layer      & $(d_o, 512, 256)$ \\
                        & Activation Function     & Leaky ReLU \\
                       & Communication Rounds & 4 per step \\
                       & Aggregation     & Sum of agent hidden state and mean of neighbours’ hidden states \\ 
\midrule
\multirow{5}{*}{DGN}   & Input Layer        & $(d_o, 512, 256)$  \\
                       & Activation Function     & Leaky ReLU \\
                       & Attention Layers   & 2 layers, 8 attention heads, key and value size 16 \\
                       & Communication Rounds & 2 per step \\
                       & Output Layer       & Concatenation of attention output and observation, size $(3 \cdot 256, D + 1)$ \\ 
\bottomrule
\end{tabular*}
\end{table*}

\clearpage 

\vspace*{-12pt}

\section{Experiments}

\subsection{Aggregation Mechanism}

\vspace{12pt}

\subsubsection{Shortest Path Regression}

\vspace{-12pt}

\begin{longtable}{M{2.75cm} M{2.75cm} M{2.75cm} M{2.75cm} M{2.75cm}}
    \caption{(MSE of aggregation mechanisms across sequence lengths on the Shortest Path Regression task. Best performance at each length is highlighted in bold.}
    \label{tab:mse} \\
    \toprule
    \textbf{Sequence Length} & \textbf{Sum} & \textbf{Mean} & \textbf{GCN} & \textbf{GAT} \\
    \midrule
    \endfirsthead
    \caption[]{(continued)} \\
    \toprule
    \textbf{Sequence Length} & \textbf{Sum} & \textbf{Mean} & \textbf{GCN} & \textbf{GAT} \\
    \midrule
    \endhead
    \midrule
    \endfoot
    \bottomrule
    \endlastfoot
    1 & 1.12 ± 0.03 & 1.10 ± 0.01 & 1.30 ± 0.08 & \textbf{0.97 ± 0.05} \\
    2 & 0.43 ± 0.01 & 0.38 ± 0.01 & 0.47 ± 0.04 & \textbf{0.33 ± 0.03} \\
    4 & 0.30 ± 0.01 & 0.25 ± 0.01 & 0.35 ± 0.04 & \textbf{0.23 ± 0.04} \\
    8 & 0.29 ± 0.01 & 0.24 ± 0.01 & 0.34 ± 0.04 & \textbf{0.22 ± 0.05} \\
    16 & 0.33 ± 0.01 & 0.27 ± 0.00 & 0.36 ± 0.04 & \textbf{0.24 ± 0.03} \\
    32 & 0.483 ± 0.03 & 0.44 ± 0.04 & 0.480 ± 0.03 & \textbf{0.39 ± 0.02} \\
    64 & 1.26 ± 0.23 & 1.12 ± 0.07 & 1.04 ± 0.15 & \textbf{0.97 ± 0.09} \\
    128 & 3.34 ± 0.87 & 3.13 ± 0.35 & 2.82 ± 0.67 & \textbf{2.64 ± 0.79} \\
    256 & 7.88 ± 2.14 & 7.32 ± 2.01 & 7.97 ± 1.98 & \textbf{5.73 ± 2.44} \\
\end{longtable}

\clearpage 

\vspace*{-30pt}

\subsubsection{Dynamic Network Packet Routing}

\begin{figure}[H]
    \centering
    \begin{minipage}[t]{0.48\linewidth}
        \centering
        \includegraphics[width=\linewidth]{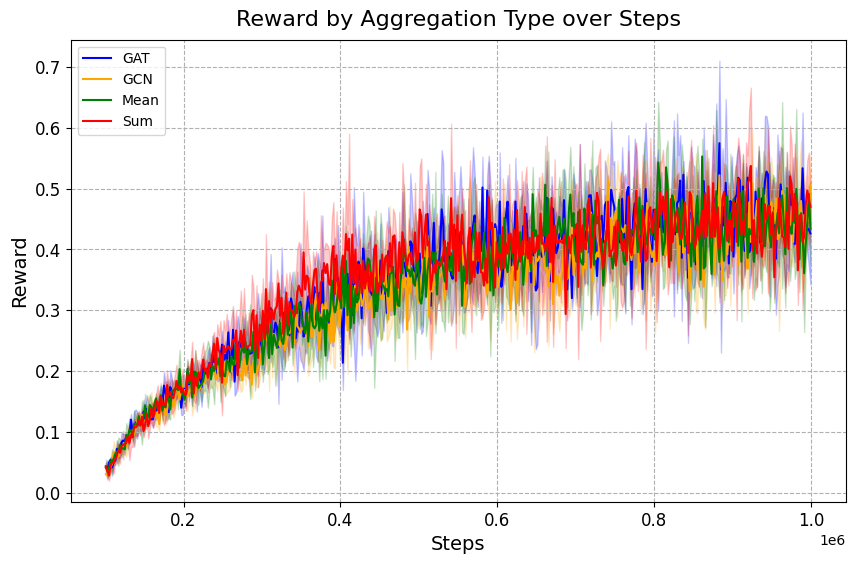}
        \caption{Rewards by Aggregation Mechanism in the Dynamic Network Packet Routing Environment. Shaded areas show standard deviation.}
        \label{fig:reward-aggregation}
    \end{minipage}
    \hfill
    \begin{minipage}[t]{0.48\linewidth}
        \centering
        \includegraphics[width=\linewidth]{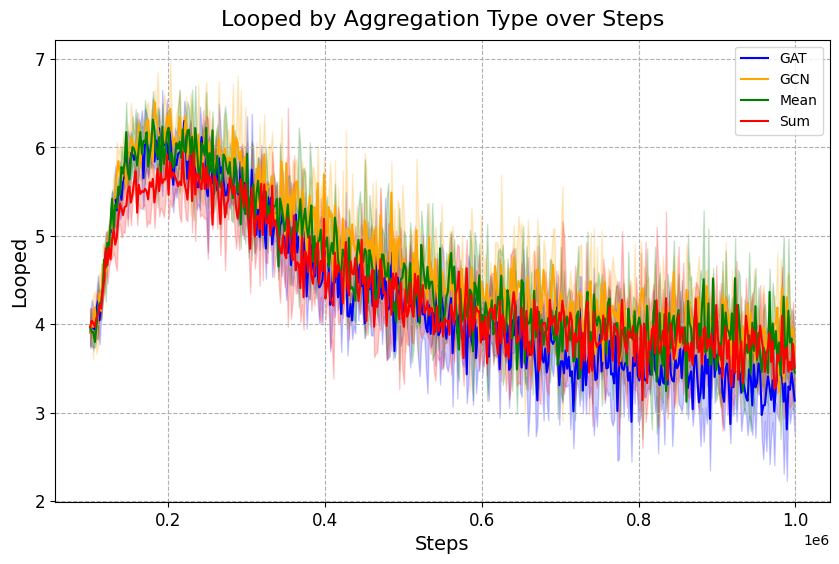}
        \caption{Looped packets by Aggregation Mechanism in the Dynamic Routing Environment. Shaded areas show standard deviation.}
        \label{fig:looped-aggregation}
    \end{minipage}
\end{figure}

\begin{figure}[H]
    \centering
    \begin{minipage}[t]{0.48\linewidth}
        \centering
        \includegraphics[width=\linewidth]{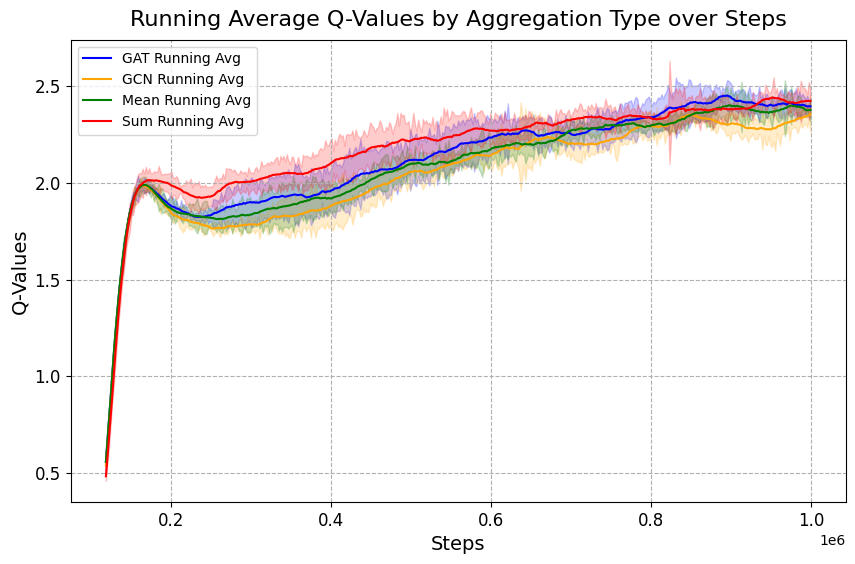}
        \caption{Running average (500-step) by Aggregation Mechanism of Q-Values in the Dynamic Routing Environment. Shaded areas show standard deviation.}
        \label{fig:q-values-aggregation}
    \end{minipage}
    \hfill
    \begin{minipage}[t]{0.48\linewidth}
        \centering
        \includegraphics[width=\linewidth]{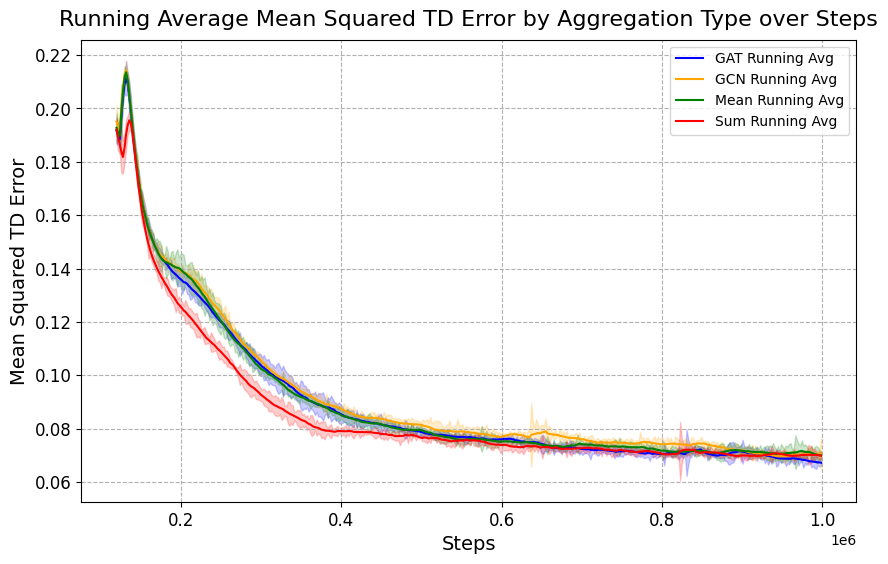}
        \caption{Running average (500-step) by Aggregation Mechanism of Loss in the Dynamic Routing Environment. Shaded areas show standard deviation.}
        \label{fig:td-error-aggregation}
    \end{minipage}
\end{figure}

\clearpage 

\vspace*{-12pt}

\subsection{Iteration Controller}

\subsubsection{Hyperparameter Testing}

\begin{figure}[H]
    \centering
    \begin{minipage}[t]{0.48\linewidth}
        \centering
        \includegraphics[width=\linewidth]{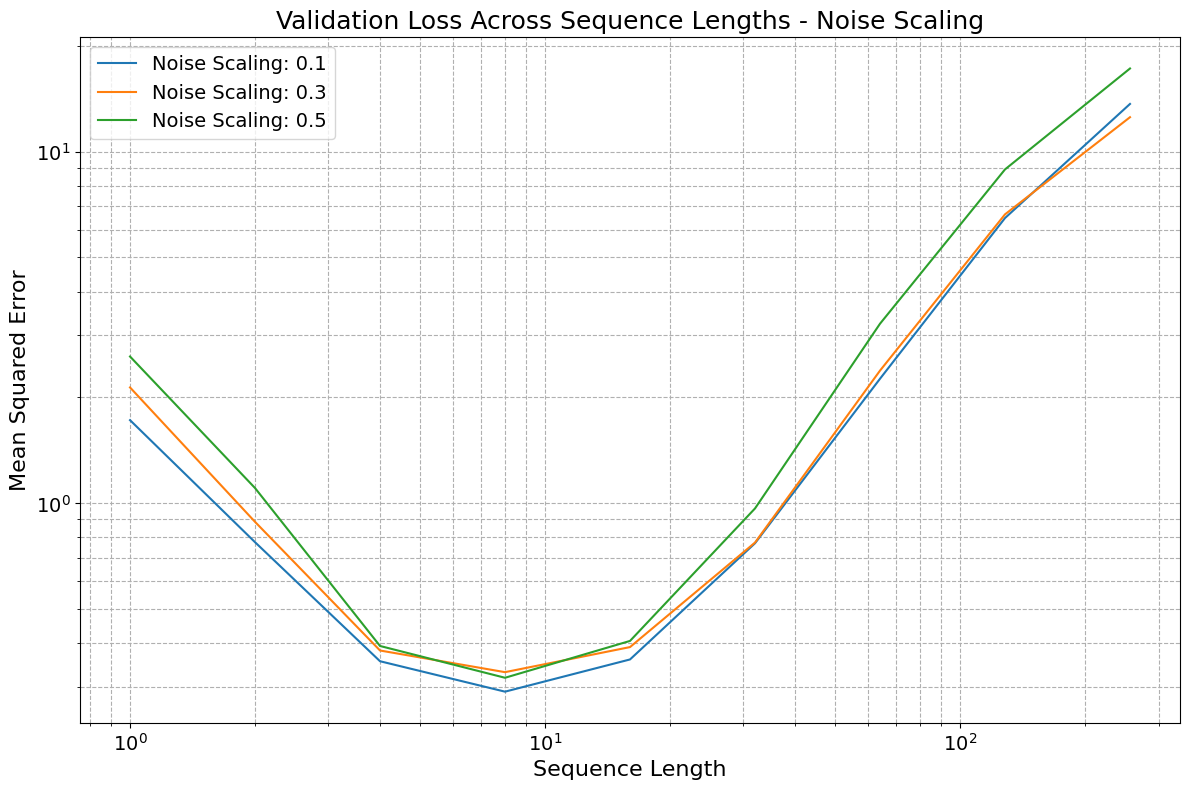}
        \caption{Validation MSE as a function of noise scaling. Lower noise scaling generally results in better performance.}
        \label{fig:test-regression-loss-noise-scaling}
    \end{minipage}
    \hfill
    \begin{minipage}[t]{0.48\linewidth}
        \centering
        \includegraphics[width=\linewidth]{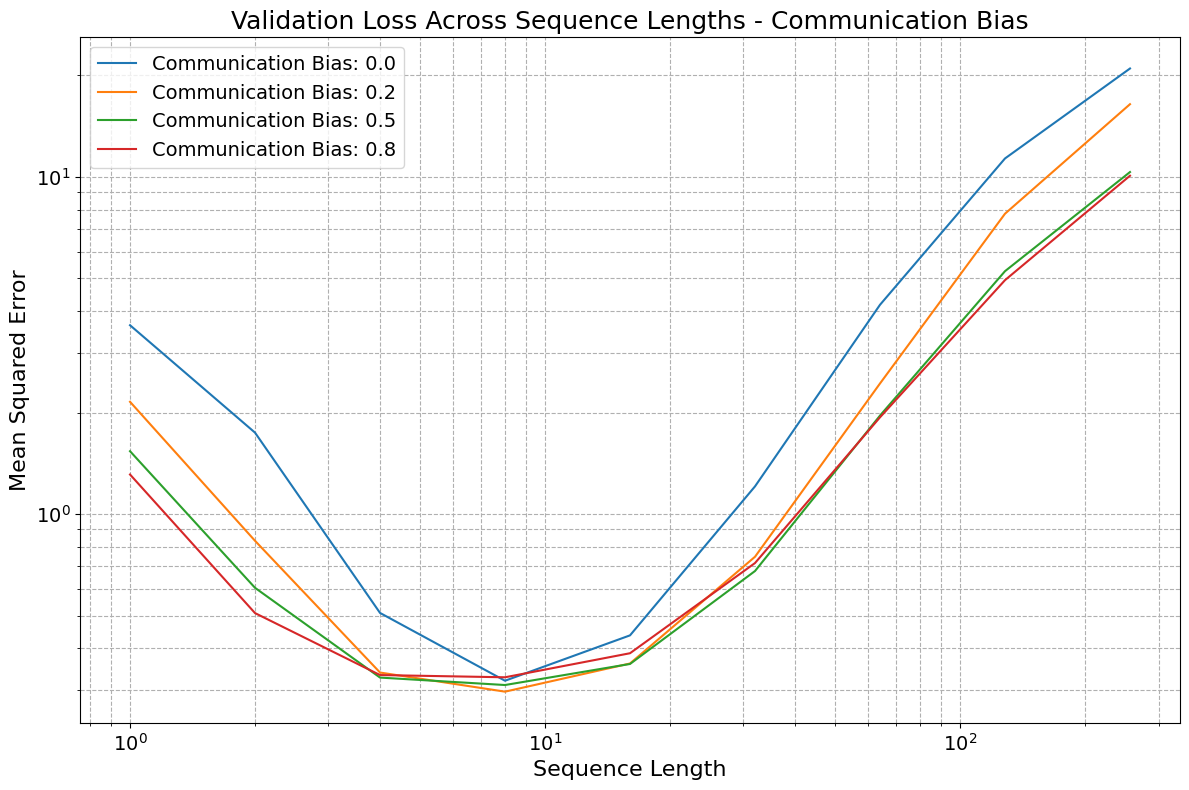}
        \caption{Validation Loss MSE as a function of communication bias. The plot illustrates the relationship between communication bias and performance.}
        \label{fig:test-regression-loss-comm-bias}
    \end{minipage}
\end{figure}

\subsubsection{Shortest Path Regression}

\begin{longtable}{M{2.5cm} M{3.7cm} M{3.7cm} M{3.7cm}}
    \caption{Mean Squared Error (MSE ± standard deviation) across sequence lengths for different communication strategies. Best performance at each sequence length is highlighted in bold.}
    \label{tab:mse_supervised_controller_comparison} \\
    \toprule
    \textbf{Sequence Length} & \textbf{Maximum Communication} & \textbf{Iteration Controller} & \textbf{Matched Communication} \\
    \midrule
    \endfirsthead
    \caption[]{(continued)} \\
    \toprule
    \textbf{Sequence Length} & \textbf{Maximum Communication} & \textbf{Iteration Controller} & \textbf{Matched Communication} \\
    \midrule
    \endhead
    \midrule
    \endfoot
    \bottomrule
    \endlastfoot
    1   & \textbf{0.99 ± 0.08}  & 1.50 ± 0.05  & 1.75 ± 0.13  \\ 
    2   & \textbf{0.30 ± 0.02}  & 0.481 ± 0.03  & 0.60 ± 0.07  \\
    4   & \textbf{0.24 ± 0.02}  & 0.26 ± 0.02  & 0.30 ± 0.05  \\
    8   & 0.24 ± 0.01  & \textbf{0.21 ± 0.02}  & 0.24 ± 0.04  \\
    16  & 0.25 ± 0.02  & \textbf{0.24 ± 0.03}  & 0.27 ± 0.03  \\ 
    32  & \textbf{0.38 ± 0.03}  & \textbf{0.38 ± 0.05}  & 0.42 ± 0.02  \\ 
    64  & 1.04 ± 0.26  & \textbf{0.83 ± 0.25}  & 0.85 ± 0.02  \\ 
    128 & 3.02 ± 0.64  & 2.22 ± 1.05  & \textbf{1.69 ± 0.01}  \\ 
    256 & 7.67 ± 1.28  & 4.28 ± 1.65  & \textbf{2.68 ± 0.12}  \\
\end{longtable}

\clearpage 

\vspace*{-12pt}

\subsubsection{Dynamic Network Packet Routing}

\begin{figure}[H]
    \centering
    \begin{minipage}[t]{0.48\linewidth}
        \centering
        \includegraphics[width=\linewidth]{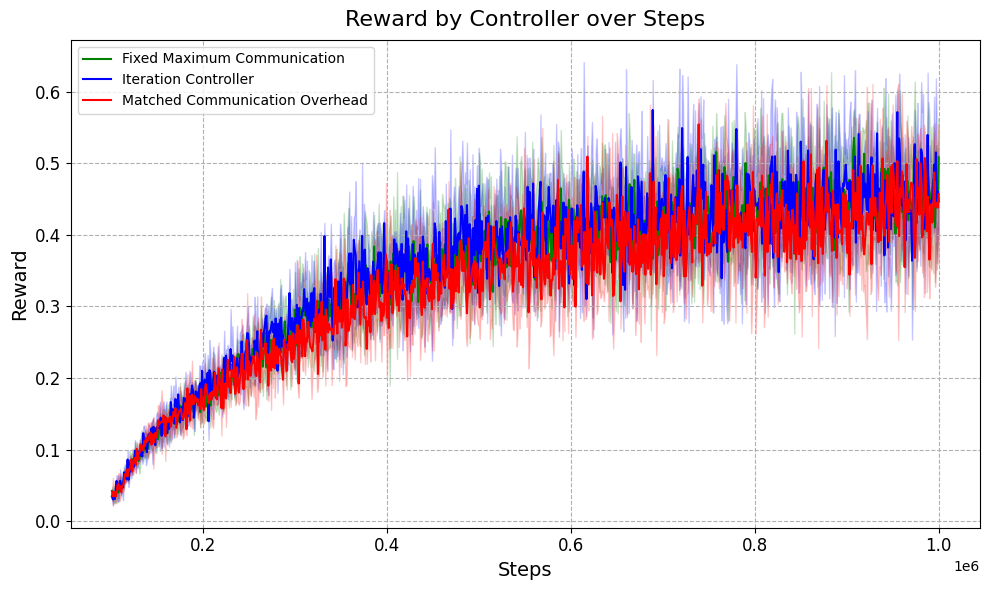}
        \caption{Rewards by Communication Type in the Dynamic Routing Environment. Shaded areas show standard deviation.}
        \label{fig:reward-routing}
    \end{minipage}
    \hfill
    \begin{minipage}[t]{0.48\linewidth}
        \centering
        \includegraphics[width=\linewidth]{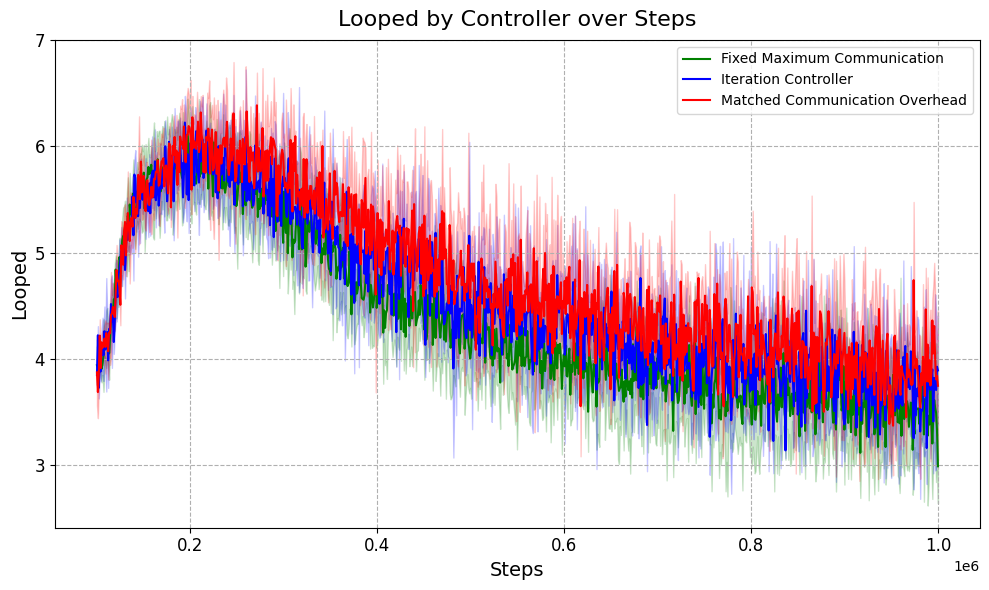}
        \caption{Looped packets by Communication Type in the Dynamic Routing Environment. Shaded areas show standard deviation.}
        \label{fig:looped-routing}
    \end{minipage}
\end{figure}

\begin{figure}[H]
    \centering
    \begin{minipage}[t]{0.48\linewidth}
        \centering
        \includegraphics[width=\linewidth]{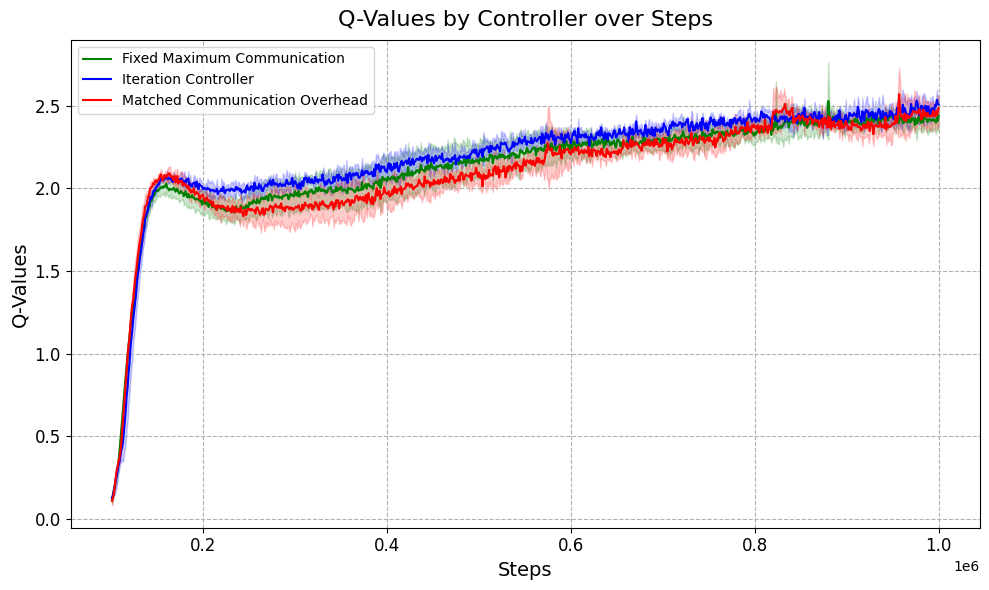}
        \caption{Running average (500-step) of Q-Values by Communication Type in the Dynamic Routing Environment. Shaded areas show standard deviation.}
        \label{fig:q-values-routing}
    \end{minipage}
    \hfill
    \begin{minipage}[t]{0.48\linewidth}
        \centering
        \includegraphics[width=\linewidth]{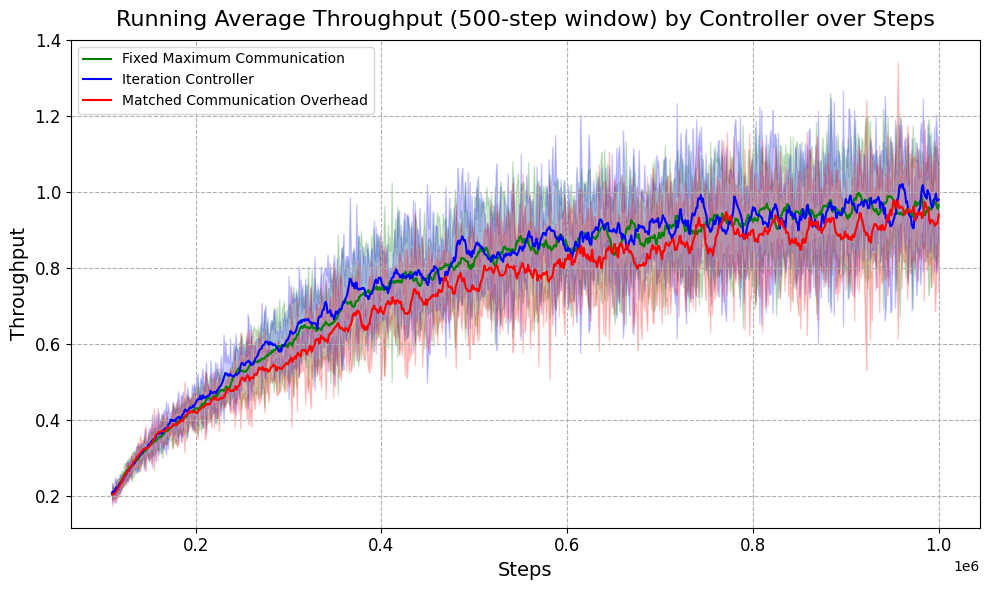}
        \caption{Running average (500-step) of Throughput by Communication Type in the Dynamic Routing Environment. Shaded areas show standard deviation.}
        \label{fig:throughput-routing}
    \end{minipage}
\end{figure}

\begin{figure}[H]
    \centering
    \includegraphics[width=0.48\linewidth]{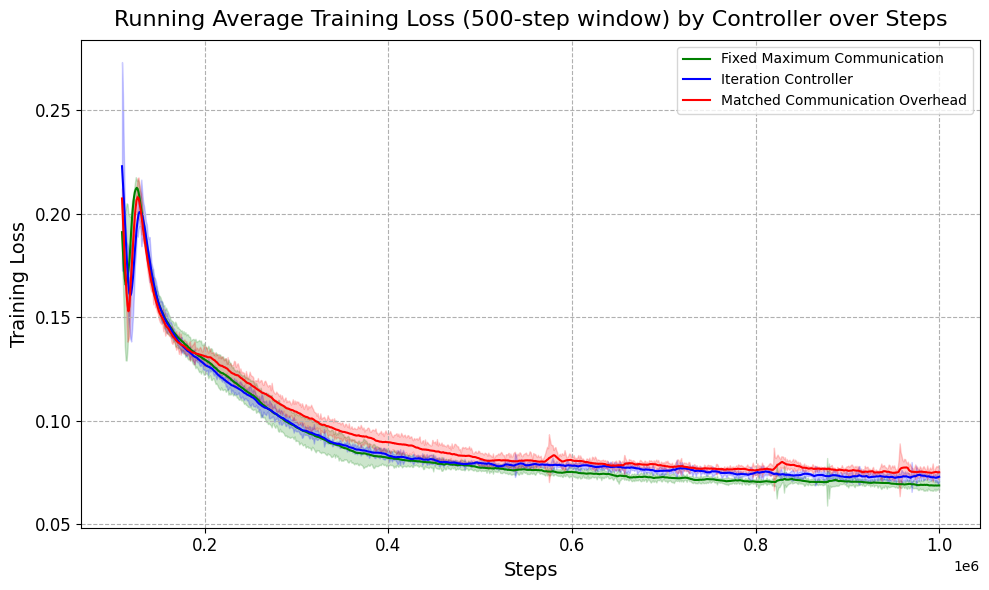}
    \caption{Running average (500-step) of Validation Loss in the Dynamic Routing Environment. Shaded areas show standard deviation.}
    \label{fig:training-loss-routing}
\end{figure}

\clearpage 

\vspace*{-12pt}

\subsection{Dynamic Communication System}

\subsubsection{Dynamic Network Packet Routing}

\begin{figure}[H]
    \centering
    \begin{minipage}[t]{0.48\linewidth}
        \centering
        \includegraphics[width=\linewidth]{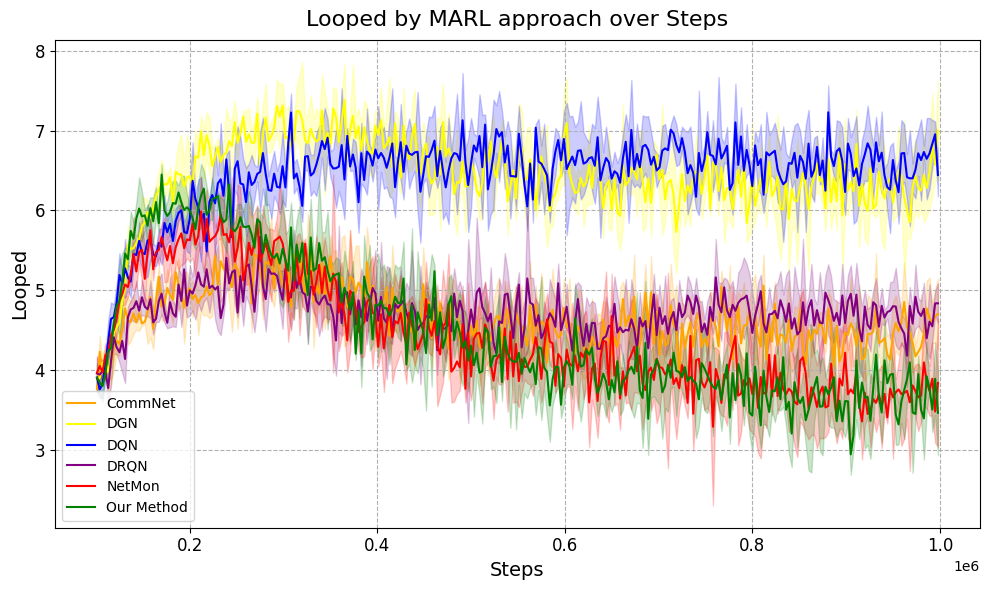}
        \caption{Overall System Looped Packets in the Dynamic Routing Environment. Shaded areas show standard deviation.}
        \label{fig:looped-final-routing}
    \end{minipage}
    \hfill
    \begin{minipage}[t]{0.48\linewidth}
        \centering
        \includegraphics[width=\linewidth]{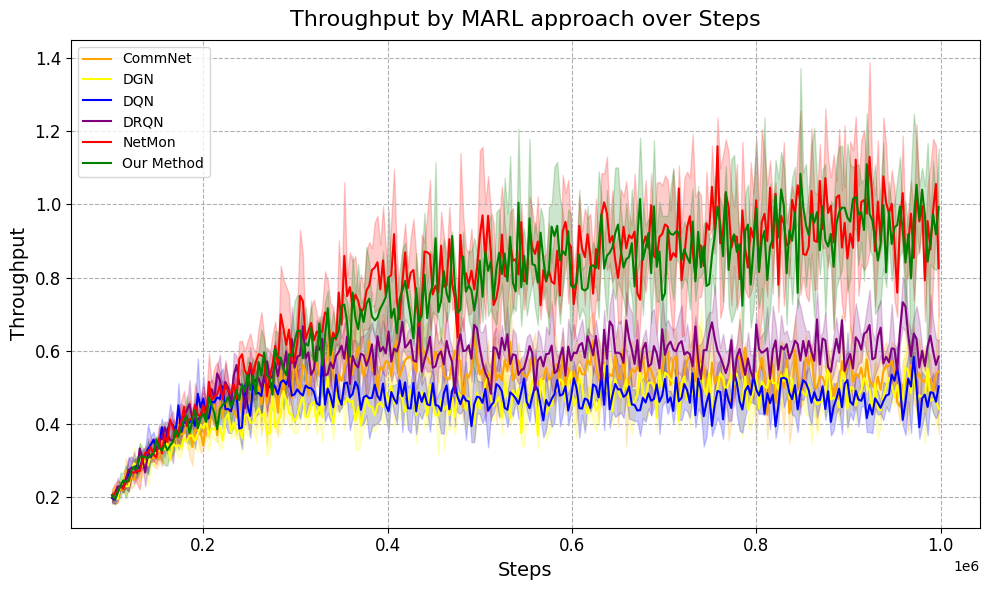}
        \caption{Overall System Throughput in the Dynamic Routing Environment. Shaded areas show standard deviation.}
        \label{fig:throughput-final-routing}
    \end{minipage}
\end{figure}

\begin{figure}[H]
    \centering
    \includegraphics[width=0.48\linewidth]{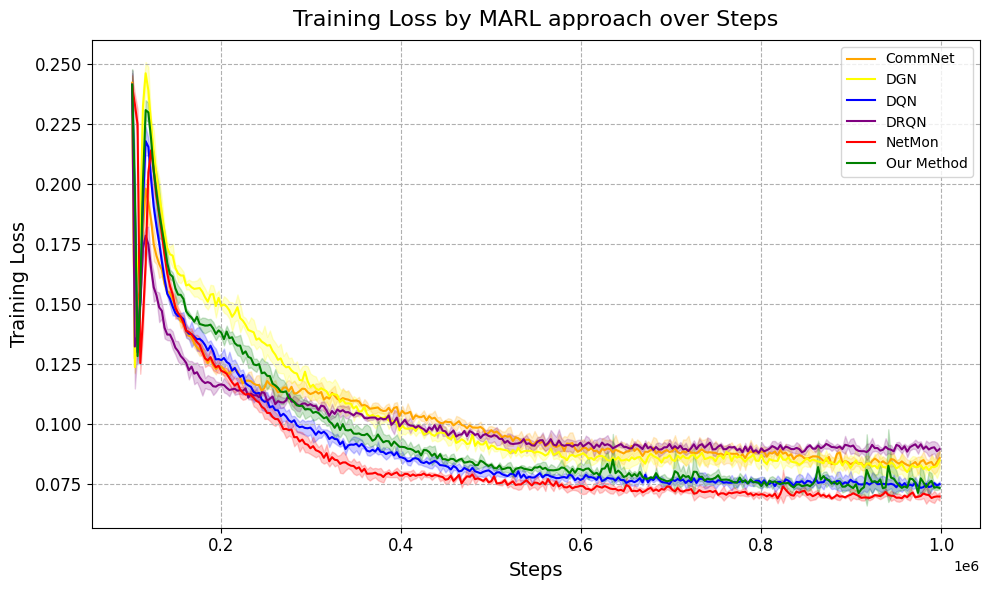}
    \caption{Overall System Validation Loss in the Dynamic Routing Environment. Shaded areas show standard deviation.}
    \label{fig:training-loss-final-routing}
\end{figure}
\end{document}